\documentclass[10pt,doublesided, doublecolumn,final]{IEEEtran}
\usepackage{float}
\usepackage{cite}
\usepackage{algorithm}
\usepackage[noend]{algorithmic}
\usepackage[geometry]{ifsym}
\usepackage {ifsym}
\usepackage{amssymb}
\usepackage{yhmath}
\usepackage{amscd}
\usepackage{dsfont}
\usepackage{amsmath}
\usepackage{epsfig}
\usepackage{graphics}
\usepackage{psfrag}
\usepackage{rotating}
\usepackage{amsmath}
\usepackage{amsfonts}
\usepackage{url}
\usepackage{color}
\usepackage{float}
\usepackage{balance} 
\usepackage{soul}

\newtheorem{lemma}{Lemma}

\newtheorem{theorem}{Theorem}
\newtheorem{definition}{Definition}

\newcommand{\bs}{\boldsymbol}

\newcommand{\ds}{\displaystyle}

\newcommand{\pr}[1]{\mathrm{Pr} \left[#1\right]}
\newcommand{\SNR}{\mathrm{SNR}}

\newcommand{\INR}{\mathrm{INR}}

\newcommand{\sfT}{\textsf{T}}

\newcommand{\Cldicnfb}{\mathcal{C} }

\newcommand{\Cgicnof}{\mathcal{C}_{\mathrm{G-IC-NF}}}
\newcommand{\agicnof}{\underline{\mathcal{C}}_{\mathrm{G-IC-NF}}}
\newcommand{\cgicnof}{\overline{\mathcal{C}}_{\mathrm{G-IC-NF}}}

\begin{document}
\title{Approximate Capacity Region of the Two-User Gaussian Interference Channel with Noisy Channel-Output Feedback
}
\author{Victor Quintero, Samir M. Perlaza, I\~naki Esnaola,  and Jean-Marie Gorce
\thanks{ Victor Quintero is with the Department of Telecommunications, Universidad del Cauca, 19003, Popay\'{a}n, Cauca,  Colombia. \newline
Samir M. Perlaza  and Jean-Marie Gorce are with the Laboratoire CITI, a joint laboratory between the Institut National de Recherche en Informatique et en Automatique (INRIA), the Universit\'{e} de Lyon and the Institut National de Sciences Apliqu\'{e}es (INSA) de Lyon. 6 Av. des Arts 69621 Villeurbanne, France. ($\lbrace$samir.perlaza, jean-marie.gorce$\rbrace$@inria.fr). \newline
I\~naki Esnaola is with the Department of Automatic Control and Systems Engineering, University of Sheffield, Mappin Street, Sheffield, S1 3JD, UK. (esnaola@sheffield.ac.uk). \newline
Samir M. Perlaza and I\~naki Esnaola are also with the Department of Electrical Engineering at Princeton University, Princeton, NJ, 08544, USA. \newline
This research was supported in part by the European Commission under Marie Sklodowska-Curie Individual Fellowship No. 659316; in part by the INSA Lyon - SPIE ICS Chair on the Internet of Things; and in part by the Administrative Department of Science, Technology, and Innovation of Colombia (Colciencias), under Grant No. 617-2013. \newline
Parts of this work were presented at the IEEE International Workshop on Information Theory (ITW), Jeju Island, South Korea, October, 2015 \cite{Quintero-ITW-2015}, and ITW, Cambridge, United Kingdom, September, 2016 \cite{Quintero-ITW-2016}. }
}
\maketitle

\begin{abstract}  
In this paper, the capacity region of the linear deterministic interference channel with noisy channel-output feedback (LD-IC-NF) is fully characterized. The proof of achievability is based on random coding arguments and rate splitting; block-Markov superposition coding; and backward decoding. The proof of the converse reuses some of the existing outer bounds and includes new ones obtained using genie-aided models.
Following the insight gained from the analysis of the LD-IC-NF, an achievability region and a converse region for the two-user Gaussian interference channel with noisy channel-output feedback (G-IC-NF) are presented.  Finally, the achievability region and the converse region are proven to approximate the capacity region of the G-IC-NF to within $4.4$ bits. 
\end{abstract}

 \section{Introduction}\label{SecIntroduction}
  
Recently, perfect feedback (PF) from the receivers to the corresponding transmitters has been shown to bring an unprecedented gain on the number of generalized degrees of freedom (GDoF) with respect to the case without feedback in the Gaussian interference channel (IC) \cite{Suh-TIT-2011}. Let $\mathcal{C}(\overrightarrow{\SNR}, \INR)$ denote a set containing all achievable rates of a symmetric Gaussian IC (G-IC) with parameters $\overrightarrow{\SNR}$ (signal to noise ratio in the forward link) and $\INR$ (interference to noise ratio). 
The number of GDoF \cite{Jafar-NOW-2010} is:
\begin{equation}
\mathrm{GDoF}(\alpha) = \hspace{-2mm} \lim_{\tiny 
\overrightarrow{\SNR} \rightarrow \infty
}
 \hspace{-2mm} 
 \frac{\sup\left\lbrace R : (R,R) \in \mathcal{C}(\overrightarrow{\SNR},  \overrightarrow{\SNR}^{\alpha}) \right\rbrace}{\log\left(\overrightarrow{\SNR}\right)} ,
\end{equation}
where $\alpha = \frac{\log(\INR)}{\log(\overrightarrow{\SNR})}$. 
In Figure~\ref{FigGDoFs}, the number of GDoF is plotted as a function of $\alpha$ when  $\mathcal{C}(\overrightarrow{\SNR}, \INR)$ is calculated without feedback \cite{Etkin-TIT-2008}; and  with PF from each receiver to their corresponding transmitters \cite{Suh-TIT-2011}. 
Note that with PF, $\mathrm{GDoF}(\alpha) \rightarrow \infty$ when $\alpha \rightarrow \infty$, which implies an arbitrarily large increment. 
Surprisingly, using only one PF link from one of the receivers to the corresponding transmitter provides the same sum-capacity as having four PF links from both receivers to both transmitters \cite{Sahai-ITW-2009, Sahai-TIT-2013, QPEG-TC-2018} in certain interference regimes.
These benefits rely on the fact that feedback provides relevant information about the interference. Hence, such information can be retransmitted to: $(a)$ perform interference cancellation at the intended receiver or $(b)$ provide an alternative communication path between the other transmitter-receiver pair.
These promising results are also observed when the system is decentralized, i.e., when each transmitter seeks to unilaterally maximize its own individual information rate \cite{Perlaza-TIT-2015, Quintero-PhD-2017}.

The capacity region of the G-IC with PF has been  approximated to within two bits in \cite{Suh-TIT-2011}. The achievability scheme presented therein is based on three well-known techniques: rate splitting \cite{Carleial-TIT-1978, Han-TIT-1981}, block-Markov superposition coding \cite{Cover-TIT-1981}, and backward decoding \cite{Willems-PhD-1982, Willems-TIT-1985}. The converse in \cite{Suh-TIT-2011} is obtained using classical tools such as cut-set bounds and genie-aided models. 
Other achievability schemes have been presented in \cite{Tuninetti-ISIT-2007} and \cite{Yang-Tuninetti-TIT-2011} using rate-splitting, block-Markov superposition coding, backward decoding, and binning/dirty paper coding in the context of a more general channel, i.e., G-IC with generalized feedback (IC-GF). 

From a system analysis perspective, PF might be an exceptionally optimistic model to study the benefits of feedback in the G-IC. Denote by $\overrightarrow{\bs{y}} = \left( \overrightarrow{y}_1, \overrightarrow{y}_2, \ldots, \overrightarrow{y}_N\right)$ a given sequence of $N$ channel outputs at a given receiver. 
A more realistic model of channel-output feedback is to consider that the feedback signal, denoted by $\overleftarrow{\bs{Y}}$, satisfies $\overleftarrow{\bs{Y}} = g\left( \overrightarrow{\bs{y}}\right)$ (random transformation in $\mathds{R}^{N}$). Hence, a relevant question is: what is a realistic assumption on $g$?
This question has been solved aiming to highlight different impairments  that feedback signals might go through. Some of these answers are discussed in the following sections.

\subsection{Rate-Limited Feedback:} Consider that the receiver produces the feedback signal using a deterministic transformation $g$, such that for a large $N$, a positive finite $C_F \in \mathds{R}$ and for all $\overrightarrow{\bs{y}} \in \mathds{R}^N$:
\begin{equation}\label{EqRateLimitedFB}
\overleftarrow{\bs{y}} = g(\overrightarrow{\bs{y}}) \in \mathcal{D}  \subseteq \mathds{R}^{N},
\end{equation}
where $\mathcal{D}$ is a codebook such that, 
\begin{equation}\label{EqC_F}
|\mathcal{D}| \leqslant 2^{N (C_F)}.
\end{equation} 
This model is known in literature as \emph{rate limited feedback}  (RLF) \cite{Vahid-TIT-2012, Ashraphijuo-Allerton-2014, Ashraphijuo-TIT-2016}, where $C_F$ is the capacity of the feedback link. The choice of the deterministic transformation $g$ subject to \eqref{EqC_F} is part of the coding scheme, i.e., the transformation $g$  takes the $N$ channel outputs observed during block $t>0$ and chooses a codeword in the codebook $\mathcal{D}$. The codeword is sent back to the transmitter during block $t+1$. From this standpoint, this model highlights the signal impairments derived from transmitting a signal with continuous support via a channel with finite-capacity.
Note that if $C_F = \infty$, then $g$ is the identity function and thus, 
\begin{equation}
\overleftarrow{\bs{y}} = g(\overrightarrow{\bs{y}}) = \overrightarrow{\bs{y}},
\end{equation} 
which is the case of PF \cite{Suh-TIT-2011}. When $C_F = 0$, then $|\mathcal{D}| = 1$ and thus, no information can be conveyed through the feedback links, which is the case studied in \cite{Han-TIT-1981, Chong-TIT-2008, Etkin-TIT-2008}. 
The main result in \cite{Vahid-TIT-2012} is twofold: first, given a fixed $C_F$, the authors provide a deterministic transformation $g$ using lattice coding  \cite{Zamir-Book-2014} and a particular power assignment such that partial or complete decoding of the interference is possible at the transmitter. 
An achievable region is presented using random coding arguments with rate splitting, block-Markov superposition coding, and backward decoding.
Second, the authors provide outer bounds that hold for any $g$ in  \eqref{EqRateLimitedFB}. This result determines a converse region whose sum-rate is shown to be within a constant gap of the achievable sum-rate, at least in the symmetric case. 
These results are generalized for the $K$-user G-IC with RLF in the symmetric case in \cite{Ashraphijuo-Allerton-2014, Ashraphijuo-TIT-2016}, where the analysis focuses on the fundamental limit of the symmetric rate. The main novelty on the extension to $K > 2$ users lies in the joint use of interference alignment and lattice codes for the proof of achievability. The proof of the converse remains an open problem when $K > 2$, even for the symmetric case.
\subsection{Intermittent Feedback} 
Assume that for all $n \in \lbrace 1,2 \ldots, N\rbrace$, the random transformation $g$ is such that given a channel output realization $\overrightarrow{y}_n$,
\begin{equation}\label{EqIFprobability}
\overleftarrow{Y}_n = \left\lbrace 
\begin{array}{rcl}
\star & \text{with probability} & 1-p\\
\overrightarrow{y}_n & \text{with probability} & p,
\end{array}
\right. 
\end{equation}
where $\star$ represents an erasure and $p \in [0,1]$. Note that the random transformation $g$ is fully  determined by the parameters of the channels, e.g., the parameter $p$. Thus, contrary to the RLF case, the transformation $g$ cannot be optimized as part of the receiver design. 
This model emphasizes the fact that the usage of the feedback link might be available only during certain channel uses, not necessarily known by the receivers with anticipation. This model is referred to as \emph{intermittent feedback} (IF) \cite{Karakus-TIT-2015}. The main result in \cite{Karakus-TIT-2015} is an approximation of the capacity region to within a constant gap. The achievability scheme is built on random coding arguments with forward decoding and a quantize-map-and-foward strategy to retransmit  the information obtained through feedback. This is because erasures might constrain either partial or complete decoding of the interference at the transmitter. Nonetheless, even a quantized version of the interference might be useful for interference cancellation or for providing an alternative path. 
\newline
\subsection{Noisy Feedback} 
Assume that for all $n \in \lbrace 1,2 \ldots, N\rbrace$, the random transformation $g$ is such that given a channel output realization $\overrightarrow{y}_n$,
\begin{equation}
\overleftarrow{Y}_n = \overleftarrow{h} \overrightarrow{y}_n + Z_n,
\end{equation}
where $\overrightarrow{h} \in \mathds{R}_+$ is a parameter of the channel and $Z_n$ is a real Gaussian random variable with zero mean and unit variance. 
This model is known in the literature as \emph{noisy feedback} (NF) or \emph{partial feedback} \cite{Gastpar-Asilomar-2006, SyQuoc-TIT-2015, Syquoc-Allerton-2012}. Note that the receiver does not apply any processing to the channel output and sends a re-scaled copy to the transmitter via a noisy channel. From this point of view,  as opposed to RLF, this model does not focus on the constraint on the number of codewords that can be used to perform feedback, but rather on the fact that the feedback channel might be noisy. Essentially, the codebook used to perform feedback in NF is $\mathds{R}^N$. 
In \cite{SyQuoc-TIT-2015}, the capacity of the G-IC with NF has been approximated to within a constant gap for the symmetric case. 
The achievable scheme in \cite{SyQuoc-TIT-2015} is a particular case of a more general achievability scheme presented in \cite{Tuninetti-ISIT-2007, Yang-Tuninetti-TIT-2011}.
An outer bound using the Hekstra-Willems dependence-balance arguments \cite{Hekstra-TIT-1989} has been introduced in \cite{Gastpar-Asilomar-2006}. These results suggest that feedback loses its efficacy on increasing the capacity region approximately when the noise variance on the feedback link is larger than on the forward link. 
Similar results have been reported in the fully decentralized IC with NF \cite{Quintero-PhD-2017, Perlaza-ISIT-2014, Quintero-ISIT-2017-1, Quintero-EW-2017}.
Inner and outer bounds on the sum-capacity using the existing connections between channel-output feedback and conferencing transmitters have been presented in \cite{Prabhakaran-TIT-2011}.
More general channel models, for instance when channel-outputs are fed back to both receivers, have been studied in \cite{Kramer-TIT-2002,  Tuninetti-ITA-2010, Tuninetti-ITW-2012, Sahai-TIT-2013}.

\subsection{A Comparison Between Feedback Models}

 \begin{figure*}[t!]
 \centerline{\epsfig{figure=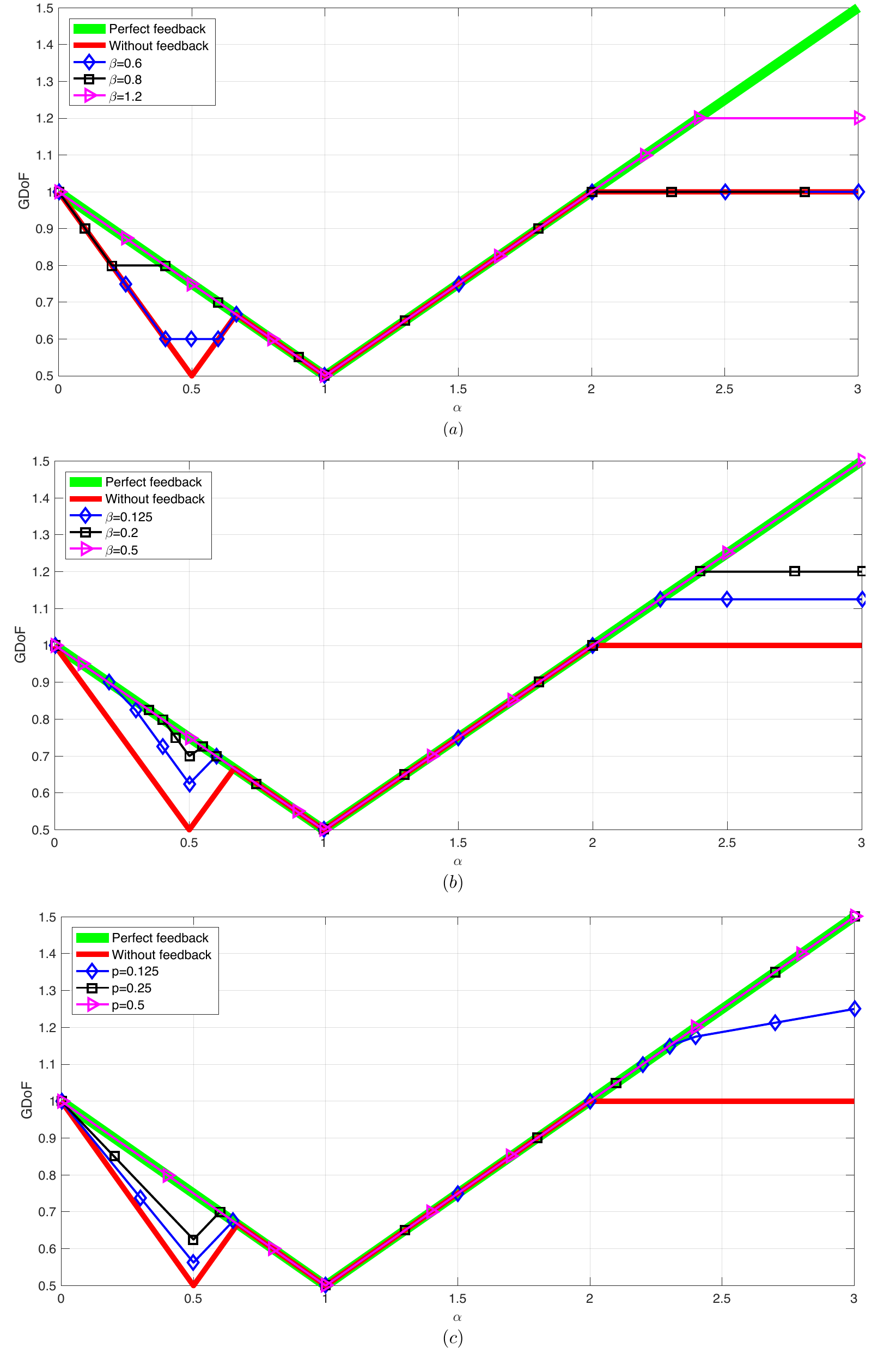,width=.85\textwidth}}
  \caption{Number of generalized degrees of freedom (GDoF) of a symmetric two-user G-IC; (a) case with NF with $\beta \in \lbrace 0.6, 0.8, 1.2\rbrace$; (b) case with RLF with $\beta \in \lbrace 0.125, 0.2, 0.5\rbrace$;  and (c) case with IF with $p \in \lbrace 0.125, 0.25, 0.5 \rbrace$.}
  \label{FigGDoFs}
\end{figure*}

In both IF and NF, the feedback signal is obtained via a random transformation. In particular, IF models the feedback link as an erasure-channel, whereas NF models the feedback link as an additive white Gaussian noise (AWGN) channel. 
Alternatively in the RLF case, the feedback signal is obtained via a deterministic transformation.
Let $\overleftarrow{\SNR}$ be the SNR in each of the feedback links from the receiver to the corresponding transmitters in the symmetric G-IC with NF (G-IC-NF) mentioned above. Let also $\beta$ and $\beta'$ be
\begin{subequations}
\begin{IEEEeqnarray}{rcl}
\label{Eqbetas}
\beta &  =  & \frac{\log\left(\overleftarrow{\SNR}\right)}{\log\left(\overrightarrow{\SNR}\right)} 
\end{IEEEeqnarray}
and
\begin{IEEEeqnarray}{rcl}
\beta' & =  & \frac{C_F}{\log\left(\overrightarrow{\SNR}\right)}.
\end{IEEEeqnarray}
\end{subequations}
These parameters approximate the ratio between the capacity of the feedback link and the capacity of the forward link in the NF case and RLF case, respectively. Hence, a fair comparison of RLF and NF must be made with $\beta = \beta'$.
The number of GDoF is plotted as a function of $\alpha$ when  $\mathcal{C}(\overrightarrow{\SNR}, \INR)$ is calculated 
with  NF for several values of $\beta$ in Figure~\ref{FigGDoFs}(a);
with RLF for different values of $\beta'$  in Figure~\ref{FigGDoFs}(b); and 
with IF for several values of  $p$  in Figure~\ref{FigGDoFs}(c).

NF is a more pessimistic channel-output feedback model than RLF in terms of the number of GDoF when $\beta=\beta'$.
When $\alpha \in \left(0,\frac{2}{3}\right)$ or $\alpha \in (2,\infty)$, RLF increases the number of GDoF for all $\beta' > 0$. Note that RLF with $\beta' = \frac{1}{2}$ achieves the same performance as PF, for all $\alpha \in [0,3]$.  
In the case of NF, there does not exist any benefit in terms of the number of GDoF for all $0<\beta < \frac{1}{2}$. A noticeable effect of NF occurs when $\alpha \in (0,\frac{2}{3})$, for all $\beta > \frac{1}{2}$; and when $\alpha \in (2,\infty)$, for all $\beta > 1$.  
This observation follows from the fact that in RLF, receivers extract  relevant information about interference and send it via a noiseless channel. Alternatively, NF requires sending to the transmitter an exact copy of the channel output via an AWGN channel. Hence with $\beta=\beta'>0$, the transmitters are always able to obtain information about the interference in RLF, whereas the same is not always true for NF. 
Finally, note that in both NF and RLF, the number of GDoF is not monotonically increasing with $\alpha$ in the interval $[2,\infty)$. Instead, it is upper-bounded by $\min\left(\frac{\alpha}{2}, \beta \right)$ in NF and by $ \min\left(\frac{\alpha}{2}, 1+ \beta \right)$ in RLF. 

The most optimistic model in terms of the number of GDoF, aside from PF, is IF. In particular because for any value of $p >0$, there always exists an improvement of the number of GDoF for all  $\alpha \in (0,\frac{2}{3})$ and $\alpha \in (2,\infty)$.
Note that, with $p \geqslant \frac{1}{2}$, IF provides the same number of GDoF as PF. Note also that the number of GDoF monotonically increases with $\alpha$ in the interval $[2,\infty)$ for any positive value of $p$ in \eqref{EqIFprobability}, which implies an arbitrarily large increment in the number of GDoF.

\subsection{Contributions} \label{SecContrib}

In this paper, the capacity of the G-IC-NF is approximated to within $4.4$ bits by a new achievable region and a new converse region. These results generalize the approximate capacity region of the G-IC-NF presented in \cite{SyQuoc-TIT-2015, Syquoc-Allerton-2012} for the symmetric case. The gap between the new achievable region and the new converse region is slightly improved with respect to the one obtained in \cite{SyQuoc-TIT-2015}.

The methodology is the same used in \cite{Etkin-TIT-2008, Suh-TIT-2011, Karakus-TIT-2015, Perlaza-TIT-2015, SyQuoc-TIT-2015}, among others, i.e., a linear deterministic (LD) approximation \cite{Avestimehr-TIT-2011} to the G-IC, referred to as LD-IC, is studied  to gain insight on the construction of both inner and outer bounds. From this perspective, a byproduct of the main results is the full characterization of the capacity region of the LD-IC with NF (LD-IC-NF).

The achievability scheme presented in this paper as well as the one in \cite{SyQuoc-TIT-2015} use a four-layer block-Markov superposition coding and backward decoding. Note that the achievability scheme used in \cite{SyQuoc-TIT-2015} is obtained as a special case of the one presented  in \cite{Tuninetti-ISIT-2007,Yang-Tuninetti-TIT-2011}. The achievability scheme presented in this paper is developed independently. 
The main difference between these achievability schemes lies on the choice of the random variables used to generate the codewords of each of the layers of the codebook. Another difference is the power optimization made to obtain the corresponding achievable regions. 

The converse region presented in this paper uses existing bounds from the case of PF in \cite{Suh-TIT-2011} and new bounds that generalize those in \cite{SyQuoc-TIT-2015}. The proof of the converse presented in  \cite{SyQuoc-TIT-2015} uses standard techniques including cut-set bounds and genie-aided channels, which are the same techniques used in this paper.
Nonetheless, such generalization is far from trivial, as suggested in \cite[Section IV-D]{SyQuoc-TIT-2015}.  

\subsection{Organization of the paper} \label{SecOrgPaper}

Section~\ref{SecNotation} introduces the notation used in this paper. Section~\ref{SecProbFNM} describes the two-user G-IC-NF. Section~\ref{SectPreResults} describes the exact capacity region of the LD-IC-NF.  Section~\ref{SectMainResults} introduces the main results, essentially, an achievable region and a converse region for the G-IC-NF. Section~\ref{SecCooperation} describes the connections between the IC-NF and an IC with conferencing transmitters (IC-CT). 
Finally, Section~\ref{SecConclusions} concludes this work and highlights some extensions. 

\section{Notation}\label{SecNotation}

Throughout this paper, sets are denoted with uppercase calligraphic letters, e.g., $\mathcal{X}$. Random variables are denoted by uppercase letters, e.g., $X$, whereas their realizations are denoted by  the corresponding lower case letter, e.g.,  $x$. The probability distribution of $X$ over the set $\mathcal{X}$ is denoted $P_{X}$. Whenever a second random variable $Y$ is involved, $P_{X \, Y}$ and $P_{Y|X}$ denote respectively the joint probability distribution of $(X, Y)$ and the conditional probability distribution of $Y$ given $X$. Let $N$ be a fixed natural number. An $N$-dimensional vector of random variables is denoted by $\bs{X} = (X_{1}, X_{2}, ..., X_{N})^\sfT$ and a corresponding realization is denoted by $\bs{x}= (x_{1}, x_{2}, ..., x_{N})^\sfT \in \mathcal{X}^{N}$.  Given $\bs{X} = (X_{1}, X_{2}, ..., X_{N})^\sfT$ and $(a,b) \in \mathds{N}^2$, with $a < b \leqslant N$,  the $(b-a+1)$-dimensional vector of random variables formed by the components $a$ to $b$ of $\bs{X}$ is denoted by ${\bs{X}_{(a:b)} = (X_a, X_{a+1}, \ldots, X_b)^\sfT}$.  The notation $(\cdot)^+$ denotes the positive part operator, i.e., $(\cdot)^+ = \max(\cdot, 0)$ and $\mathbb{E}_{X}[ \cdot ]$ denotes the expectation with respect to the distribution $P_{X}$ of the random variable $X$. The logarithm function log is assumed to be base 2.  

\section{Problem Formulation} \label{SecProbFNM}
\begin{figure}[t!]
\vspace{13mm}
 \centerline{\epsfig{figure=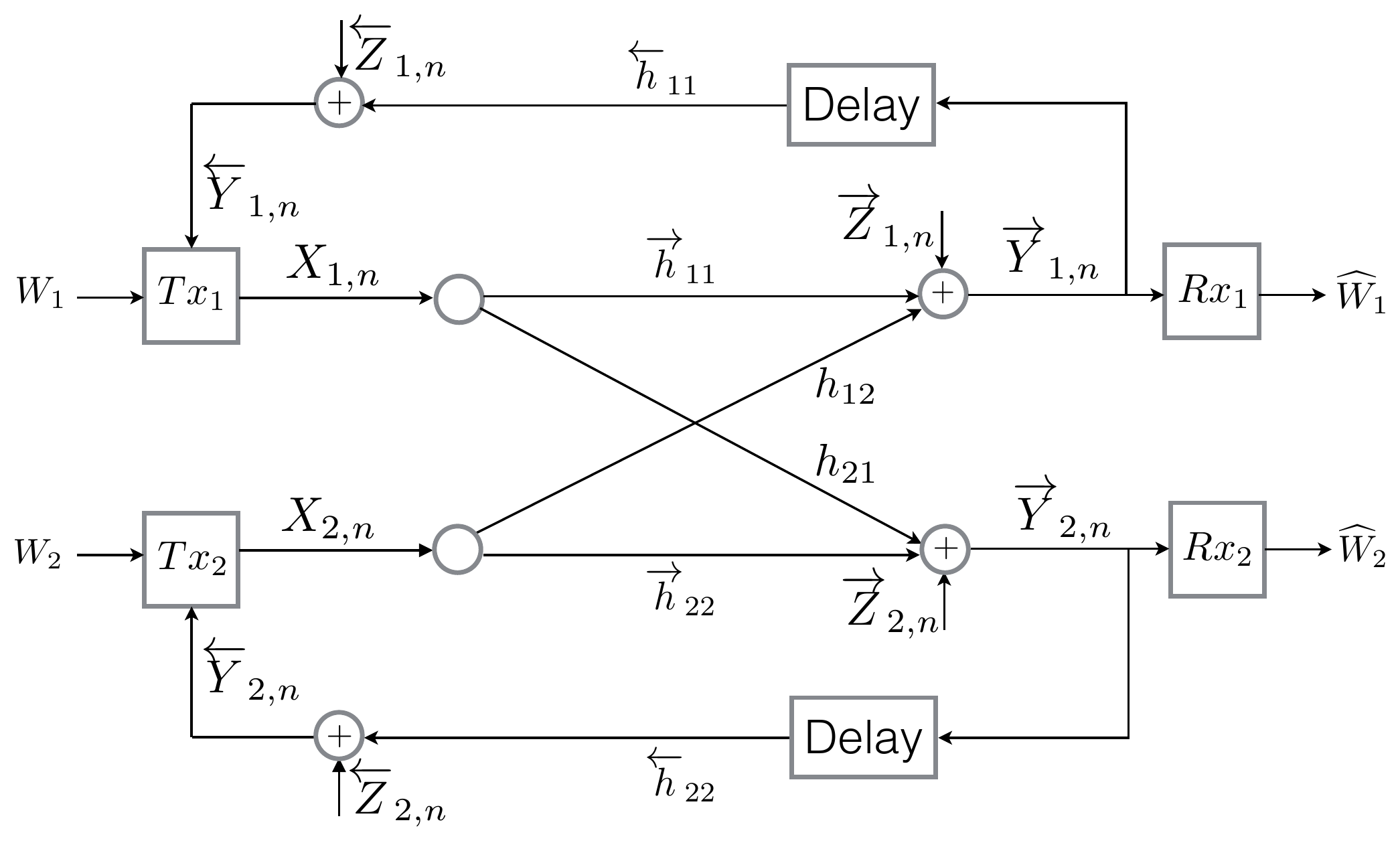,width=0.5\textwidth}}
  \caption{Gaussian interference channel with noisy channel-output feedback at channel use~$n$.}
  \label{Fig:G-IC-NF}
\end{figure}
Consider the two-user G-IC-NF depicted in Figure~\ref{Fig:G-IC-NF}. Transmitter $i$, with $i \in \{1,2\}$, communicates with receiver $i$ subject to the interference produced by transmitter $j$, with $j \in \{1,2\} \backslash \{i\}$. There are two independent and uniformly distributed messages, $W_i \in \mathcal{W}_i$, with $\mathcal{W}_i=\{1, 2,  \ldots, 2^{NR_i}\}$, where $N \in \mathds{N}$ denotes the block-length in channel uses and $R_i \geqslant 0$ is the transmission rate in bits per channel use. For transmitting a message index $W_i$, transmitter $i$  sends a given codeword denoted by  $\bs{X}_{i}=\left(X_{i,1}, X_{i,2}, \ldots, X_{i,N}\right)^\sfT \in \mathds{R}^N$. 
The channel coefficient from transmitter $j$ to receiver $i$ is denoted by $h_{ij}$; the channel coefficient from transmitter $i$ to receiver $i$ is denoted by $\overrightarrow{h}_{ii}$; and the channel coefficient from channel-output $i$ to transmitter $i$ is denoted by $\overleftarrow{h}_{ii}$. All channel coefficients are assumed to be non-negative real numbers.
At a given channel use $n \in \{1, 2, \ldots, N\}$, the channel output realization at receiver $i$ is denoted by $\overrightarrow{Y}_{i,n}$.  
During channel use $n$, the input-output relation of the channel model is given by
\begin{IEEEeqnarray}{rcl}
\label{Eqsignalyif}
\overrightarrow{Y}_{i,n}& = & \overrightarrow{h}_{ii}X_{i,n} + h_{ij}X_{j,n}+\overrightarrow{Z}_{i,n},
\end{IEEEeqnarray}
where $\overrightarrow{Z}_{i,n}$ is a real Gaussian random variable with zero mean and unit variance that represents the noise at the input of receiver $i$.
Let $d>0$ be the finite feedback delay measured in channel uses. At the end of channel use $n$, transmitter $i$ observes $\overleftarrow{Y}_{i,n}$, which consists of a scaled and noisy version of $\overrightarrow{Y}_{i,n-d}$.
More specifically,
\begin{IEEEeqnarray}{rcl}
\label{Eqsignalyib}
\overleftarrow{Y}_{i,n}  & = & 
\begin{cases}
 \overleftarrow{Z}_{i,n} &  \textrm{for } n \! \in \lbrace \! 1, \! 2,  \ldots, d \rbrace  \\ 
\overleftarrow{h}_{ii}\overrightarrow{Y}_{i,n-d} \! + \! \overleftarrow{Z}_{i,n} \!  &  \textrm{for } n \! \in \lbrace  d \! + \! 1, \! d \! + \! 2, \ldots, \! N \rbrace,
 \end{cases} \quad
\end{IEEEeqnarray}
where $\overleftarrow{Z}_{i,n}$ is a real Gaussian random variable with zero mean and unit variance that represents the noise in the feedback link of transmitter-receiver pair  $i$. The random variables $\overrightarrow{Z}_{i,n}$ and $\overleftarrow{Z}_{i,n}$ are assumed to be independent.
In the following, without loss of generality, the feedback delay is assumed to be equal to one channel use, i.e., $d=1$. 
The encoder of transmitter $i$ is defined by a set of deterministic functions $\Big \lbrace f_i^{(1)}, f_i^{(2)}, \ldots, f_i^{(N)} \Big \rbrace$, with $f_i^{(1)}:\mathcal{W}_i \rightarrow \mathds{R}$ and for all $n \in \{2, 3, \ldots, N\}$, $f_i^{(n)}:\mathcal{W}_i\times\mathds{R}^{n-1} \rightarrow \mathds{R}$, such that
\begin{subequations}
\label{Eqencod}
\begin{IEEEeqnarray}{rcl}
\label{Eqencodi1}
X_{i,1}&  = &f_i^{(1)}\left(W_i\right)  \mbox{ and }  \\
\label{Eqencodit}
X_{i,n}& = &f_i^{(n)}\left(W_i,\overleftarrow{Y}_{i,1}, \overleftarrow{Y}_{i,2}, \ldots,\overleftarrow{Y}_{i,n-1}\right).
\end{IEEEeqnarray}
\end{subequations}

The components of the input vector $\bs{X}_{i}$ are real numbers subject to an average power constraint:
\begin{equation}
\label{Eqconstpow}
\frac{1}{N}\sum_{n=1}^{N} \mathbb{E}_{X_{i,n}}\left[X_{i,n}^2\right] \leq 1.
\end{equation}
The decoder of receiver $i$ is defined by a deterministic function ${\psi_i^{(N)}: \mathds{R}_i^{N} \rightarrow \mathcal{W}_i}$.
At the end of the communication, receiver $i$ uses the vector $\Big(\overrightarrow{Y}_{i,1}$, $\overrightarrow{Y}_{i,2}$, $\ldots$, $\overrightarrow{Y}_{i,N}\Big)^\sfT$ to obtain an estimate of the message index:
\begin{IEEEeqnarray}{rcl}
\label{Eqdecoder}
\widehat{W}_i &  = & \psi_i^{(N)} \left(\overrightarrow{Y}_{i,1}, \overrightarrow{Y}_{i,2}, \ldots, \overrightarrow{Y}_{i,N}\right), \qquad
\end{IEEEeqnarray} 
where $\widehat{W}_i$ is an estimate of the message index $W_i$.
The decoding error probability in the two-user G-IC-NF of a codebook of block-length $N$, denoted by $P_{e} (N)$, is given by   
\begin{IEEEeqnarray}{rcl}
\label{EqDecErrorProb}
 P_{e} ( N ) &  = &   \max   \Bigg(  \pr{ \widehat{W_1}  \neq   W_1  }  ,  \pr{ \widehat{W_2}  \neq   W_2 }\Bigg).  \qquad
\end{IEEEeqnarray}

The definition of an achievable rate pair $(R_1,R_2) \in \mathds{R}_+^{2}$ is given below. 
\begin{definition}[Achievable Rate Pairs]\label{DefAchievableRatePairs}\emph{
A rate pair $(R_1,R_2) \in \mathds{R}_+^{2}$ is achievable if there exist encoding functions $f_i^{(1)}, f_i^{(2)}, \ldots, f_i^{(N)}$ and decoding functions $\psi_i^{(1)}$, $\psi_i^{(2)},$ $\ldots,$ $\psi_i^{(N)}$ for all $i \in \lbrace 1,2 \rbrace$ such that the decoding error probability $P_{e}(N)$ can be made arbitrarily small by letting the block-length $N$ grow to infinity.
 }
\end{definition}
In the next sections, it is shown that the capacity region of the two-user G-IC-NF in Figure~\ref{Fig:G-IC-NF} can be described by six parameters: $\overrightarrow{\SNR}_i$, $\overleftarrow{\SNR}_i$, and $\INR_{ij}$, with $i \in \{1,2\}$ and $j \in \{1,2\} \backslash \{i\}$, which are defined as follows:
\begin{IEEEeqnarray}{rcl}
\label{EqSNRifwd}
\overrightarrow{\SNR}_i & = & \overrightarrow{h}_{ii}^2, \\
\label{EqINRij}
\INR_{ij}& = & h_{ij}^2, \mbox{ and } \\
\label{EqSNRibwd}
\overleftarrow{\SNR}_i&  = & \overleftarrow{h}_{ii}^2\left(\overrightarrow{h}_{ii}^2 + 2\overrightarrow{h}_{ii}h_{ij}+h_{ij}^2+1\right). \quad
\end{IEEEeqnarray}

\section{Linear Deterministic Channels} \label{SectPreResults}

This section describes the two-user LD-IC-NF and its exact capacity region. The relevance of this result is that it provides the main insight used to obtain the approximate capacity of the G-IC-NF in Section~\ref{SectMainResults}.  
\begin{figure}[t!]
 \centerline{\epsfig{figure=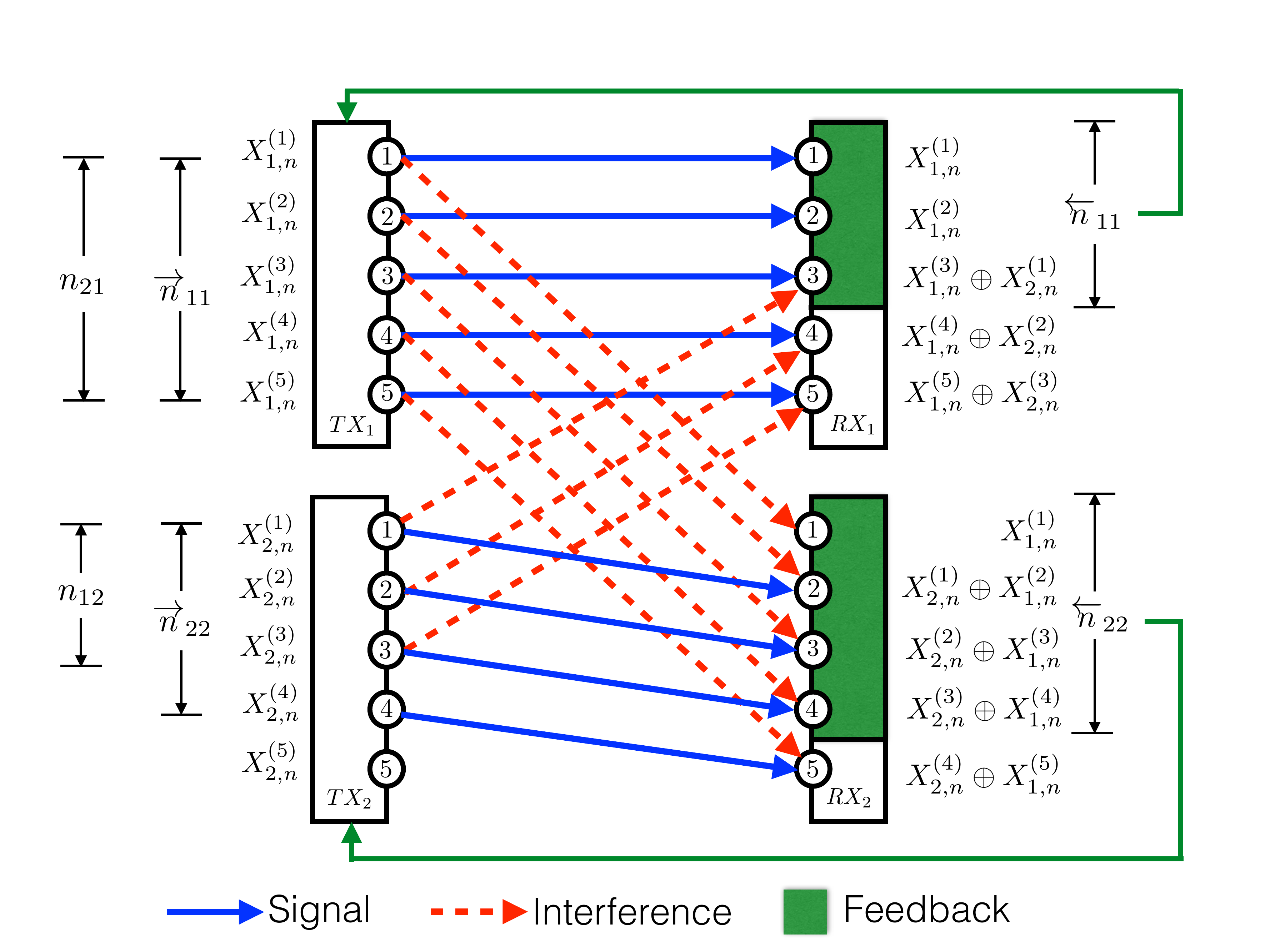,width=0.5\textwidth}}
\caption{Two-user linear deterministic interference channel with noisy channel-output feedback at channel use $n$.} 
  \label{FigLD-IC-NF}
\end{figure}

\subsection{Two-User Linear Deterministic Interference Channel with Noisy Feedback}\label{SecNMLDICNF}

Consider the two-user LD-IC-NF depicted in Figure~\ref{FigLD-IC-NF}. 
For all $i \in \lbrace 1, 2 \rbrace$, with $j\in \lbrace 1, 2 \rbrace\setminus \lbrace i \rbrace$, the number of bit-pipes between transmitter $i$ and its  intended receiver  is denoted by $\overrightarrow{n}_{ii}$; the number of bit-pipes between transmitter $i$ and its  non-intended receiver is denoted by $n_{ji}$; and the number of bit-pipes between receiver $i$ and its corresponding transmitter is denoted by  $\overleftarrow{n}_{ii}$. 

At transmitter $i$, the channel-input $\bs{X}_{i,n}$ at channel use $n$, with $n \in \lbrace 1, 2,  \ldots, N \rbrace$, is a $q$-dimensional binary vector ${\bs{X}_{i,n} = \left(X_{i,n}^{(1)}, X_{i,n}^{(2)}, \ldots, X_{i,n}^{(q)}\right)^{\sfT}}$, with
\begin{equation}\label{Eqq}
q=\ds\max \left(\overrightarrow{n}_{11}, \overrightarrow{n}_{22}, n_{12}, n_{21}\right),
\end{equation}
and $N$ the block-length. 
At receiver $i$, the channel-output $\overrightarrow{\bs{Y}}_{i,n}$ at channel use $n$ is also a $q$-dimensional binary vector ${\overrightarrow{\bs{Y}}_{i,n} = \left(\overrightarrow{Y}_{i,n}^{(1)}, \overrightarrow{Y}_{i,n}^{(2)}, \ldots, \overrightarrow{Y}_{i,n}^{(q)}\right)^{\sfT}}$. 
The input-output relation during channel use $n$ is given by 
\begin{IEEEeqnarray}{rcl}
\label{EqLDICsignals}
\overrightarrow{\bs{Y}}_{i,n}& = & \bs{S}^{q - \overrightarrow{n}_{ii}} \bs{X}_{i,n} + \bs{S}^{q - n_{ij}} \bs{X}_{j,n},
\end{IEEEeqnarray}
where, $\bs{S}$ is a $q\times q$ lower shift matrix.

The feedback signal $\bs{\overleftarrow{Y}}_{i,n}$ available at transmitter $i$ at the end of channel use $n$ satisfies
\begin{IEEEeqnarray}{rcl}\label{EqLDICsignalsc}
\left(\left(0, \ldots, 0\right), \bs{\overleftarrow{Y}}_{i,n}^{\sfT} \right)^{\sfT} & = & \bs{S}^{\left(\max(\overrightarrow{n}_{ii},n_{ij})-\overleftarrow{n}_{ii}\right)^+} \, \bs{\overrightarrow{Y}}_{i,n-d}, \quad
\end{IEEEeqnarray}
where $d$ is a finite delay, additions and multiplications are defined over the binary field.
The dimension of the vector $\left(0,  \ldots,  0\right)$ in \eqref{EqLDICsignalsc} is $q  -  \min \big(\overleftarrow{n}_{ii}, \max(\overrightarrow{n}_{ii}, n_{ij}) \big)$ and the vector $\bs{\overleftarrow{Y}}_{i,n}$ represents the $\min \big(\overleftarrow{n}_{ii}, \max(\overrightarrow{n}_{ii},n_{ij}) \big)$ least significant bits of $\bs{S}^{\left(\max(\overrightarrow{n}_{ii},n_{ij})-\overleftarrow{n}_{ii}\right)^+} \, \bs{\overrightarrow{Y}}_{i,n-d}$.
\begin{figure*}[t]
\begin{theorem}\label{TheoremANFBLDMCap} \emph{
The capacity region $\Cldicnfb(\overrightarrow{n}_{11}, \overrightarrow{n}_{22}, n_{12}, n_{21}, \overleftarrow{n}_{11} , \overleftarrow{n}_{22})$ of the two-user LD-IC-NF is the set of non-negative rate pairs $(R_1,R_2)$ that satisfy for all $i \in \lbrace 1, 2 \rbrace$, with $j\in\lbrace 1, 2 \rbrace\setminus\lbrace i \rbrace$:
\begin{subequations}
\begin{IEEEeqnarray}{rcl}
\label{EqRiV2}
R_{i}& \leqslant  & \min\left(\max\left(\overrightarrow{n}_{ii},n_{ji}\right),\max\left(\overrightarrow{n}_{ii},n_{ij}\right)\right), \\
\label{EqRi-2-V2}
R_i&  \leqslant &\min\left(\max\left(\overrightarrow{n}_{ii},n_{ji}\right),\max\left(\overrightarrow{n}_{ii},\overleftarrow{n}_{jj}-\left(\overrightarrow{n}_{jj}-n_{ji}\right)^+\right)\right),\\
\label{EqRi+Rj-1-V2}
R_1+R_2  &  \leqslant & \min \left(\max\left(\overrightarrow{n}_{22},n_{12}\right)+\left(\overrightarrow{n}_{11}-n_{12}\right)^+, \max\left(\overrightarrow{n}_{11},n_{21}\right)+\left(\overrightarrow{n}_{22}-n_{21}\right)^+\right), \\
\nonumber
R_1+R_2 &  \leqslant & \max\Big(\left(\overrightarrow{n}_{11}-{n}_{12} \right)^+, n_{21}, \overrightarrow{n}_{11}-\left(\max\left(\overrightarrow{n}_{11},n_{12}\right)-\overleftarrow{n}_{11}\right)^+\Big)\\
  \label{EqRi+Rj-2-V2}
 & &+\max\Big(\left(\overrightarrow{n}_{22}-{n}_{21} \right)^+, n_{12}, \overrightarrow{n}_{22}-\left(\max\left(\overrightarrow{n}_{22},n_{21}\right)-\overleftarrow{n}_{22}\right)^+\Big),\\
\label{Eq2Ri+Rj-V2}
2R_i+R_j  &  \leqslant & \max\left(\overrightarrow{n}_{ii},{n}_{ji} \right) \! + \! \left(\overrightarrow{n}_{ii} \! - \! {n}_{ij} \right)^+ \! + \! \max\Big(\! \left(\overrightarrow{n}_{jj} \! - \! {n}_{ji} \right)^+, n_{ij}, \overrightarrow{n}_{jj} \! - \! \left(\max\left(\overrightarrow{n}_{jj},{n}_{ji} \right)-\overleftarrow{n}_{jj}\right)^+ \! \Big). \qquad
\end{IEEEeqnarray}
\end{subequations}
}
\end{theorem} 
\noindent\rule{16.5cm}{0.6pt}
\end{figure*}

The feedback delay is assumed to be equal to one channel use, i.e., $d=1$. 
The encoding and decoding operations are performed analogously to those in the G-IC case.
The decoding error probability is calculated following \eqref{EqDecErrorProb}. 
Similarly, a rate pair $(R_1, R_2) \in \mathds{R}_{+}^{2}$ is said to be achievable if it satisfies Definition~\ref{DefAchievableRatePairs}.

\subsection {Capacity Region of the  Two-User Linear Deterministic Interference Channel with Noisy Channel-Output Feedback} \label{SectCapLDICNF}

Denote by $\Cldicnfb(\overrightarrow{n}_{11}, \overrightarrow{n}_{22}, n_{12}, n_{21}$, $ \overleftarrow{n}_{11} , \overleftarrow{n}_{22})$ the capacity region of the LD-IC-NF with parameters $\overrightarrow{n}_{11}$, $\overrightarrow{n}_{22}$, $n_{12}$, $n_{21}$, $\overleftarrow{n}_{11}$, and $\overleftarrow{n}_{22}$. Theorem~\ref{TheoremANFBLDMCap} (in the top of next page) fully characterizes this capacity region. 
The proof of Theorem~\ref{TheoremANFBLDMCap} is divided into two parts. The first part describes the proof of achievability and is presented in Appendix \ref{AppAch-IC-NF}. The second part describes the proof of the converse and is presented in  Appendix~\ref{App-C-LD-IC-NF}. 
 
\subsubsection{Comments on the Achievability Scheme}\label{SecCommentsAchievabilityLD}
\begin{figure*}[t!]
\vspace{+5mm}
 \centerline{\epsfig{figure=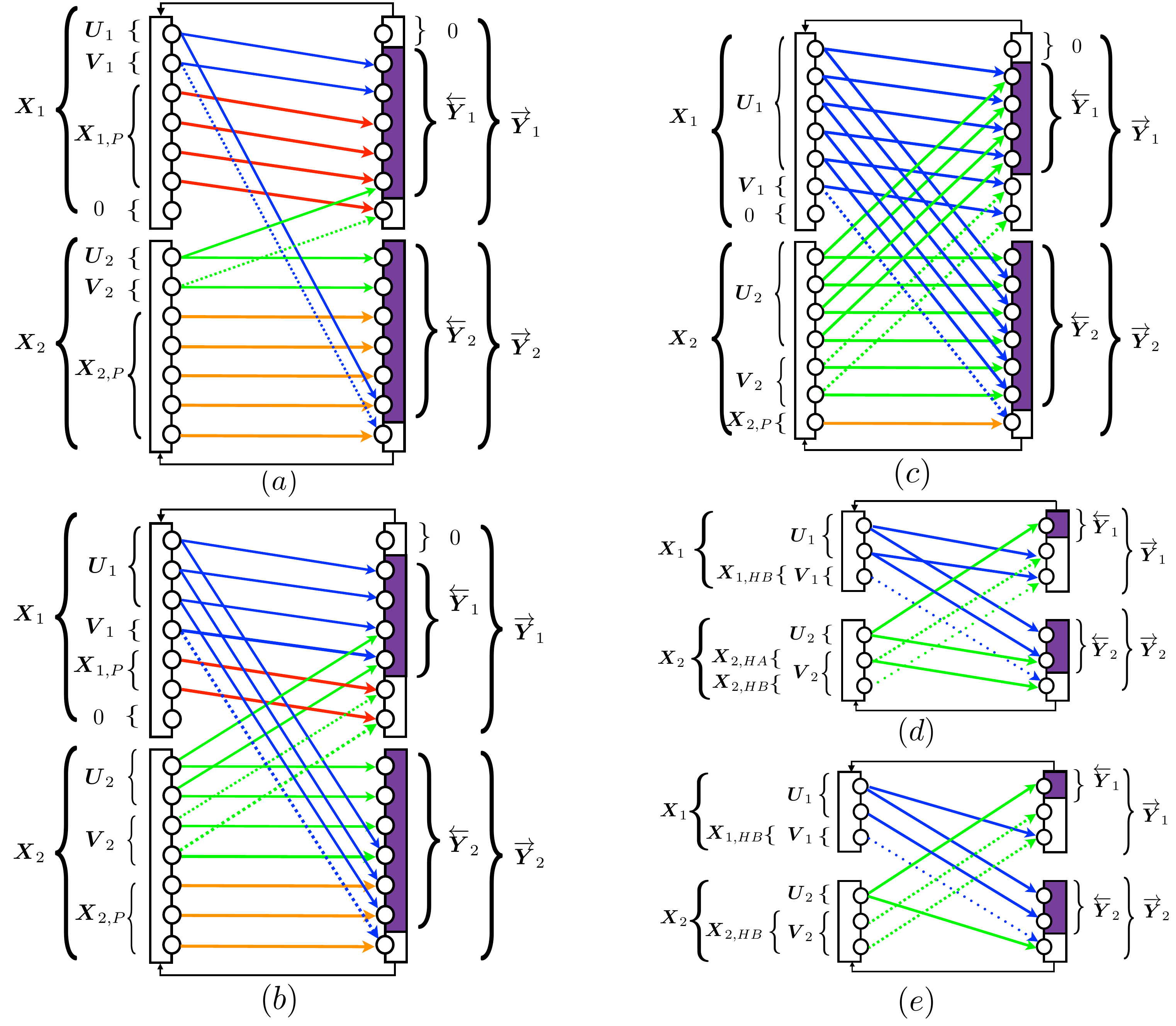,width=1\textwidth}}
 \caption{Concatenation of binary vectors $\bs{U}_{i,n} \in \lbrace 0,1 \rbrace^{q_{i,1}}$, with $q_{i,1}$ defined in \eqref{Eqqi1}; $\bs{V}_{i,n}\in \lbrace 0,1 \rbrace^{q_{i,2}}$, with $q_{i,2}$ defined in \eqref{Eqqi2}; and $\bs{X}_{i,P,n} \in \lbrace 0,1 \rbrace^{q_{i,3}}$, with $q_{i,3}$  defined in \eqref{Eqqi3} to form the input symbol $\bs{X}_{i,n}$ in $(a)$ the very weak interference regime, $(b)$ the weak interference regime, $(c)$ the moderate interference regime, $(d)$ the strong interference regime, and $(e)$ the very strong interference regime.}
\label{FigachievabilityMessagesLDICNFB}
\end{figure*}
 \begin{figure*}[t!]
 \centerline{\epsfig{figure=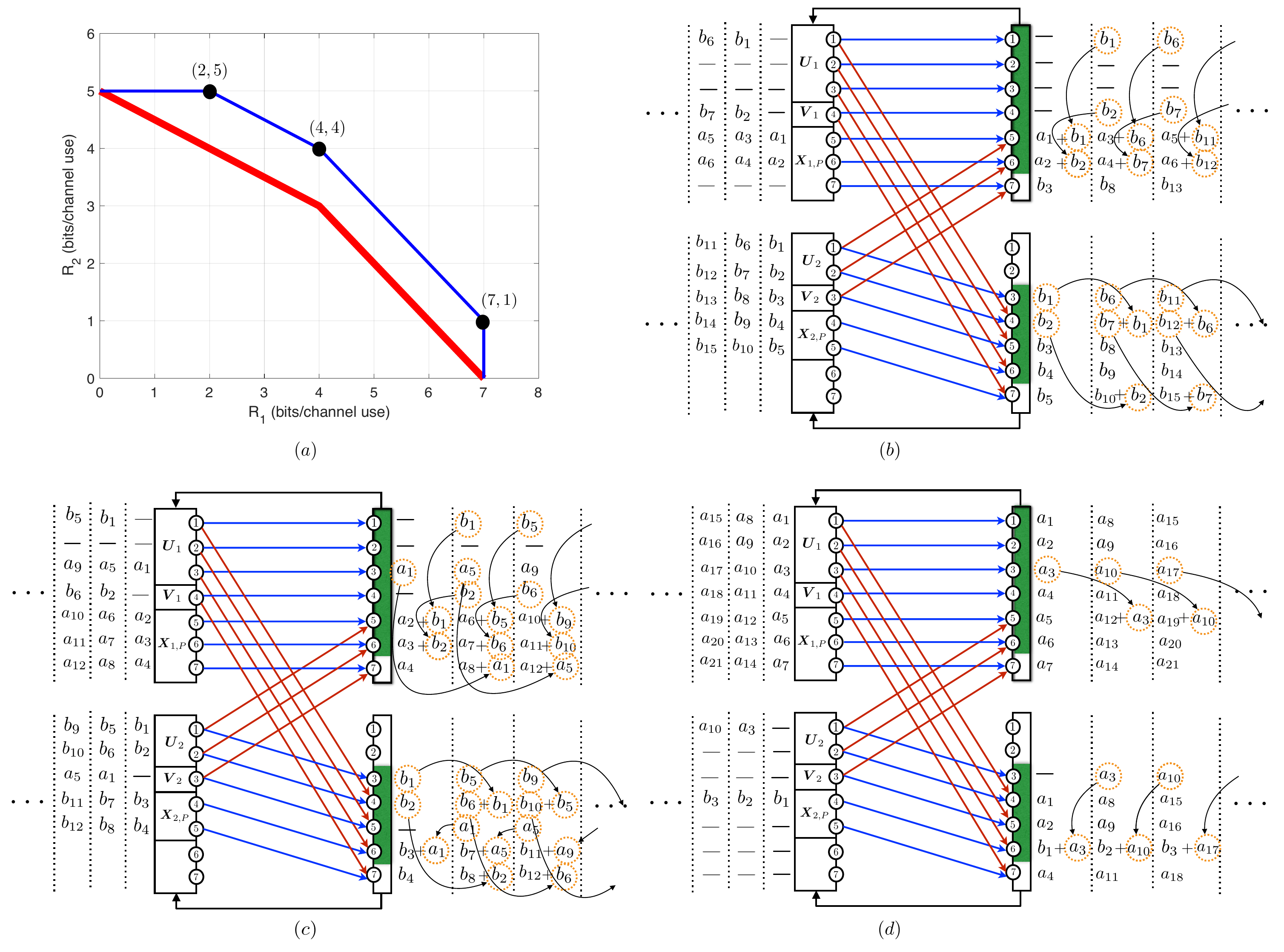,width=1.02\textwidth}}
  \caption{(a) Capacity regions of $\mathcal{C}(7,5,3,4,0,0)$ (thick red line) and $\mathcal{C}(7,5,3,4,6,4)$ (thin blue line). (b) Achievability of the rate pair $(2,5)$ in $\mathcal{C}(7,5,3,4,6,4)$. (c) Achievability of the rate pair $(4,4)$ in an $\mathcal{C}(7,5,3,4,6,4)$. (d) Achievability of the rate pair $(7,1)$ in an $\mathcal{C}(7,5,3,4,6,4)$.} 
  \label{FigachievabilityExamples}
\end{figure*}

Let the channel input of transmitter $i$ during channel use $n$ be $\bs{X}_{i,n} = \left( X_{i,n}^{(1)}, X_{i,n}^{(2)}, \ldots, X_{i,n}^{(q)}\right)^{\sf{T}} \in \lbrace 0,1 \rbrace^q$.
The achievability scheme is a sequel to the following observation: feedback allows the transmitters to obtain information about the interference produced by its counterpart at the intended receiver.  This information could be retransmitted aiming either at performing interference cancellation at the intended receiver or at providing an alternative communication path to the other transmitter-receiver pair \cite{Suh-TIT-2011, Perlaza-TIT-2015, SyQuoc-TIT-2015}. 
From this standpoint, there are three types of bit-pipes that start at transmitter $i$ that are particularly relevant: 
\newline
$(i)$ the set of bit-pipes that are observed above the noise level by both receiver $j$ and transmitter $j$, i.e., 
\begin{equation}\label{Equ}
\bs{U}_{i,n} = \left( X_{i,n}^{(1)}, X_{i,n}^{(2)}, \ldots, X_{i,n}^{(q_{i,1})}\right)^{\sf{T}},
\end{equation}
where, 
\begin{IEEEeqnarray}{rcl}
\label{Eqqi1}
q_{i,1} &=& \left(n_{ji}-\left(\max\left(\overrightarrow{n}_{jj},n_{ji}\right)-\overleftarrow{n}_{jj}\right)^+\right)^+;
\end{IEEEeqnarray}
$(ii)$ the set of bit-pipes that are observed above the noise level by receiver $j$ and below the (feedback) noise level by transmitter $j$, i.e., 
\begin{equation}
\bs{V}_{i,n} = \left( X_{i,n}^{(q_{i,1}+1)}, X_{i,n}^{(q_{i,1}+ 2)}, \ldots, X_{i,n}^{(q_{i,1} + q_{i,2})}\right)^{\sf{T}},
\end{equation}
where, 
\begin{IEEEeqnarray}{rcl}
\label{Eqqi2}
q_{i,2} &=& \min \left(n_{ji},\left(\max\left(\overrightarrow{n}_{jj},n_{ji}\right)-\overleftarrow{n}_{jj}\right)^+\right); \mbox{ and } 
\end{IEEEeqnarray}
$(iii)$ the set of bit-pipes that are exclusively observed above the noise level by receiver $i$, i.e.,
\begin{equation}
\! \bs{X}_{i,P,n} \! = \! \left(  \! X_{i,n}^{(q_{i,1} +q_{i,2}+1)}  \! ,   X_{i,n}^{(q_{i,1} +q_{i,2}+2)}  \! ,   \ldots  \!,    X_{i,n}^{(q_{i,1} +q_{i,2}+q_{i,3})} \! \right)^{\sf{T}}  \! ,
\end{equation}
where, 
\begin{IEEEeqnarray}{rcl}
\label{Eqqi3}
q_{i,3} &=&\left(\overrightarrow{n}_{ii}-n_{ji}\right)^+.
\end{IEEEeqnarray}
Note that 
\begin{equation}\label{Eqq}
q_{i,1} +q_{i,2} + q_{i,3} =  \max\left(\overrightarrow{n}_{ii},n_{ji}\right) \leqslant q,
\end{equation}
and thus,  
\begin{equation*}
\! \left(  \! X_{i,n}^{(q_{i,1} +q_{i,2}+q_{i,3}+1)}  \! ,   X_{i,n}^{(q_{i,1} +q_{i,2}+q_{i,3}+2)}  \!,   \ldots  \!,   X_{i,n}^{(q)}  \! \right)  \! =  \! \left( 0,   \ldots  \!,   0   \right)  \! .
\end{equation*}
Hence, for all $n \in \lbrace 1, 2, \ldots, N \rbrace$,
\begin{equation}
\bs{X}_{i,n} = \left( \bs{U}_{i,n}^{\sf{T}}, \bs{V}_{i,n}^{\sf{T}}, \bs{X}_{i,P,n}^{\sf{T}},  \left( 0, \ldots, 0\right) \right)^{\sf{T}}.
\end{equation}
An example of the concatenation of $\bs{U}_{i,n}$, $\bs{V}_{i,n}$ and $\bs{X}_{i,P,n}$ to form the input symbol $\bs{X}_{i,n}$ is presented in Figure~\ref{FigachievabilityMessagesLDICNFB}. 
Note that the vector $\left( 0, \ldots, 0\right)$ exists only when \eqref{Eqq} holds with strict inequality. 

Within this context, some key observations are worth highlighting:
\newline
$(a)$ The interference produced by the bits $\bs{U}_{j,n}$ on $\bs{X}_{i,n}$ at receiver $i$ can be eliminated in two cases: when $\bs{U}_{j,n}$ consists of bits previously transmitted by transmitter $i$; or when the bits $\bs{U}_{j,n}$ are retransmitted by transmitter $i$ at a later channel use such that they can be reliably decoded by receiver $i$. 
In both cases, receiver $i$ is able to implement interference cancellation.
\newline
$(b)$ The interference produced by the bits $\bs{V}_{j,n}$ on $\bs{X}_{i,n}$ at receiver $i$ can be eliminated in a single case: when $\bs{V}_{j,n}$ consists of bits previously transmitted by transmitter $i$.
\newline
$(c)$ The top
\begin{equation*}
 \! \min  \!\left( \! ( \!\overrightarrow{n}_{ii} - n_{ij} \!)^+  \!,  \!  \left(  \! \min  \! \left( \! \overleftarrow{n}_{jj}  ,  \! \max  \! \left( \! \overrightarrow{n}_{jj}  \! ,  \! n_{ji}  \right)  \right)  \!-  \!\left(  \! \overrightarrow{n}_{jj}  \!-  \!n_{ji}  \!\right)^+ \! \right)^+  \! \right)
\end{equation*}
bits in $\bs{U}_{i,n}$ are observed interference-free at receiver $i$ and thus, they are reliably decoded at the end of channel use $n$. Note also that these top bits in $\bs{U}_{i,n}$ produce interference in receiver $j$. However, thanks to feedback, transmitter $j$ can retransmit these bits in $\bs{U}_{j,n+m}$ or $\bs{V}_{j,n+m}$, for some  $m > 0$. Interestingly, the interference produced by these retransmissions can be eliminated at receiver $i$ at the end of channel use $n+m$.
\newline
$(d)$ The bottom
\begin{IEEEeqnarray}{l}
 \nonumber
 \min\Bigg(\! \left(n_{ji} \! - \! \overrightarrow{n}_{ii}\right)^+ \!,\Big(\overleftarrow{n}_{jj}-\overrightarrow{n}_{ii}-\min\left(\left(\overrightarrow{n}_{jj}-n_{ji}\right)^+,n_{ij}\right)\\
 \nonumber
 -\left(\left(\overrightarrow{n}_{jj}-n_{ij}\right)^+-n_{ji}\right)^+\Big)^+\Bigg) 
\end{IEEEeqnarray}
bits of $\bs{U}_{i,n}$ are not observed at receiver $i$ but receiver $j$. Thus, they can be fed back to transmitter $j$. If transmitter $j$ retransmits these bits in any of the bits $\bs{U}_{j,n+m}$ or $\bs{V}_{j,n+m}$, for some $m>0$, these bits might be reliably decoded at receiver $i$ forming an alternative communication path to transmitter-receiver pair $i$.
\newline
Taking into account the facts $(a)$ - $(d)$, simple coding schemes can be constructed as shown in the examples in Figure~\ref{FigachievabilityExamples}.
A complete proof of achievability is presented in Appendix \ref{AppAch-IC-NF}.
 \subsubsection{Comments on the Converse Region}
\begin{figure*}[t!]
 \centerline{\epsfig{figure=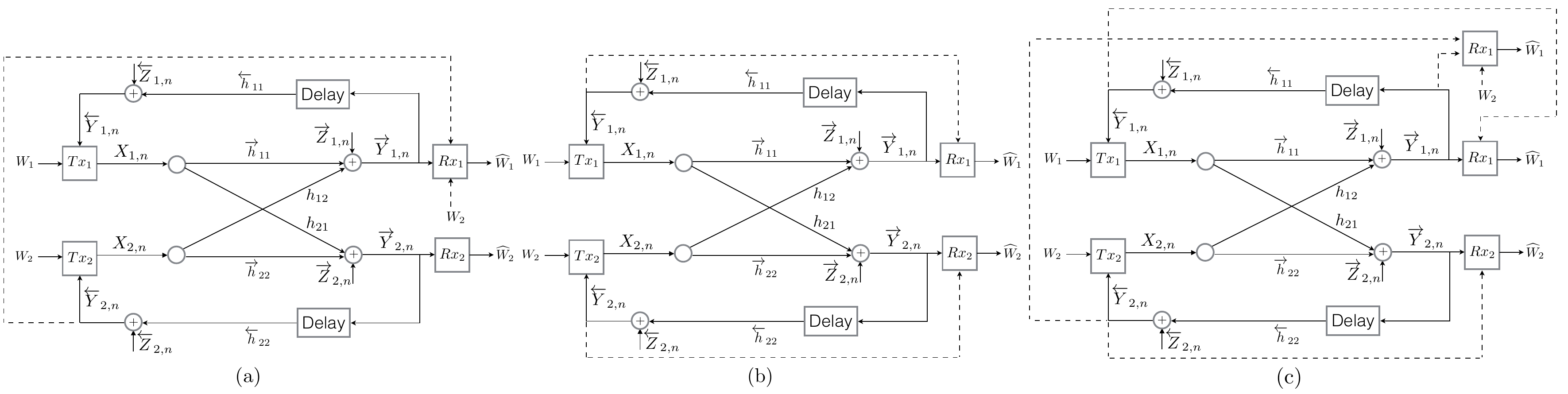,width=1.0\textwidth}}
  \caption{Genie-aided G-IC-NF models for channel use~$n$: $(a)$ model used to calculate the outer bound on $R_1$; $(b)$ model used to calculate the outer bound on $R_1+R_2$; and  $(c)$ model used to calculate the outer bound on $2 R_1+R_2$.}
  \label{Fig:G-IC-NF-Conv}
\end{figure*}
The outer bounds \eqref{EqRiV2} and \eqref{EqRi+Rj-1-V2} are  cut-set bounds and were first reported in \cite{Bresler-ETT-2008} for the case without feedback. These outer bounds are still useful in the case of perfect channel-output feedback \cite{Suh-TIT-2011}. 
The outer bounds  \eqref{EqRi-2-V2}, \eqref{EqRi+Rj-2-V2} and \eqref{Eq2Ri+Rj-V2} are new.

The outer bound \eqref{EqRi-2-V2}  is on the individual rate of  transmitter-receiver $i$ when at channel use $n$, receiver $i$ is granted with the knowledge of the message index $W_j$  and all previous  feedback signals observed at transmitter $j$, i.e., $\bs{\overleftarrow{Y}}_{j,(1:n-1)}$. See for instance Figure \ref{Fig:G-IC-NF-Conv}(a) for the case $i=1$.
A complete proof of \eqref{EqRi-2-V2} is presented in Appendix \ref{App-C-LD-IC-NF}.

The outer bound \eqref{EqRi+Rj-2-V2} is  on the sum-rate of both transmitter-receiver pairs with enhanced versions of the receivers. More specifically,  for all $i \in \lbrace 1, 2 \rbrace$, receiver $i$ is granted at channel use $n$ with the knowledge of all previous feedback signals observed at transmitter $i$, i.e., $\bs{\overleftarrow{Y}}_{i,(1:n-1)}$. See for instance Figure \ref{Fig:G-IC-NF-Conv}(b).
A complete proof of \eqref{EqRi+Rj-2-V2} is presented in Appendix \ref{App-C-LD-IC-NF}.

The outer bounds \eqref{Eq2Ri+Rj-V2} are on the sum-rate of three receivers:  for all $i \in \lbrace 1, 2 \rbrace$, receiver $i$ is granted at channel use $n$ with all previous feedback signals observed at transmitter $i$, i.e., $\bs{\overleftarrow{Y}}_{i,(1:n-1)}$. For at most one $i \in \lbrace 1,2 \rbrace$, a third receiver has input $\overrightarrow{\bs{Y}}_{i,n}$ and it is granted with the knowledge of the message index  $W_j$ and  all previous feedback signals observed at transmitter $j$, i.e., $\bs{\overleftarrow{Y}}_{j,(1:n-1)}$. See for instance Figure \ref{Fig:G-IC-NF-Conv}(c) for the case $i =1$.
A complete proof of \eqref{Eq2Ri+Rj-V2} is presented in Appendix \ref{App-C-LD-IC-NF}.

\subsection{Connections with Previous Results}

Theorem~\ref{TheoremANFBLDMCap} generalizes previous results on the capacity region of the LD-IC with channel-output feedback. For instance, when $\overleftarrow{n}_{11} = 0$ and $\overleftarrow{n}_{22} = 0$, Theorem~\ref{TheoremANFBLDMCap} describes the capacity region of the LD-IC without feedback (Lemma~$4$ in \cite{Bresler-ETT-2008}); when  $\overleftarrow{n}_{11} \geqslant \max\left( \overrightarrow{n}_{11}, n_{12} \right)$ and  $\overleftarrow{n}_{22} \geqslant \max\left( \overrightarrow{n}_{22}, n_{21} \right)$, Theorem~\ref{TheoremANFBLDMCap} describes the capacity region of the LD-IC with perfect channel output feedback (Corollary $1$ in \cite{Suh-TIT-2011}); when  $\overrightarrow{n}_{11}=\overrightarrow{n}_{22}$, $ n_{12}=n_{21}$ and $\overleftarrow{n}_{11}=\overleftarrow{n}_{22}$,  Theorem~\ref{TheoremANFBLDMCap} describes the capacity region of the symmetric LD-IC with noisy channel output feedback (Theorem~$1$ in \cite{SyQuoc-TIT-2015}  and Theorem~$4.1$, case $1001$ in \cite{Sahai-TIT-2013}); and when $\overrightarrow{n}_{11}=\overrightarrow{n}_{22}$, $ n_{12}=n_{21}$, $\overleftarrow{n}_{ii}\geqslant \max\left( \overrightarrow{n}_{ii}, n_{ij} \right)$ and $\overleftarrow{n}_{jj}=0$, with $i \in \lbrace1, 2 \rbrace$ and $j \in \lbrace 1, 2 \rbrace\setminus\lbrace i \rbrace$, Theorem~\ref{TheoremANFBLDMCap} describes the capacity region of the symmetric LD-IC when only one of the receivers provides perfect channel-output feedback (Theorem~$4.1$, cases $1000$ and $0001$ in \cite{Sahai-TIT-2013}).

\section{Gaussian Channels} \label{SectMainResults}

This section introduces an achievable region (Theorem~\ref{TheoremA-G-IC-NF})  and a converse region (Theorem~\ref{TheoremC-G-IC-NF}), denoted by $\agicnof$ and $\cgicnof$ respectively, for a two-user G-IC-NF with fixed parameters $\overrightarrow{\SNR}_{1}$, $\overrightarrow{\SNR}_{2}$, $\INR_{12}$, $\INR_{21}$, $\overleftarrow{\SNR}_{1}$, and $\overleftarrow{\SNR}_{2}$.
In general, the capacity region of a given multi-user channel is said to be approximated to within a constant gap according to the following definition.

\begin{definition}[Approximation to within $\xi$ units]\label{DefGap}\emph{  A closed and convex set $\mathcal{T}\subset\mathbb{R}_{+}^{m}$ is approximated to within $\xi$ units by the sets $\underline{\mathcal{T}}$ and $\overline{\mathcal{T}}$ if  $\underline{\mathcal{T}} \subseteq \mathcal{T} \subseteq \overline{\mathcal{T}}$ and for all $\bs{t}=(t_1$, $t_2, \ldots$, $t_m)\in \overline{\mathcal{T}}$, $\Big(\left(t_1-\xi\right)^+$, $\left(t_2-\xi\right)^+, \ldots$, $\left(t_m-\xi\right)^+\Big) \in \underline{\mathcal{T}}$.}
\end{definition}

Denote by $\Cgicnof$ the capacity region of the 2-user G-IC-NF.  The achievable region $\agicnof$ (Theorem~\ref{TheoremA-G-IC-NF})  and the converse region $\cgicnof$ (Theorem~\ref{TheoremC-G-IC-NF}) approximate the capacity region $\Cgicnof$ to within $4.4$ bits (Theorem~\ref{TheoremGAP-G-IC-NF}).

\subsection{An Achievable Region for the Two-User G-IC-NF}
The description of the achievable region $\agicnof$ is presented using the constants $a_{1,i}$; the functions $a_{2,i}:[0,1] \rightarrow \mathds{R}_{+}$,  $a_{l,i}:[0,1]^2\rightarrow \mathds{R}_{+}$, with $l \in \lbrace 3, \ldots, 6 \rbrace$; and $a_{7,i}:[0,1]^3\rightarrow \mathds{R}_{+}$, which are defined as follows, for all $i \in \lbrace 1, 2 \rbrace$, with $j \in \lbrace 1, 2 \rbrace \setminus \lbrace i \rbrace$:

\begin{subequations}
\label{Eq-a}
\begin{IEEEeqnarray}{rcl}
\label{Eq-a1}
a_{1,i}  & = &  \frac{1}{2}\log \left(2+\frac{\overrightarrow{\SNR_{i}}}{\INR_{ji}}\right)-\frac{1}{2}, \\
\label{Eq-a2}
a_{2,i}(\rho) & = & \frac{1}{2}\log \Big(b_{1,i}(\rho)+1\Big)-\frac{1}{2}, \\
\nonumber
a_{3,i}(\rho,\mu) & = & \!  \frac{1}{2}\!\log\! \left(\! \frac{\! \overleftarrow{\SNR}_i \! \Big(b_{2,i}(\rho)+2\Big)+b_{1,i}(1)+1}{\overleftarrow{\SNR}_i\Big( \! \left(1\!-\!\mu\right) \! b_{2,i}( \rho )\!+\!2\Big)\!+\! b_{1,i}( 1  ) \!+ \!1} \! \right) \!, \\
\label{Eq-a3}\\
\label{Eq-a4}
a_{4,i}(\rho,\mu) & = & \frac{1}{2}\log \bigg(\Big(1-\mu\Big)b_{2,i}(\rho)+2 \bigg)-\frac{1}{2}, \\
\nonumber
a_{5,i}(\rho,\mu) & = & \frac{1}{2}\log \left(2+\frac{\overrightarrow{\SNR}_{i}}{\INR_{ji}}+\Big(1-\mu\Big)b_{2,i}(\rho)\right)-\frac{1}{2},\\
\label{Eq-a5} \\
\nonumber
a_{6,i}(\rho,\mu) & = & \frac{1}{2}\!\log\! \left(\!\frac{\overrightarrow{\SNR}_{i}}{\INR_{ji}}\bigg(\Big(1\!-\!\mu\Big)b_{2,j}(\rho)\!+\!1\bigg)\!+\!2\right)\!-\!\frac{1}{2}, \\
\label{Eq-a6}
\end{IEEEeqnarray}
and
\begin{IEEEeqnarray}{rcl}
\nonumber
a_{7,i}(\rho,\!\mu_1\!,\!\mu_2\!) & = & \! \frac{1}{2}\!\log \! \Bigg(\!\frac{\overrightarrow{\SNR}_{i}}{\INR_{ji}}\bigg(\Big(1\!-\!\mu_i\Big)b_{2,j}(\rho)\!+\!1\bigg) \\
\label{Eq-a7}
& & +\Big(1\!-\!\mu_j\Big)b_{2,i}(\rho)+2\Bigg)\!-\!\frac{1}{2}, 
\end{IEEEeqnarray}
\end{subequations}
where the functions $b_{l,i}:[0,1]\rightarrow \mathds{R}_{+}$, with $(l,i) \in \lbrace1, 2 \rbrace^2$ are defined as follows: 
\begin{subequations}
\label{Eqfnts}
\begin{IEEEeqnarray}{rcl}
\label{Eqb1i}
b_{1,i}(\rho)& = &\overrightarrow{\SNR}_{i}+2\rho\sqrt{\overrightarrow{\SNR}_{i}\INR_{ij}}+\INR_{ij} \mbox{ and } \\
\label{Eqb5i}
b_{2,i}(\rho)& = &\Big(1-\rho\Big)\INR_{ij}-1,
\end{IEEEeqnarray}
\end{subequations}
with $j \in \lbrace 1, 2 \rbrace \setminus \lbrace i \rbrace$.

\begin{figure*}[t]
\begin{theorem} \label{TheoremA-G-IC-NF} \emph{
The capacity region $\Cgicnof$ contains the region $\agicnof$ given by the closure of all non-negative rate pairs $(R_1,R_2)$ that satisfy
\begin{subequations}
\label{EqRa-G-IC-NF}
\begin{IEEEeqnarray}{rcl}
\label{EqR1a-G-IC-NF}
R_{1}  & \leqslant  & \min\Big(a_{2,1}(\rho),a_{6,1}(\rho,\mu_1)+a_{3,2}(\rho,\mu_1), a_{1,1}+a_{3,2}(\rho,\mu_1)+a_{4,2}(\rho,\mu_1)\Big),  \\ 
\label{EqR2a-G-IC-NF}
R_{2}   &  \leqslant  & \min\Big(a_{2,2}(\rho),a_{3,1}(\rho,\mu_2)+a_{6,2}(\rho,\mu_2), a_{3,1}(\rho,\mu_2)+a_{4,1}(\rho,\mu_2)+a_{1,2}\Big),   \\
\nonumber
R_{1}& + &R_{2}  \leqslant  \min\Big(a_{2,1}(\rho)+a_{1,2}, a_{1,1}+a_{2,2}(\rho), a_{3,1}(\rho,\mu_2)+a_{1,1}+a_{3,2}(\rho,\mu_1)+a_{7,2}(\rho,\mu_1,\mu_2), \\
\label{EqR1+R2a-G-IC-NF}
& & a_{3,1}(\rho,\mu_2) \! + \! a_{5,1}(\rho,\mu_2)+a_{3,2}(\rho,\mu_1)+a_{5,2}(\rho,\mu_1), a_{3,1}(\rho,\mu_2)+a_{7,1}(\rho,\mu_1,\mu_2)+a_{3,2}(\rho,\mu_1)+a_{1,2}\Big),  \\
\label{Eq2R1+R2a-G-IC-NF}
2R_{1}& + &R_{2}  \leqslant  \min\Big(a_{2,1}(\rho)+a_{1,1}+a_{3,2}(\rho,\mu_1)+a_{7,2}(\rho,\mu_1,\mu_2),  \\
\nonumber
& &  a_{3,1}(\rho,\mu_2)+a_{1,1}+a_{7,1}(\rho,\mu_1,\mu_2)+2a_{3,2}(\rho,\mu_1)+a_{5,2}(\rho,\mu_1), a_{2,1}(\rho)+a_{1,1}+a_{3,2}(\rho,\mu_1)+a_{5,2}(\rho,\mu_1)\Big), \\
\nonumber
R_{1}& + &2R_{2}  \leqslant  \min\Big(a_{3,1}(\rho,\mu_2)+a_{5,1}(\rho,\mu_2)+a_{2,2}(\rho)+a_{1,2}, a_{3,1}(\rho,\mu_2)+a_{7,1}(\rho,\mu_1,\mu_2)+a_{2,2}(\rho)+a_{1,2}, \\
\label{EqR1+2R2a-G-IC-NF}
& &  2a_{3,1}(\rho,\mu_2)+a_{5,1}(\rho,\mu_2)+a_{3,2}(\rho,\mu_1)+a_{1,2}+a_{7,2}(\rho,\mu_1,\mu_2)\Big),
\end{IEEEeqnarray}
\end{subequations}
with $\left(\rho, \mu_1, \mu_2\right) \in \left[0,\left(1-\max\left(\frac{1}{\INR_{12}},\frac{1}{\INR_{21}}\right) \right)^+\right]\times[0,1]\times[0,1]$.
}
\end{theorem}
\noindent\rule{16.5cm}{0.6pt}
\vspace{-5mm}
\end{figure*}
Finally, using this notation, Theorem~\ref{TheoremA-G-IC-NF} presents a new achievability region.

The proof of achievability is based on random coding arguments that make use of the insight obtained in the analysis of the LD-IC-NF. The techniques are rate splitting, block-Markov superposition coding, and backward decoding. The complete proof is described in Appendix \ref{AppAch-IC-NF}.  However, a brief description of the ideas leading to the construction of the achievability scheme are discussed hereunder. 

One of the central observations in the examples presented for the LD-IC-NF is that transmitters use correlated codewords.
This stems from the fact that a fraction of the bits sent by transmitter $i$ are received by receiver $j$; fed back to transmitter $j$ and finally, retransmitted by transmitter $j$. See for instance \cite{Suh-TIT-2011, Perlaza-TIT-2015, SyQuoc-TIT-2015, Yang-Tuninetti-TIT-2011}, and \cite{Tuninetti-ITA-2010}.
This observation is the driving idea in the construction of the achievability schemes presented in previous works and it is central in the proof of Theorem \ref{TheoremA-G-IC-NF}.

Let the message index sent by transmitter $i$ during the $t$-th block be denoted by $W_i^{(t)} \in \lbrace1, 2,  \ldots, 2^{N R_i}\rbrace$. Following a rate-splitting argument, assume that $W_i^{(t)}$ is represented by three subindices $(W_{i,C1}^{(t)}, W_{i,C2}^{(t)}, W_{i,P}^{(t)}) \in \lbrace 1, 2,  \ldots, 2^{NR_{i,C1}} \rbrace \times \lbrace 1, 2,  \ldots, 2^{NR_{i,C2}} \rbrace \times \lbrace 1, 2, \ldots, 2^{NR_{i,P}} \rbrace$, where $R_{i,C1} + R_{i,C2}+R_{i,P} = R_{i}$. The number of rate splits is based on the fact that each symbol in the LD-IC-NF was divided into three parts.

The codeword generation follows a four-level superposition coding scheme. The number of layers is the number of rate splits plus an additional common layer that accounts for the correlation between codewords. This correlation is induced as follows.
The index  $W_{i,C1}^{(t-1)}$ is assumed to be decoded at transmitter $j$ via the feedback link of transmitter-receiver pair $j$ at the end of the transmission of block $t-1$. Therefore, at the beginning of block $t$, each transmitter possesses the knowledge of the indices $W_{1,C1}^{(t-1)}$ and $W_{2,C1}^{(t-1)}$. 
Using these indices both transmitters are able to identify the same codeword in the first code-layer. 
In the case of the first block $t = 1$ and the last block $t = T$, the indices  $W_{1,C1}^{(0)}$, $W_{2,C1}^{(0)}$, $W_{1,C1}^{(T)}$, and $W_{2,C1}^{(T)}$ correspond to indices assumed to be known by all transmitters and receivers.

The first code-layer is a sub-codebook of $2^{N(R_{1,C1} + R_{2,C1})}$ codewords (see Figure~\ref{FigSuperpos}). Denote by $\bs{u}\Big(W_{1,C1}^{(t-1)}$, $W_{2,C1}^{(t-1)}\Big)$ the corresponding codeword in the first code-layer.  
The second codeword used by transmitter $i$ is selected using $W_{i,C1}^{(t)}$ from the second code-layer, which is a sub-codebook of $2^{N\,R_{i,C1}}$ codewords specific to $\bs{u}\left(W_{1,C1}^{(t-1)},W_{2,C1}^{(t-1)}\right)$ as shown in Figure~\ref{FigSuperpos}. Denote by $\bs{u}_i\left(W_{1,C1}^{(t-1)},W_{2,C1}^{(t-1)},W_{i,C1}^{(t)}\right)$ the corresponding codeword in the second code-layer.  
The third codeword used by transmitter $i$ is selected using  $W_{i,C2}^{(t)}$ from the third code-layer, which is a sub-codebook of $2^{N\,R_{i,C2}}$ codewords specific to $\bs{u}_i\left(W_{1,C1}^{(t-1)},W_{2,C1}^{(t-1)},W_{t,C1}^{(t)}\right)$ as shown in Figure~\ref{FigSuperpos}. Denote by $\bs{v}_i\left(W_{1,C1}^{(t-1)},W_{2,C1}^{(t-1)},W_{i,C1}^{(t)},W_{i,C2}^{(t)}\right)$ the corresponding codeword in the third code-layer.  
The fourth codeword used by transmitter $i$ is selected using $W_{i,P}^{(t)}$ from the fourth code-layer, which is a sub-codebook of $2^{N\,R_{i,P}}$ codewords specific to $\bs{v}_i\left(W_{1,C1}^{(t-1)},W_{2,C1}^{(t-1)},W_{i,C1}^{(t)},W_{i,C2}^{(t)}\right)$ as shown in Figure~\ref{FigSuperpos}. Denote by $\bs{x}_{i,P}\Big(W_{1,C1}^{(t-1)}$, $W_{2,C1}^{(t-1)}$, $W_{i,C1}^{(t)}$, $W_{i,C2}^{(t)}$, $W_{i,P}^{(t)}\Big)$ the corresponding codeword in the fourth code-layer.   
Finally, the channel input sequence at transmitter $i$ during block $t$ is denoted by $\bs{x}_{i,t} = \left( x_{i,t,1}, x_{i,t,2}, \ldots, x_{i,t,N}\right)$, with $t \in \lbrace 1, 2,  \ldots, T \rbrace$, and it is a weighted sum of the codewords $\bs{u}\Big(W_{1,C1}^{(t-1)},W_{2,C1}^{(t-1)}\Big)$, $\bs{u}_i\Big(W_{1,C1}^{(t-1)},W_{2,C1}^{(t-1)},W_{i,C1}^{(t)}\Big)$, $\bs{v}_i\Big(W_{1,C1}^{(t-1)},W_{2,C1}^{(t-1)},W_{i,C1}^{(t)},W_{i,C2}^{(t)}\Big)$ and $\bs{x}_{i,P}\Big(W_{1,C1}^{(t-1)}$, $W_{2,C1}^{(t-1)}$, $W_{i,C1}^{(t)}$, $W_{i,C2}^{(t)}$, $W_{i,P}^{(t)}\Big)$.

At receiver $i$, the decoding follows a backward decoding approach as in previous works \cite{Suh-TIT-2011, Perlaza-TIT-2015, SyQuoc-TIT-2015}, among others.

\subsection{A Converse Region for the Two-User G-IC-NF}

The description of the converse region $\cgicnof$ is determined by two events denoted by $S_{l_{1},1}$ and $S_{l_{2},2}$, where $(l_{1},l_{2}) \in \lbrace 1, \ldots, 5 \rbrace^2$. The events are defined as follows:
 \begin{subequations}
\label{EqSi}
\begin{IEEEeqnarray}{rcl}
\label{EqS1i}  
 S_{1,i}&: \quad & \overrightarrow{\SNR}_{j} < \min\left(\INR_{ij},\INR_{ji}\right), \\ 
\label{EqS2i}
 S_{2,i}&: \quad & \INR_{ji} \leqslant  \overrightarrow{\SNR}_{j} < \INR_{ij},\\
 \label{EqS3i}
 S_{3,i}&: \quad & \INR_{ij} \leqslant \overrightarrow{\SNR}_{j} < \INR_{ji}, \\
 \label{EqS4i}
 S_{4,i}&: \quad & \max\left(\INR_{ij}, \INR_{ji}\right) \leqslant \overrightarrow{\SNR}_{j} < \INR_{ij}\INR_{ji}, \qquad \\
 \label{EqS5i}
 S_{5,i}&: \quad & \overrightarrow{\SNR}_{j} \geqslant  \INR_{ij}\INR_{ji}.
\end{IEEEeqnarray}
\end{subequations}
Note that for all $i \in \lbrace 1, 2 \rbrace$, the events $S_{1,i}$, $S_{2,i} $, $S_{3,i}$, $S_{4,i}$, and $S_{5,i}$ are mutually exclusive. This observation shows that given any $4$-tuple $(\overrightarrow{\SNR}_{1}, \overrightarrow{\SNR}_{2}, \INR_{12}, \INR_{21})$, there always exists one and only one pair of events  $(S_{l_{1},1}, S_{l_{2},2})$, with $(l_{1},l_{2}) \in \lbrace 1, \ldots, 5 \rbrace^2$, that determines a unique scenario. Note also that the pairs of events $(S_{2,1}, S_{2,2})$ and $(S_{3,1}, S_{3,2})$ are not feasible. In view of this, twenty-three different scenarios can be identified using the events in \eqref{EqSi}.
Once the exact scenario is identified, the converse region is described using the functions $\kappa_{l,i}: [0,1]\rightarrow \mathds{R}_{+}$, with $l \in \lbrace 1, \ldots, 3\rbrace$; $\kappa_{l}: [0,1]\rightarrow \mathds{R}_{+}$, with $l \in \lbrace 4, 5 \rbrace$; $\kappa_{6,l}: [0,1]\rightarrow \mathds{R}_{+}$, with $l \in \lbrace 1, \ldots, 4\rbrace$; and $\kappa_{7,i,l}:[0,1]\rightarrow \mathds{R}_{+}$, with $l \in \lbrace 1, 2 \rbrace$. These functions are defined as follows, for all $i \in \lbrace 1, 2 \rbrace$, with $j \in \lbrace 1, 2 \rbrace\setminus\lbrace i \rbrace$:
\begin{subequations}
\label{Eqconv}
\begin{IEEEeqnarray}{rcl}
\label{Eqconv1}
\kappa_{1,i}(\rho)  &  =  & \frac{1}{2}\log \Big(b_{1,i}(\rho)+1\Big), \\
\label{Eqconv2}
\kappa_{2,i}(\rho)  &  =  & \frac{1}{2}\log \Big(1+b_{5,j}(\rho)\Big) \!+\! \frac{1}{2}\log \Bigg(1\!+\! \frac{b_{4,i}(\rho)}{1+b_{5,j}(\rho)}\Bigg)\!, \, \qquad  \\
\nonumber
\kappa_{3,i}(\rho)  &  =  & \frac{1}{2} \! \log \! \left(\frac{\bigg(b_{4,i}(\rho)+b_{5,j}(\rho)+1\bigg)\overleftarrow{\SNR}_j}{\bigg(b_{1,j}(1) \! + \! 1\bigg)\bigg(b_{4,i}(\rho) \! + 1 \! \bigg)} \! + \! 1 \! \right) \\
\label{Eqconv3}
& & +\frac{1}{2}\log\Big(b_{4,i}(\rho)+1\Big), \\
\label{Eqconv4}
\kappa_{4}(\rho)  & = & \frac{1}{2}\log \! \Bigg(1+\frac{b_{4,1}(\rho)}{1+b_{5,2}(\rho)}\Bigg) \! + \! \frac{1}{2}\log \! \Big(b_{1,2}(\rho)\!+\!1\!\Big)\!, \qquad\\
\label{Eqconv5}
\kappa_{5}(\rho)  & = & \frac{1}{2}\log \Bigg(1 \! + \! \frac{b_{4,2}(\rho)}{1 \!+ \! b_{5,1}(\rho)}\Bigg) \! + \! \frac{1}{2}\log \Big(b_{1,1}(\rho) \! + \! 1\Big), \\
\label{Eqk6} 
\kappa_{6}(\rho) &  = & \begin{cases} \kappa_{6,1}(\rho) & \textrm{if }  (S_{1,2} \lor S_{2,2} \lor S_{5,2} ) \\
                                                                               & \hspace{4mm} \land (S_{1,1} \lor S_{2,1} \lor S_{5,1} )\\
                                                \kappa_{6,2} (\rho) & \textrm{if }(S_{1,2} \lor S_{2,2} \lor S_{5,2} ) \\
                                                                               & \hspace{4mm}  \land (S_{3,1} \lor S_{4,1} )\\
                                                \kappa_{6,3}(\rho) & \textrm{if } (S_{3,2} \lor S_{4,2}) \\
                                                                              & \hspace{4mm}  \land (S_{1,1} \lor S_{2,1} \lor S_{5,1})\\
                                                \kappa_{6,4}(\rho) & \textrm{if } (S_{3,2} \lor S_{4,2} ) \\
                                                                              & \hspace{4mm}  \land ( S_{3,1} \lor S_{4,1}),
                                           \end{cases}  \\                                         
\label{Eqk77}
\kappa_{7,i}(\rho) & = & \begin{cases} \kappa_{7,i,1}(\rho) & \textrm{if } (S_{1,i} \lor S_{2,i} \lor S_{5,i})\\
                                                  \kappa_{7,i,2}(\rho) & \textrm{if } (S_{3,i} \lor S_{4,i}),
                                                \end{cases} 
\end{IEEEeqnarray}
\end{subequations}
where,
\begin{subequations}
\label{Eqconv6}
\begin{IEEEeqnarray}{rcl}
\nonumber
\kappa_{6,1}&(&\rho)  =  \frac{1}{2} \! \log \! \Big(b_{1,1}(\rho) \! + \! b_{5,1}(\rho)\INR_{21}\Big) \! - \! \frac{1}{2}\log\Big(1 \! + \! \INR_{12}\Big) 
\end{IEEEeqnarray}
\begin{IEEEeqnarray}{rcl}\nonumber
& & +\frac{1}{2}\log\left(1+\frac{b_{5,2}(\rho)\overleftarrow{\SNR}_2}{b_{1,2}(1)+1}\right)\\
\nonumber
& & +\frac{1}{2}\log\Big(b_{1,2}(\rho)+b_{5,1}(\rho)\INR_{21}\Big) -\frac{1}{2}\log\Big(1 \! + \! \INR_{21}\Big) \! \\
\label{Eqconv61}
& & + \! \frac{1}{2}\log\left(\! 1 \! + \! \frac{b_{5,1}(\rho)\overleftarrow{\SNR}_1}{b_{1,1}(1)+1}\right)+\log(2 \pi e), \\
\nonumber   
\kappa_{6,2}&(&\rho)  =    \frac{1}{2}\log\left(b_{6,2}(\rho)+\frac{b_{5,1}(\rho)\INR_{21}}{\overrightarrow{\SNR}_2}\Big(\overrightarrow{\SNR}_2+b_{3,2}\Big)\right) \\
\nonumber
& & - \frac{1}{2}\log\Big(1 \!+ \! \INR_{12}\Big) +\frac{1}{2}\log\left(1+\frac{b_{5,1}(\rho)\overleftarrow{\SNR}_1}{b_{1,1}(1)+1}\right) \\
\nonumber
& & + \frac{1}{2} \! \log\Big(b_{1,1}(\rho) \! + \! b_{5,1}(\rho)\INR_{21}\Big) \! -\frac{1}{2}\log\Big(1+\INR_{21}\Big)\\
\nonumber
& & +\frac{1}{2}\log\Bigg(1+\frac{b_{5,2}(\rho)}{\overrightarrow{\SNR}_2}\left(\INR_{12}+\frac{b_{3,2} \overleftarrow{\SNR}_2}{b_{1,2}(1)+1}\right)\Bigg)\\
\label{Eqconv62}
& & -\frac{1}{2}\log\left(1+\frac{b_{5,1}(\rho)\INR_{21}}{\overrightarrow{\SNR}_2}\right)+\log(2 \pi e), \\
\nonumber
\kappa_{6,3}&(&\rho)  =  \frac{1}{2}\log\Bigg(b_{6,1}(\rho)+\frac{b_{5,1}(\rho)\INR_{21}}{\overrightarrow{\SNR}_1}\Big(\overrightarrow{\SNR}_1+b_{3,1}\Big)\Bigg)\\
\nonumber
& & -\frac{1}{2}\log\Big(1+\INR_{12}\Big)+\frac{1}{2}\log\left(1+\frac{b_{5,2}(\rho)\overleftarrow{\SNR}_2}{b_{1,2}(1)+1}\right) \\ 
\nonumber
& & +\frac{1}{2}\log\Big(b_{1,2}(\rho) \! + \! b_{5,1}(\rho)\INR_{21}\Big) \! - \! \frac{1}{2}\log\Big(1 \! + \! \INR_{21}\Big)\\
\nonumber
& & +\frac{1}{2}\log\Bigg(1+\frac{b_{5,1}(\rho)}{\overrightarrow{\SNR}_1}\left(\INR_{21}+\frac{b_{3,1} \overleftarrow{\SNR}_1}{b_{1,1}(1)+1}\right)\Bigg)\\
\label{Eqconv63}
& & -\frac{1}{2}\log\left(1+\frac{b_{5,1}(\rho)\INR_{21}}{\overrightarrow{\SNR}_1}\right) +\log(2 \pi e), \\
\nonumber
\kappa_{6,4}&(&\rho)=\frac{1}{2}\log\left(b_{6,1}(\rho)+\frac{b_{5,1}(\rho)\INR_{21}}{\overrightarrow{\SNR}_1}\Big(\overrightarrow{\SNR}_1+b_{3,1}\Big)\right)\\
\nonumber
& & -\frac{1}{2}\log\Big(1+\INR_{12}\Big)-\frac{1}{2}\log\Big(1+\INR_{21}\Big)\\
\nonumber
& & +\frac{1}{2}\log\left(1+\frac{b_{5,2}(\rho)}{\overrightarrow{\SNR}_2}\left(\INR_{12}+\frac{b_{3,2}\overleftarrow{\SNR}_2}{b_{1,2}(1)+1}\right)\right)\\
\nonumber
& & -\frac{1}{2}\log\left(1+\frac{b_{5,1}(\rho)\INR_{21}}{\overrightarrow{\SNR}_2}\right)\\
\nonumber
& & -\frac{1}{2}\log\left(1+\frac{b_{5,1}(\rho)\INR_{21}}{\overrightarrow{\SNR}_1}\right) \\
\nonumber
& & +\frac{1}{2}\log\left(b_{6,2}(\rho)+\frac{b_{5,1}(\rho)\INR_{21}}{\overrightarrow{\SNR}_2}\Big(\overrightarrow{\SNR}_2+b_{3,2}\Big)\right)\\
\nonumber
& & +\frac{1}{2}\log\Bigg(1+\frac{b_{5,1}(\rho)}{\overrightarrow{\SNR}_1}\left(\INR_{21}+\frac{b_{3,1} \overleftarrow{\SNR}_1}{b_{1,1}(1)+1}\right)\Bigg)\\
\label{Eqconv64}
& & +\log(2 \pi e), 
\end{IEEEeqnarray}
\end{subequations}
and
\begin{subequations}
\label{Eqconv7i}
\begin{IEEEeqnarray}{rcl}
\nonumber
\kappa_{7,i,1}&(&\rho) =\frac{1}{2}\log\Big(b_{1,i}(\rho)+1\Big)-\frac{1}{2}\log\Big(1+\INR_{ij}\Big)\\
\nonumber
& & \! + \!\frac{1}{2}\log\left(1+\frac{b_{5,j}(\rho)\overleftarrow{\SNR}_j}{b_{1,j}(1)+1}\right) + \! 2\log(2 \pi e) \\
\nonumber
& & \! + \! \frac{1}{2}\log\Big(b_{1,j}(\rho)+b_{5,i}(\rho)\INR_{ji}\Big) \! 
\end{IEEEeqnarray}
\begin{IEEEeqnarray}{rcl}
\label{Eqconv7i1}
& & + \! \frac{1}{2}\log\Big(1 \! + \! b_{4,i}(\rho) \!+ \! b_{5,j}(\rho) \Big) \! - \! \frac{1}{2}\log\Big(1 \! + \! b_{5,j}(\rho)\Big) \! \\
\nonumber
\kappa_{7,i,2}&(&\rho) = \frac{1}{2}\log\Big(b_{1,i}(\rho)+1\Big)-\frac{1}{2}\log\Big(1+\INR_{ij}\Big)\\
\nonumber
& & -\frac{1}{2}\log\Big(1+b_{5,j}(\rho)\Big) \! + \! \frac{1}{2}\log\Big(1+b_{4,i}(\rho)+b_{5,j}(\rho) \Big) \quad\\
\nonumber
& & +\frac{1}{2} \! \log\Bigg( \!1\!+\!\Big(1 \! - \! \rho^2\Big)\frac{\INR_{ji}}{\overrightarrow{\SNR}_j}\Bigg( \! \INR_{ij} \! + \! \frac{b_{3,j}\overleftarrow{\SNR}_j}{b_{1,j}(1)+1}\Bigg)\Bigg)\\
\nonumber
& & -\frac{1}{2}\log\left(1+\frac{b_{5,i}(\rho)\INR_{ji}}{\overrightarrow{\SNR}_j}\right)\\
\nonumber
& & +\frac{1}{2}\!\log\!\left(\!b_{6,j}(\rho)\!+\!\frac{b_{5,i}(\rho)\INR_{ji}}{\overrightarrow{\SNR}_j}\Big(\overrightarrow{\SNR}_j+b_{3,j}\Big)\right)\\
\label{Eqconv7i2}
& & +2\log(2 \pi e).
\end{IEEEeqnarray}
\end{subequations}
The functions $b_{l,i}$, with $(l,i) \in \lbrace1, 2 \rbrace^2$ are defined in \eqref{Eqfnts}; $b_{3,i}$ are constants; and the functions $b_{l,i}:[0,1]\rightarrow \mathds{R}_{+}$, with $(l,i) \in \lbrace 4, 5, 6 \rbrace\times\lbrace1, 2 \rbrace$ are defined as follows, with $j \in \lbrace 1, 2 \rbrace \setminus \lbrace i \rbrace$:
\begin{subequations}
\label{Eqfnts2}
\begin{IEEEeqnarray}{rcl}
\label{Eqb2i}
b_{3,i}& = &\overrightarrow{\SNR}_i-2\sqrt{\overrightarrow{\SNR}_i\INR_{ji}}+\INR_{ji}, \\
\label{Eqb3i}
b_{4,i}(\rho)& = &\Big(1-\rho^2\Big)\overrightarrow{\SNR}_{i}, \\
\label{Eqb4i}
b_{5,i}(\rho)& = &\Big(1-\rho^2\Big)\INR_{ij},\\
\nonumber
b_{6,i}(\rho)& = &\overrightarrow{\SNR}_i \! + \! \INR_{ij} \!+ \! 2\rho\sqrt{\INR_{ij}}\left(\sqrt{\overrightarrow{\SNR}_i}-\sqrt{\INR_{ji}}\right) \! \\
\label{Eqb6i}
& & + \frac{\INR_{ij}\sqrt{\INR_{ji}}}{\overrightarrow{\SNR}_i} \left(\sqrt{\INR_{ji}}-2\sqrt{\overrightarrow{\SNR}_i}\right). \qquad
\end{IEEEeqnarray}
\end{subequations}
Finally, using this notation, Theorem~\ref{TheoremC-G-IC-NF} is presented below.
\begin{theorem} \label{TheoremC-G-IC-NF} \emph{
The capacity region $\Cgicnof$ is contained within the region $\cgicnof$ given by the closure of the set of non-negative rate pairs  $(R_1,R_2)$ that for all $i \in \lbrace 1, 2 \rbrace$, with $j\in\lbrace 1, 2 \rbrace\setminus\lbrace i \rbrace$, satisfy:
\begin{subequations}\label{EqRic-G-IC-NF}
\begin{IEEEeqnarray}{rcl}
\label{EqRic-12-G-IC-NF}
R_{i}  &  \leqslant  & \min\Big(\kappa_{1,i}(\rho), \kappa_{2,i}(\rho)\Big), \\ 
\label{EqRic-3-G-IC-NF}
R_{i}  &  \leqslant  & \kappa_{3,i}(\rho), \\
\label{EqR1+R2c-12-G-IC-NF}
R_{1}+R_{2}  &  \leqslant  & \min\Big(\kappa_{4}(\rho), \kappa_{5}(\rho)\Big),\\
\label{EqR1+R2c-3g-G-IC-NF}
R_{1}+R_{2}  &  \leqslant  &\kappa_{6}(\rho),\\
\label{Eq2Ri+Rjc-g-G-IC-NF}
2R_i+R_j&  \leqslant &  \kappa_{7,i}(\rho),
\end{IEEEeqnarray}
\end{subequations}
with $\rho \in [0,1]$.
}
\end{theorem} 
\begin{IEEEproof}
The proof of Theorem~\ref{TheoremC-G-IC-NF} is presented in Appendix \ref{App-C-G-IC-NF}.
\end{IEEEproof}

The outer bounds \eqref{EqRic-12-G-IC-NF} and \eqref{EqR1+R2c-12-G-IC-NF} play the same role as the outer bounds \eqref{EqRiV2} and \eqref{EqRi+Rj-1-V2} in the linear deterministic model and have been previously reported in \cite{Suh-TIT-2011} for the case of perfect channel-output feedback.
The bounds \eqref{EqRic-3-G-IC-NF}, \eqref{EqR1+R2c-3g-G-IC-NF}, and \eqref{Eq2Ri+Rjc-g-G-IC-NF} correspond to new outer bounds.    
The intuition for deriving these outer bounds follows along the same lines of the outer bounds \eqref{EqRi-2-V2}, \eqref{EqRi+Rj-2-V2}, and \eqref{Eq2Ri+Rj-V2} in the LD-IC-NF, respectively. 

\subsection{Connections with Previous Results}

Theorem~\ref{TheoremA-G-IC-NF} generalizes previous results on the achievable region of the two-user G-IC with channel-output feedback. For instance, when $\overleftarrow{\SNR}_{1} = 0$, $\overleftarrow{\SNR}_{2} = 0$, and $\rho=0$, Theorem~\ref{TheoremA-G-IC-NF} describes the achievable region of the G-IC without feedback \cite{Han-TIT-1981, Chong-ITA-2006, Chong-TIT-2008}; when  $\overleftarrow{\SNR}_{1} \rightarrow \infty$ and $\overleftarrow{\SNR}_{2} \rightarrow \infty$, Theorem~\ref{TheoremA-G-IC-NF} describes the achievable region of the G-IC-PF (Theorem $2$ in \cite{Suh-TIT-2011}); when  $\overrightarrow{\SNR}_{1}=\overrightarrow{\SNR}_{2}$, $\INR_{12}=\INR_{21}$ and $\overleftarrow{\SNR}_{1}=\overleftarrow{\SNR}_{2}$,  Theorem~\ref{TheoremA-G-IC-NF} describes the achievable region of the symmetric G-IC-NF (Theorem~$3$ in \cite{SyQuoc-TIT-2015}). Theorem~\ref{TheoremC-G-IC-NF} generalizes previous results on the converse region of the two-user G-IC with channel-output feedback. For instance, when $\overleftarrow{\SNR}_{1} = 0$, $\overleftarrow{\SNR}_{2} = 0$, and $\rho=0$, Theorem~\ref{TheoremC-G-IC-NF} describes the converse region of the G-IC without feedback \cite{Etkin-TIT-2008}; when  $\overleftarrow{\SNR}_{1} \rightarrow \infty$ and $\overleftarrow{\SNR}_{2} \rightarrow \infty$, Theorem~\ref{TheoremC-G-IC-NF} describes the converse region of the G-IC-PF (Theorem $3$ in \cite{Suh-TIT-2011}); when  $\overrightarrow{\SNR}_{1}=\overrightarrow{\SNR}_{2}$, $\INR_{12}=\INR_{21}$ and $\overleftarrow{\SNR}_{1}=\overleftarrow{\SNR}_{2}$,  Theorem~\ref{TheoremC-G-IC-NF} describes the converse region of the symmetric G-IC-NF (Theorem~$2$ in \cite{SyQuoc-TIT-2015}).

\subsection{A Gap Between the Achievable Region and the Converse Region}

Theorem~\ref{TheoremGAP-G-IC-NF} describes the gap between the achievable region $\agicnof$ and the converse region $\cgicnof$ (Definition \ref{DefGap}).

\begin{theorem} \label{TheoremGAP-G-IC-NF} \emph{The capacity region of the two-user G-IC-NF is approximated to within $4.4$ bits by the achievable region $\agicnof$ and the converse region $\cgicnof$.
}
\end{theorem} 

\begin{figure}[t]
\centerline{\epsfig{figure=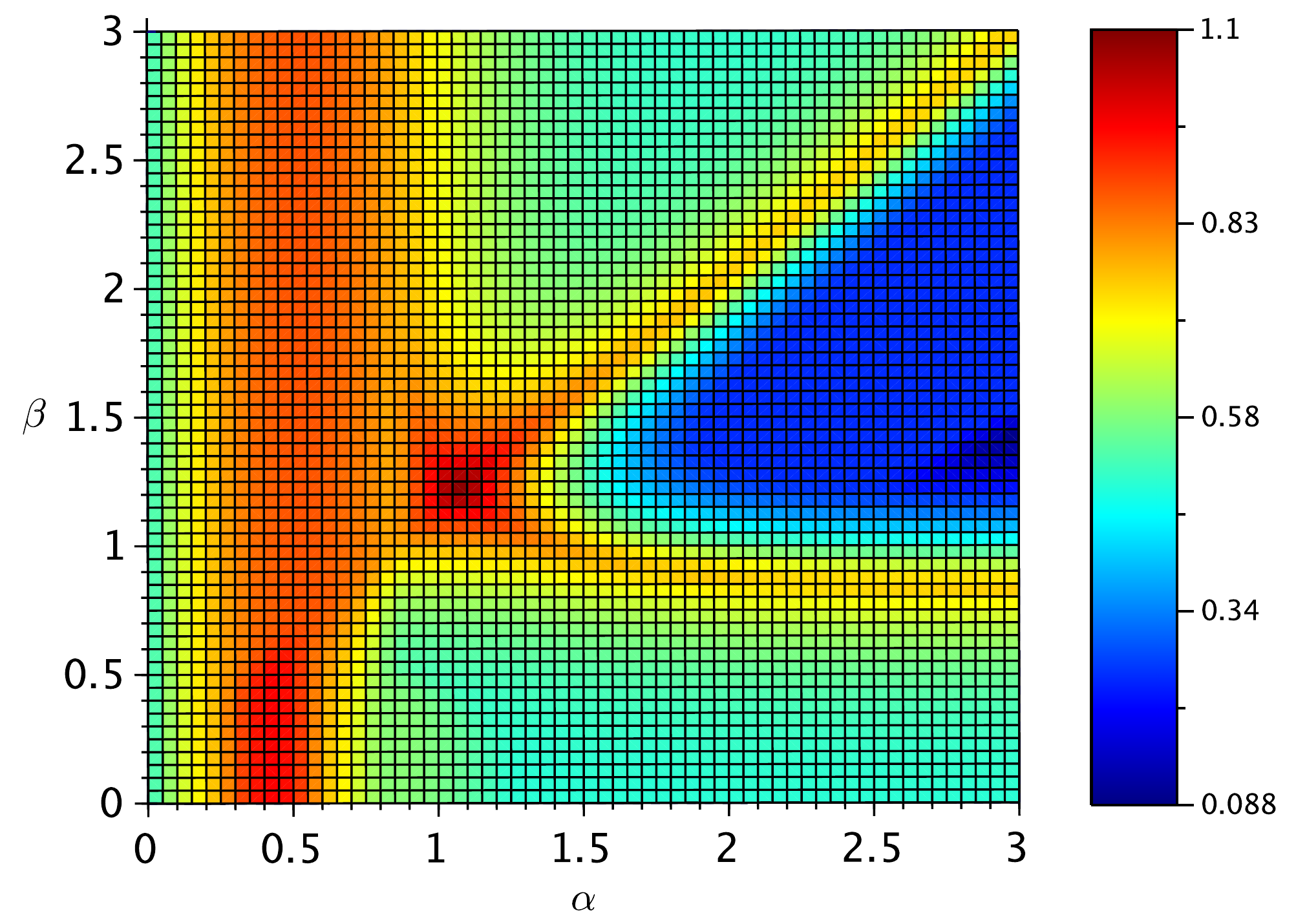,width=0.5\textwidth}}
\caption{Gap between the converse region $\cgicnof$ and the achievable region $\agicnof$ of the two-user G-IC-NF, under symmetric channel conditions, i.e., $\protect\overrightarrow{\SNR}_1=\protect\overrightarrow{\SNR}_2=\protect\overrightarrow{\SNR}$, $\INR_{12}=\INR_{21}=\INR$, and $\protect\overleftarrow{\SNR}_1=\protect\overleftarrow{\SNR}_2=\protect\overleftarrow{\SNR}$, as a function of $\alpha=\frac{\log \INR}{\log{\protect\overrightarrow{\SNR}}}$ and $\beta=\frac{\log\protect\overleftarrow{\SNR}}{\log\protect\overrightarrow{\SNR}}$.}
\label{FigGapGICNF}
\end{figure}

\begin{IEEEproof}
The proof of Theorem~\ref{TheoremGAP-G-IC-NF} is presented in Appendix \ref{AppG-Gap}.
\end{IEEEproof}
To the extent of the knowledge of the authors, this approximation to the capacity region of the G-IC-NF is the most general with respect to existing literature and the one that guarantees the smallest gap between the achievable and converse regions when feedback links are subject to Gaussian additive noise.
Figure~\ref{FigGapGICNF} presents the exact gap existing between the achievable region $\agicnof$ and the converse region $\cgicnof$ for the case in which $\overrightarrow{\SNR}_1=\overrightarrow{\SNR}_2=\overrightarrow{\SNR}$, $\INR_{12}=\INR_{21}=\INR$, and $\overleftarrow{\SNR}_1=\overleftarrow{\SNR}_2=\overleftarrow{\SNR}$ as a function of $\alpha=\frac{\log \INR}{\log{\overrightarrow{\SNR}}}$ and $\beta=\frac{\log\overleftarrow{\SNR}}{\log\overrightarrow{\SNR}}$. Note that in this case, the maximum gap is $1.1$ bits and occurs when $\alpha=1.05$ and $\beta=1.2$.

\section{Transmitter Cooperation via Feedback}\label{SecCooperation}
 
Despite the fundamental differences discussed in \cite{SyQuoc-TIT-2015}, there exist several similarities between the G-IC-NF and the G-IC with conferencing transmitters (G-IC-CT).  For instance, there exist cases in which the side information obtained is the same for both the transmitter cooperation case and the  channel-output feedback case.

\begin{figure*}[t]
 \centerline{\epsfig{figure=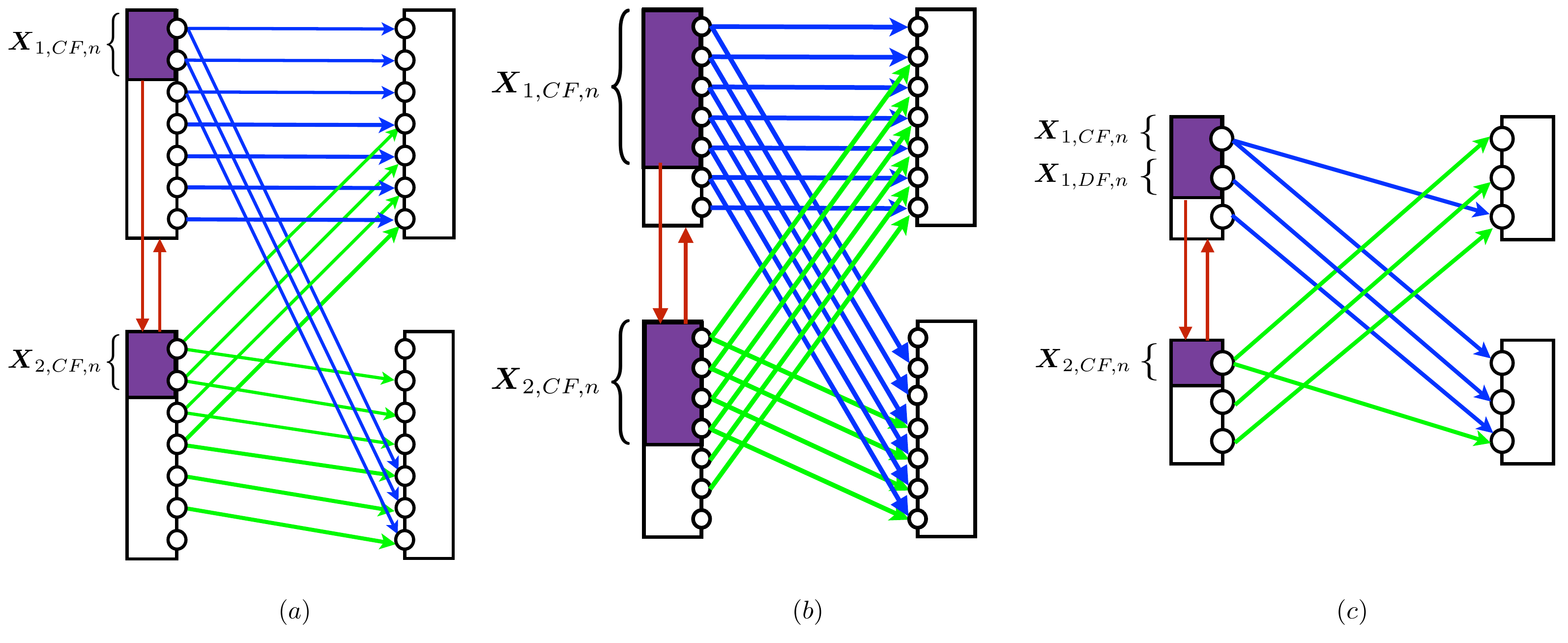,width=0.9\textwidth}}
 \caption{Example of a LD-IC with Conferencing Transmitters.}
\label{FigMessagesLDICNFBCT}
\end{figure*}

\begin{figure*}[t]
\vspace{+5mm}
 \centerline{\epsfig{figure=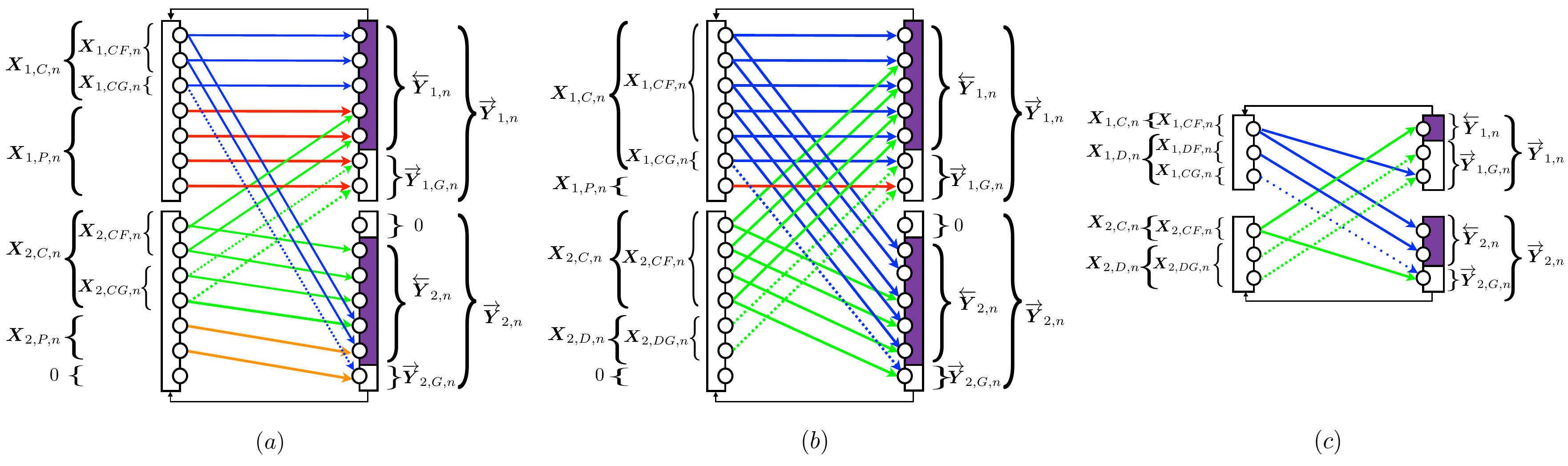,width=1.0\textwidth}}
 \caption{Example of the notation used in Appendix \ref{App-C-LD-IC-NF}.}
\label{FigMessagesLDICNFB}
\end{figure*}

Consider for instance the LD-IC with conferencing-transmitters (LD-IC-CT) depicted in Figure~\ref{FigMessagesLDICNFBCT} and the LD-IC-NF depicted in Figure~\ref{FigMessagesLDICNFB}. Note that the noise level in the link from transmitter $i$ to transmitter $j$ is such that transmitter $j$ observes only the bits sent by transmitter $i$ through the bit-pipes $\bs{X}_{i,CF,n}$ and $\bs{X}_{i,DF,n}$ (see Figure \ref{FigMessagesLDICNFBCT} and Figure~\ref{FigMessagesLDICNFB}), for all $i \in \lbrace 1,2 \rbrace$ and $j \in \lbrace 1, 2 \rbrace \setminus \lbrace i \rbrace$. Then, in this particular case, subject to a finite delay, in both channel models (Figure~\ref{FigMessagesLDICNFBCT} and Figure~\ref{FigMessagesLDICNFB}) the corresponding transmitters possess the same side information and the corresponding receivers observe the same channel outputs. From this point of view, any outer bound on the individual or weighted sum-rate for the case of conferencing transmitters is also an outer bound for the case of channel-output feedback and vice-versa.
 A similar observation can be made for the case of a G-IC-NF and a  G-IC with conferencing transmitters (G-IC-CT).

Outer bounds on the sum-rate for the LD-IC-CT and the G-IC-CT have been reported in \cite{Prabhakaran-TIT-2011}. The gap between the achievable sum-rate and the corresponding outer bound in the G-IC-CT is $20$ bits in \cite{Prabhakaran-TIT-2011}. Other outer bounds have been presented in \cite{Host-Madsen-TIT-2006}, but the gap between the achievable and converse regions is not reported. 


\section{Concluding Remarks} \label{SecConclusions}

In this paper, the exact capacity region of the LD-IC-NF has been fully characterized (Theorem~\ref{TheoremANFBLDMCap}). Exploiting the insight obtained in the LD-IC-NF, an achievability region (Theorem~\ref{TheoremA-G-IC-NF}) and a converse region (Theorem~\ref{TheoremC-G-IC-NF}) have been presented for the two-user G-IC-NF. These two regions approximate the capacity region of the G-IC-NF to within $4.4$ bits (Theorem~\ref{TheoremGAP-G-IC-NF}).  
 
Despite the contributions made in this paper, several questions remain unsolved in the understanding of the benefits of channel-output feedback in the G-IC-NF. For instance, the case in which the channel-output feedback is observed by both transmitters is still an open problem. Only the case with symmetric channels has been fully studied.  Another case in which very little is known about the benefits of channel-output feedback is that of a large number of users (more than two) and large number of antennas (more than one) at each network component.  
 
\begin{appendices}

\section{Proof of Achievability} \label{AppAch-IC-NF}

This appendix describes an achievability scheme for the IC-NF based on a three-part message splitting, block-Markov superposition coding, and backward decoding. This coding scheme is general and thus, it holds for the two-user LD-IC-NF and the two-user G-IC-NF.

\textbf{Codebook Generation}: Fix a strictly positive joint probability distribution  
\begin{IEEEeqnarray}{rcl}
\nonumber
P_{U\, U_1\,U_2\, V_1\,V_2\, X_{1,P}\, X_{2,P}}&(&u, u_1,u_2, v_1,v_2, x_{1,P}, x_{2,P})=P_U(u) \\
\nonumber
& & \hspace {-25mm} P_{U_1|U}(u_1|u) P_{U_2|U}(u_2|u) P_{V_1|U\,U_1}(v_1|u,u_1)\\
\nonumber
& &  \hspace {-25mm} P_{V_2|U\,U_2}(v_2|u,u_2)P_{X_{1,P}|U\,U_1\,V_1}(x_{1,P}|u,u_1,v_1) \\
\label{Eqprobdist}
& &  \hspace {-25mm} P_{X_{2,P}|U\,U_2\,V_2}(x_{2,P}|u,u_2,v_2),
\end{IEEEeqnarray}
for all $\left(u,u_1, u_2, v_1, v_2, x_{1,P}, x_{2,P}\right) \in \left(\mathcal{X}_1\cap \mathcal{X}_2\right) \times \left(\mathcal{X}_1 \times \mathcal{X}_2 \right)^3$.

Let $R_{1,C1}$, $R_{1,C2}$, $R_{2,C1}$, $R_{2,C2}$, $R_{1,P}$, and $R_{2,P}$ be non-negative real numbers. Let also ${R_{1,C}=R_{1,C1}+R_{1,C2}}$, ${R_{2,C}=R_{2,C1}+R_{2,C2}}$, $R_{1}=R_{1,C}+R_{1,P}$, and ${R_{2}=R_{2,C}+R_{2,P}}$. 

Generate $2^{N(R_{1,C1} + R_{2,C1})}$ i.i.d. $N$-dimensional codewords denoted by ${\bs{u}(s,r) = \big(u_{1}(s,r), u_{2}(s,r), \ldots, u_{N}(s,r)\big)}$ according to 

\begin{equation}
P_{\bs{U}}\big(\bs{u}(s,r)\big) = \ds\prod_{n =1}^N P_{U}(u_{n}(s,r)),
\end{equation}
with $s \in \lbrace 1, 2,  \ldots, 2^{NR_{1,C1}}\rbrace$ and $r \in \lbrace 1, 2,  \ldots, 2^{NR_{2,C1}}\rbrace$. 

For encoder $1$, generate for each codeword $\bs{u}(s,r)$, $2^{NR_{1,C1}}$ i.i.d. $N$-dimensional codewords denoted by $\bs{u}_1(s,r,k) = \big(u_{1,1}(s,r,k)$, $u_{1,2}(s,r,k), \ldots$, $u_{1,N}(s,r,k)\big)$  according to 
\begin{equation}
P_{\bs{U}_1|\bs{U}}\big(\bs{u}_1(s,r,k)|\bs{u}(s,r) \! \big) \! = \!  \ds\prod_{n =1}^N \! P_{U_{1}|U}\big(u_{1,n}(s,r,k)|u_{n}(s,r)\big),
\end{equation}
with $k \in \lbrace 1, 2,  \ldots, 2^{NR_{1,C1}}\rbrace$.  For each pair of codewords $\big(\bs{u}(s,r),\bs{u}_{1}(s,r,k)\big)$, generate $2^{NR_{1,C2}}$ i.i.d. $N$-dimensional codewords denoted by $\bs{v}_1(s,r,k,l) = \big(v_{1,1}(s,r,k,l), v_{1,2}(s,r,k,l), \ldots, v_{1,N}(s,r,k,l)\big)$  according to 
\begin{IEEEeqnarray}{rcl}
\nonumber
&P&_{\bs{V}_1 | \bs{U}\,\bs{U}_1}\big(\bs{v}_1(s,r,k,l)|\bs{u}(s,r) ,\bs{u}_1(s,r,k)\big) \\
& & = \ds\prod_{n =1}^N P_{V_{1}|U\,U_{1}}\big(v_{1,n}(s,r,k,l)|u_{n}(s,r),u_{1,n}(s,r,k)\big), \quad
\end{IEEEeqnarray}
with $l \in \lbrace 1, 2, \ldots, 2^{NR_{1,C2}}\rbrace$. For each tuple of codewords $\big(\bs{u}(s,r)$, $\bs{u}_{1}(s,r,k)$, $\bs{v}_{1}(s,r,k,l)\big)$, generate  $2^{NR_{1,P}}$ i.i.d. $N$-dimensional codewords denoted by $\bs{x}_{1,P}(s,r,k,l,q) = \big(x_{1,P,1}(s,r,k,l,q), x_{1,P,2}(s,r,k,l,q), \ldots$, $x_{1,P,N}(s,r,k,l,q)\big)$ according to 
\begin{IEEEeqnarray}{rcl}
\nonumber
&P\!&_{\bs{X}_{1,P} | \bs{U}\, \bs{U}_{1} \!\, \!\bs{V}_{1}}\!\big(\!\bs{x}_{1,P}(s,r,k,l,q)\! | \! \bs{u}(s,r),\!\bs{u}_{1}(s,r,k),\!\bs{v}_{1}(s,r,k,l)\!\big)\!\\
\nonumber
& & =\ds\prod_{n =1}^N \! P_{X_{1,P}|U\,U_{1}\,V_{1}}\Big(x_{1,P,n}(s,r,k,l,q)|u_{n}(s,r),u_{1,n}(s,r,k), \\
&& v_{1,n}(s,r,k,l)\Big), 
\end{IEEEeqnarray}
with $q \in \lbrace 1, 2, \ldots, 2^{NR_{1,P}}\rbrace$. 

\begin{figure*}[t]
 \centerline{\epsfig{figure=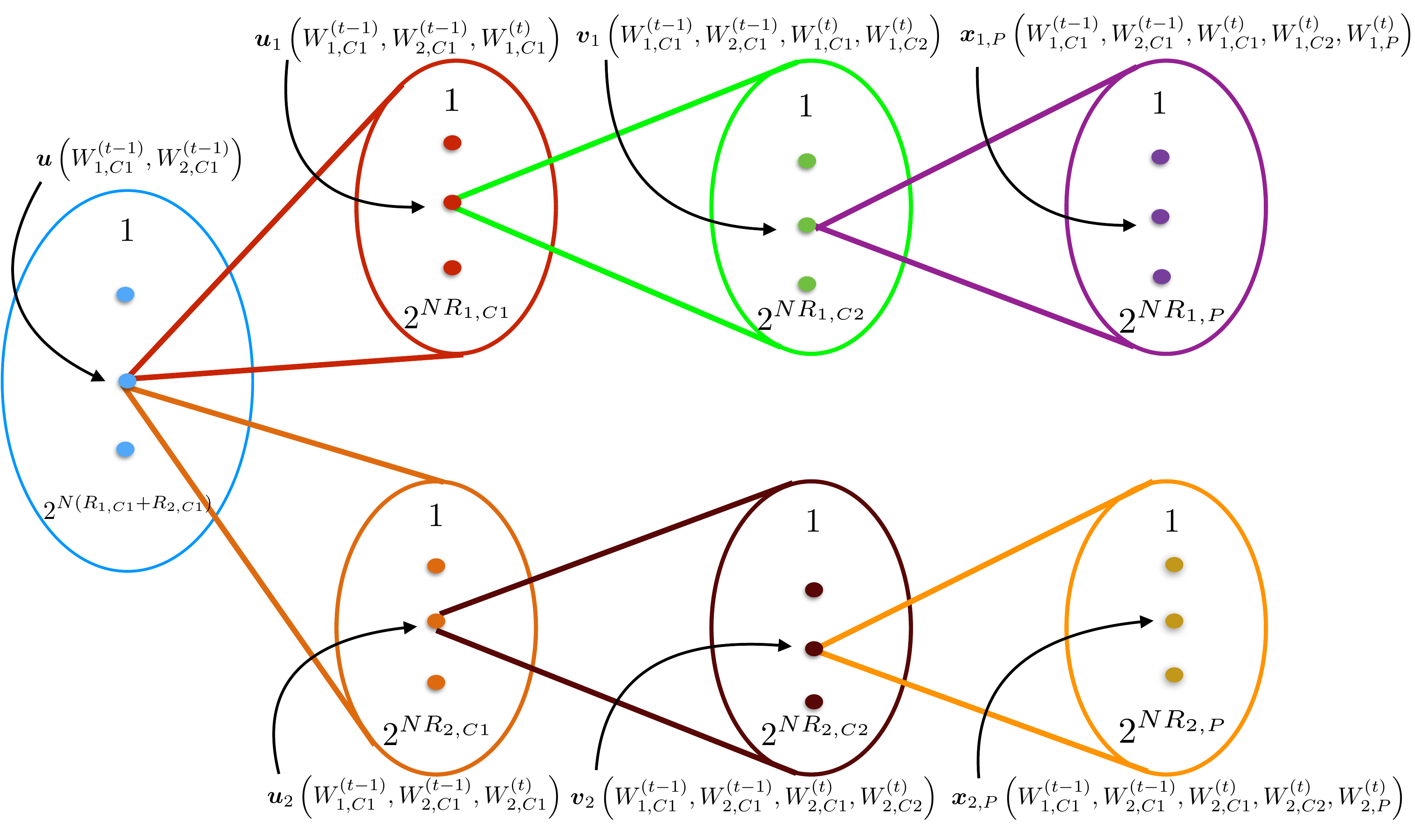,width=1\textwidth}}
 \caption{Structure of the superposition code. The codewords corresponding to the message indices $W_{1,C1}^{(t-1)}, W_{2,C1}^{(t-1)},W_{i,C1}^{(t)},W_{i,C2}^{(t)},W_{i,P}^{(t)}$ with $i \in \lbrace 1, 2 \rbrace$ as well as the block index $t$ are both highlighted.} 
\label{FigSuperpos}
\end{figure*}

For encoder $2$, generate for each codeword $\bs{u}(s,r)$,  $2^{NR_{2,C1}}$ i.i.d. $N$-dimensional codewords denoted by $\bs{u}_2(s,r,j) = \big(u_{2,1}(s,r,j)$, $u_{2,2}(s,r,j), \ldots$, $u_{2,N}(s,r,j)\big)$  according to 
\begin{equation}
P_{\bs{U}_2|\bs{U}}\big(\bs{u}_2(s,r,j)|\bs{u}(s,r)\big) \! = \! \ds\prod_{n =1}^N \! P_{U_{2}|U}\big(u_{2,n}(s,r,j)|u_{n}(s,r)\big), 
\end{equation}
with $j \in \lbrace 1, 2,  \ldots, 2^{NR_{2,C1}}\rbrace$. For each pair of codewords $\big(\bs{u}(s,r),\bs{u}_{2}(s,r,j)\big)$, generate $2^{NR_{2,C2}}$ i.i.d. $N$-dimensional codewords denoted by $\bs{v}_2(s,r,j,m)=\big(v_{2,1}(s,r,j,m), v_{2,2}(s,r,j,m),  \ldots, v_{2,N}(s,r,j,m)\big)$ according to 
\begin{IEEEeqnarray}{rcl}
\nonumber
&P&_{\bs{V}_2 | \bs{U}\, \bs{U}_2}\big(\bs{v}_2(s,r,j,m) | \bs{u}(s,r), \bs{u}_2(s,r,j)\big) \\
& & = \ds\prod_{n =1}^N P_{V_{2} |  U\, U_{2}}(v_{2,n}(s,r,j,m) |  u_{n}(s,r), u_{2,n}(s,r,j)), \quad
\end{IEEEeqnarray}
with $m \in \lbrace 1, 2,  \ldots, 2^{NR_{2,C2}}\rbrace$. 
\noindent
For each tuple of codewords $\big(\bs{u}(s,r)$, $\bs{u}_{2}(s,r,j),\bs{v}_{2}(s,r,j,m)\big)$, generate  $2^{NR_{2,P}}$ i.i.d. $N$-dimensional codewords denoted by $\bs{x}_{2,P}(s,r,j,m,b)$ $=$ $\big(x_{2,P,1}(s,r,j,m,b),$ $x_{2,P,2}(s,r,j,m,b),$ $\ldots,$ $ x_{2,P,N}(s,r,j,m,b) \big)$ according to 
\begin{IEEEeqnarray}{rcl}
\nonumber
&\!P\!&_{\bs{X}_{2,P}   |   \bs{U} \bs{U}_{2} \bs{V}_{2}} \big( \!  \bs{x}_{2,P}(s \!,r \!,j\!,m\!,b\!)  | \!  \bs{u}(s,r) \!, \!  \bs{u}_{2}(s,r,j) \!, \!  \bs{v}_{2}(s,r,j,m) \! \big) \\
\nonumber
& & = \ds\prod_{n =1}^N \! P_{X_{2,P}  | U U_{2} V_{2}}\Big(x_{2,P,n}(s,r,j,m,b) | u_{n}(s,r), u_{2,n}(s,r,j), \\
& & v_{2,n}(s,r,j,m)\Big), 
\end{IEEEeqnarray}
with $b \in \lbrace 1, 2,  \ldots, 2^{NR_{2,P}}\rbrace$.  
The resulting code structure is shown in Figure~\ref{FigSuperpos}.

\textbf{Encoding}: Denote by $W_{i}^{(t)} \in \lbrace 1, 2, \ldots, 2^{NR_{i}} \rbrace$  the message index of transmitter $i \in \lbrace 1,2 \rbrace$ during block $t \in \lbrace 1, 2,  \ldots, T \rbrace$, with $T$ the total number of blocks. Let $W_{i}^{(t)}$ be composed by the message index $W_{i,C}^{(t)} \in \lbrace 1, 2,  \ldots, 2^{NR_{i,C}} \rbrace$ and message index $W_{i,P}^{(t)} \in \lbrace 1$, $2$, $  \ldots, 2^{NR_{i,P}} \rbrace$. The message index $W_{i,P}^{(t)}$ must be reliably decoded at receiver $i$. Let also $W_{i,C}^{(t)}$ be composed by the message indices $W_{i,C1}^{(t)} \in \lbrace 1, 2,  \ldots, 2^{NR_{i,C1}} \rbrace$ and $W_{i,C2}^{(t)} \in \lbrace 1, 2,  \ldots, 2^{NR_{i,C2}} \rbrace$.  The message index $W_{i,C1}^{(t)}$ must be reliably decoded by the other transmitter (via feedback) and by both receivers. The message index $W_{i,C2}^{(t)}$ must be reliably decoded by both receivers, but not by transmitter $j$. 

Consider block-Markov encoding over $T$ blocks. At encoding step $t$, with $t \in \lbrace 1, 2,  \ldots, T \rbrace$, transmitter $1$ sends the codeword:
 \begin{IEEEeqnarray}{rcl}
 \nonumber
 \bs{x}_1^{(t)} & = & \Theta_1 \Bigg(\! \bs{u}\Big(\!W_{1,C1}^{(t-1)}, W_{2,C1}^{(t-1)} \!\Big), \! \bs{u}_1\Big(\! W_{1,C1}^{(t-1)}, W_{2,C1}^{(t-1)},W_{1,C1}^{(t)} \!\Big), \\
 \nonumber
 & & \bs{v}_1\Big(W_{1,C1}^{(t-1)}, W_{2,C1}^{(t-1)},W_{1,C1}^{(t)}, W_{1,C2}^{(t)}\Big), \\
\label{Eqtransmittercodeword}
 & & \bs{x}_{1,P}\Big(W_{1,C1}^{(t-1)}, W_{2,C1}^{(t-1)}, W_{1,C1}^{(t)}, W_{1,C2}^{(t)},W_{1,P}^{(t)}\Big)\Bigg), 
 \end{IEEEeqnarray}
 where,  $\Theta_1: \left(\mathcal{X}_1\cap \mathcal{X}_2\right)^{N} \times \mathcal{X}_1^{3N} \rightarrow \mathcal{X}_1^{N}$ is a function that transforms the codewords $\bs{u}\Big(W_{1,C1}^{(t-1)}$, $W_{2,C1}^{(t-1)}\Big)$, $\bs{u}_1\Big(W_{1,C1}^{(t-1)}, W_{2,C1}^{(t-1)},W_{1,C1}^{(t)}\Big)$, $\bs{v}_1\Big(W_{1,C1}^{(t-1)}, W_{2,C1}^{(t-1)},W_{1,C1}^{(t)}, W_{1,C2}^{(t)}\Big) \ $, and $\bs{x}_{1,P}\Big(W_{1,C1}^{(t-1)}$, $W_{2,C1}^{(t-1)}$, $W_{1,C1}^{(t)}$, $W_{1,C2}^{(t)}$, $W_{1,P}^{(t)}\Big)$ into the $N$-dimensional vector of channel inputs denoted by $\bs{x}_{1}^{t}$. The indices $W_{1,C1}^{(0)} = W_{1,C1}^{(T)} = s^*$ and  $W_{2,C1}^{(0)} = W_{2,C1}^{(T)} = r^*$, and the pair $(s^*,r^*) \in  \lbrace 1, 2,  \ldots, 2^{N \, R_{1,C1}} \rbrace \times \lbrace 1, 2,  \ldots, 2^{NR_{2,C1}} \rbrace$  are pre-defined and known by both receivers and transmitters. It is worth noting that the message index  $W_{2,C1}^{(t-1)}$ is obtained by transmitter $1$ from the feedback signal $\overleftarrow{\bs{y}}_{1}^{(t-1)}$ at the end of the previous encoding step $t-1$.

Transmitter $2$ follows a similar encoding scheme.

\textbf{Decoding}: Both receivers decode their message indices at the end of block $T$ in a backward decoding fashion. At each decoding step $t$, with $t \in \lbrace 1, 2,  \ldots, T \rbrace$, receiver $1$ obtains the message indices $\big(\widehat{W}_{1,C1}^{(T-t)}$, $\widehat{W}_{2,C1}^{(T-t)}$, $\widehat{W}_{1,C2}^{(T-(t-1))}$, $\widehat{W}_{1,P}^{(T-(t-1))}$, $\widehat{W}_{2,C2}^{(T-(t-1))}\big) \in \lbrace 1$, $2,  \ldots $, $ 2^{NR_{1,C1}}\rbrace \times \lbrace 1$, $ 2,  \ldots, 2^{NR_{2,C1}} \rbrace  \times  \lbrace 1$, $ 2,  \ldots, 2^{NR_{1,C2}} \rbrace \times  \lbrace 1$, $ 2,  \ldots, 2^{NR_{1,P}}\rbrace \times  \lbrace 1$, $2,  \ldots, 2^{NR_{2,C2}} \rbrace $ from the channel output $\overrightarrow{\bs{y}}_1^{(T-(t-1))}$. The tuple $\Big(\widehat{W}_{1,C1}^{(T-t)}$, $\widehat{W}_{2,C1}^{(T-t)}$, $\widehat{W}_{1,C2}^{(T-(t-1))}$, $\widehat{W}_{1,P}^{(T-(t-1))}$, $\widehat{W}_{2,C2}^{(T-(t-1))}\Big)$ is the unique tuple that satisfies

\begin{IEEEeqnarray}{ll}
\nonumber
\Big( & \bs{u}\left(\widehat{W}_{1,C1}^{(T-t)}, \widehat{W}_{2,C1}^{(T-t)}\right), \bs{u}_1\left(\widehat{W}_{1,C1}^{(T-t)}, \widehat{W}_{2,C1}^{(T-t)}, W_{1,C1}^{(T-(t-1))}\right), \\
\nonumber
& \bs{v}_1 \left(\widehat{W}_{1,C1}^{(T-t)}, \widehat{W}_{2,C1}^{(T-t)}, W_{1,C1}^{(T-(t-1))}, \widehat{W}_{1,C2}^{(T-(t-1))} \right), \\
\nonumber
& \bs{x}_{1,P}\! \Big(\! \widehat{W}_{1,C1}^{(T-t)}\!\!, \widehat{W}_{2,C1}^{(T-t)}\!\!, W_{1,C1}^{(T-(t-1))}\!,\!\widehat{W}_{1,C2}^{(T-(t-1))}\!, \!\widehat{W}_{1,P}^{(T-(t-1))} \! \Big), \\
\nonumber
& \bs{u}_2\left(\widehat{W}_{1,C1}^{(T-t)}, \widehat{W}_{2,C1}^{(T-t)}, W_{2,C1}^{(T-(t-1))}\right), \\
\nonumber
& \bs{v}_2 \! \left(\! \widehat{W}_{1,C1}^{(T-t)}\!, \widehat{W}_{2,C1}^{(T-t)}\!, W_{2,C1}^{(T-(t-1))}\!,\widehat{W}_{2,C2}^{(T-(t-1))}\! \right)\!, \overrightarrow{\bs{y}}_1^{(T-(t-1))} \! \Big) \\
\label{EqDecodingW1cW2cW1p}
& \in \mathcal{T}_{\big[U \ U_1 \ V_1  \ X_{1,P} \ U_2 \ V_2  \ \overrightarrow{Y}_1\big]}^{(N, e)},
\end{IEEEeqnarray}
where $W_{1,C1}^{(T-(t-1))}$ and $W_{2,C1}^{(T-(t-1))}$ are assumed to be perfectly decoded in the previous decoding step $t-1$. The set $\mathcal{T}_{\big[U \ U_1 \ V_1  \ X_{1,P} \ U_2 \ V_2  \ \overrightarrow{Y}_1\big]}^{(N, e)}$ represents the set of jointly typical sequences of the random variables $U, U_1, V_1, X_{1,P}, U_2, V_2$, and $\overrightarrow{Y}_1$, with $e>0$.
Receiver $2$ follows a similar decoding scheme.

\textbf{Error Probability Analysis}: An error occurs during encoding step $t$ if the message index $W_{2,C1}^{(t-1)}$ is not correctly decoded at transmitter $1$. From the asymptotic equipartion property (AEP) \cite{Cover-Book-1991}, it follows that the message index $W_{2,C1}^{(t-1)}$ can be reliably decoded at transmitter $1$ during encoding step $t$, under the condition:
\begin{eqnarray}
\nonumber
R_{2,C1} & \leqslant & I\left( \overleftarrow{Y}_1 ; U_2  | U, U_1, V_1, X_1 \right) \\
\label{EqConditionNoError1}
&=& I\left( \overleftarrow{Y}_1 ; U_2  | U, X_1  \right).
\end{eqnarray} 
An error occurs during the (backward) decoding step $t$ if the message indices $W_{1,C1}^{(T-t)}$, $W_{2,C1}^{(T-t)}$, $W_{1,C2}^{(T-(t-1))}$, $W_{1,P}^{(T-(t-1))}$, and $W_{2,C2}^{(T-(t-1))}$ are not decoded correctly given that the message indices $W_{1,C1}^{(T-(t-1))}$ and $W_{2,C1}^{(T-(t-1))}$ were correctly decoded in the previous decoding step $t-1$. 
These errors might arise for two reasons: $(i)$ there does not exist a tuple $\Big(\widehat{W}_{1,C1}^{(T-t)}$, $\widehat{W}_{2,C1}^{(T-t)}, \widehat{W}_{1,C2}^{(T-(t-1))},\widehat{W}_{1,P}^{(T-(t-1))},\widehat{W}_{2,C2}^{(T-(t-1))}\Big)$  that satisfies \eqref{EqDecodingW1cW2cW1p}, or $(ii)$ there exist several tuples $\Big(\widehat{W}_{1,C1}^{(T-t)}, \widehat{W}_{2,C1}^{(T-t)}, \widehat{W}_{1,C2}^{(T-(t-1))},\widehat{W}_{1,P}^{(T-(t-1))}, \widehat{W}_{2,C2}^{(T-(t-1))}\Big)$ that simultaneously satisfy \eqref{EqDecodingW1cW2cW1p}. 
From the AEP, the probability of an error due to $(i)$ tends to zero when $N$ grows to infinity. Consider the error due to $(ii)$ and define the event $E_{(s, r, l, q, m)}$ that describes the case in which the codewords $\big(\bs{u}(s,r)$, $\bs{u}_1(s,r,W_{1,C1}^{(T-(t-1))})$, $\bs{v}_1(s,r,W_{1,C1}^{(T-(t-1))},l)$, $\bs{x}_{1,P}(s,r,W_{1,C1}^{(T-(t-1))},l,q)$, $\bs{u}_2(s,r,W_{2,C1}^{(T-(t-1))})$, and $\bs{v}_2(s,r,W_{2,C1}^{(T-(t-1))},m)\big)$ are jointly typical with $ \overrightarrow{\bs{y}}_1^{(T-(t-1))}$ during decoding step $t$. 
Assume now that the codeword to be decoded at decoding step $t$ corresponds to the indices $(s,r,l,q,m) = (1,1,1,1,1)$. This assumption does not incur in any loss of generality due to the symmetry of the code. Then, the probability of error due to $(ii)$ during decoding step $t$, is upper-bounded as follows:
\begin{IEEEeqnarray}{rcl}
\nonumber
P_e (N)&  =  &\pr{\ds\bigcup_{(s,r,l,q,m) \neq (1,1,1,1,1)} E_{(s,r,l,q,m)} }\\
\label{EqConditionNoError2a}
& \leqslant & \ds\sum_{\scriptscriptstyle (s, r, l, q, m) \in \mathcal{T}} \pr{ E_{(s,r,l,q,m)}}, 
\end{IEEEeqnarray}
with $\mathcal{T}=\Big\lbrace \lbrace 1, 2,  \ldots 2^{NR_{1,C1}} \rbrace \times \lbrace 1, 2,  \ldots 2^{NR_{2,C1}} \rbrace \times \lbrace 1, 2,  \ldots 2^{NR_{1,C2}} \rbrace \times \lbrace 1, 2,  \ldots 2^{NR_{1,P}} \rbrace \times \lbrace 1, 2,  \ldots 2^{NR_{2,C2}} \rbrace \Big\rbrace \setminus \lbrace (1,1,1,1,1) \rbrace$. 

From the AEP, it follows that
\begin{IEEEeqnarray}{rcl}
\nonumber
P_e(N) & \leqslant  & 2^{N (R_{2,C2} - I(\overrightarrow{Y}_1;V_2 | U, U_1, U_2, V_1, X_1) + 2\epsilon) } \\
\nonumber
& &+2^{N (R_{1,P} - I(\overrightarrow{Y}_1;X_1 | U, U_1, U_2, V_1, V_2) + 2\epsilon) }\\
\nonumber
& & +2^{N (R_{2,C2} +R_{1,P} - I(\overrightarrow{Y}_1;V_2, X_1 | U, U_1, U_2, V_1) + 2\epsilon) }  \\
\nonumber
& & +2^{N (R_{1,C2} - I(\overrightarrow{Y}_1; V_1, X_1 | U, U_1, U_2, V_2) + 2\epsilon) } \\
\nonumber
 & & +2^{N (R_{1,C2} +R_{2,C2} - I(\overrightarrow{Y}_1;V_1,V_2, X_1 | U, U_1, U_2) + 2\epsilon) } \\
 \nonumber
 & & +2^{N (R_{1,C2} +R_{1,P} - I(\overrightarrow{Y}_1; V_1, X_1 | U, U_1, U_2, V_2) + 2\epsilon) }\\
 \nonumber
 & & +2^{N (R_{1,C2} + R_{1,P}+R_{2,C2} - I(\overrightarrow{Y}_1;V_1, V_2, X_1 | U, U_1, U_2) + 2\epsilon) } \\
 \nonumber
 & & +2^{N (R_{2,C1}  - I(\overrightarrow{Y}_1; U, U_1, U_2, V_1, V_2, X_1) + 2\epsilon) }\\
\nonumber
 & & +2^{N (R_{2,C}  - I(\overrightarrow{Y}_1; U, U_1, U_2, V_1, V_2, X_1) + 2\epsilon) } \\
 \nonumber
 & & +2^{N (R_{2,C1} +R_{1,P} - I(\overrightarrow{Y}_1; U, U_1, U_2, V_1, V_2, X_1) + 2\epsilon) } \\
  \nonumber
& & +2^{N (R_{2,C} + R_{1,P} - I(\overrightarrow{Y}_1; U, U_1, U_2, V_1, V_2, X_1) + 2\epsilon) } \\
\nonumber
& & +2^{N (R_{2,C1} +R_{1,C2}  - I(\overrightarrow{Y}_1; U, U_1, U_2, V_1, V_2, X_1) + 2\epsilon) } 
   \end{IEEEeqnarray}
\begin{IEEEeqnarray}{rcl} 
\nonumber
& & +2^{N (R_{2,C} +R_{1,C2} - I(\overrightarrow{Y}_1; U, U_1, U_2, V_1, V_2, X_1) + 2\epsilon) } \\
\nonumber
& & +2^{N (R_{2,C1} +R_{1,C2}+R_{1,P}  - I(\overrightarrow{Y}_1; U, U_1, U_2, V_1, V_2, X_1) + 2\epsilon) } \\
\nonumber
& & +2^{N (R_{2,C} +R_{1,C2}+R_{1,P} - I(\overrightarrow{Y}_1; U, U_1, U_2, V_1, V_2, X_1) + 2\epsilon) } \\
\nonumber
& & +2^{N (R_{1,C1} - I(\overrightarrow{Y}_1; U, U_1, U_2, V_1, V_2, X_1) + 2\epsilon) } \\
\nonumber
& & +2^{N (R_{1,C1} +R_{2,C2} - I(\overrightarrow{Y}_1; U, U_1, U_2, V_1, V_2, X_1) + 2\epsilon) } \\
\nonumber
& & +2^{N (R_{1,C1} +R_{1,P} - I(\overrightarrow{Y}_1; U, U_1, U_2, V_1, V_2, X_1) + 2\epsilon) }\\
\nonumber
& & +2^{N (R_{1,C1} +R_{1,P}+R_{2,C2} - I(\overrightarrow{Y}_1; U, U_1, U_2, V_1, V_2, X_1) + 2\epsilon) } \\
\nonumber
& & +2^{N (R_{1,C}  - I(\overrightarrow{Y}_1; U, U_1, U_2, V_1, V_2, X_1) + 2\epsilon) } \\
\nonumber
& & +2^{N (R_{1,C}+R_{2,C2} - I(\overrightarrow{Y}_1; U, U_1, U_2, V_1, V_2, X_1) + 2\epsilon) } \\
\nonumber
& & +2^{N (R_{1} - I(\overrightarrow{Y}_1; U, U_1, U_2, V_1, V_2, X_1) + 2\epsilon) }\\
\nonumber
& & +2^{N (R_{1} +R_{2,C2} - I(\overrightarrow{Y}_1; U, U_1, U_2, V_1, V_2, X_1) + 2\epsilon) } \\
\nonumber
& & +2^{N (R_{1,C1} +R_{2,C1} - I(\overrightarrow{Y}_1; U, U_1, U_2, V_1, V_2, X_1) + 2\epsilon) } \\
\nonumber
& & +2^{N (R_{1,C1} +R_{2,C}  - I(\overrightarrow{Y}_1; U, U_1, U_2, V_1, V_2, X_1) + 2\epsilon) } \\
\nonumber
& & +2^{N (R_{1,C1} +R_{2,C1}+R_{1,P} - I(\overrightarrow{Y}_1; U, U_1, U_2, V_1, V_2, X_1) + 2\epsilon) } \\
\nonumber
& & +2^{N (R_{1,C1} +R_{2,C}+R_{1,P} - I(\overrightarrow{Y}_1; U, U_1, U_2, V_1, V_2, X_1) + 2\epsilon) } \\
\nonumber
& & +2^{N (R_{1,C} +R_{2,C1} - I(\overrightarrow{Y}_1; U, U_1, U_2, V_1, V_2, X_1) + 2\epsilon) } \\
\nonumber
& & +2^{N (R_{1,C} +R_{2,C}  - I(\overrightarrow{Y}_1; U, U_1, U_2, V_1, V_2, X_1) + 2\epsilon) }\\
\nonumber
& & +2^{N (R_{1} +R_{2,C1} - I(\overrightarrow{Y}_1; U, U_1, U_2, V_1, V_2, X_1) + 2\epsilon) }\\
\label{EqConditionNoError2b}
& & +2^{N (R_{1} +R_{2,C} - I(\overrightarrow{Y}_1; U, U_1, U_2, V_1, V_2, X_1) + 2\epsilon) }. 
\end{IEEEeqnarray}
The same analysis of the probability of error holds for transmitter-receiver pair $2$.
Hence, in general, from \eqref{EqConditionNoError1} and \eqref{EqConditionNoError2b}, reliable decoding holds under the following conditions for transmitter $i \in \lbrace1,2 \rbrace$, with $j \in  \lbrace1,2 \rbrace\setminus\lbrace i \rbrace$:
\begin{subequations}
\label{EqRateRegion-z}
\begin{IEEEeqnarray}{rcl}
\nonumber
R_{j,C1}  &  \leqslant  &  I\left( \overleftarrow{Y}_i ; U_j  | U, U_i, V_i,X_i  \right) \\ 
\nonumber
&=& I\left( \overleftarrow{Y}_i ; U_j  | U, X_i  \right)\\
\label{EqRateRegion0}
& \triangleq &  \theta_{1,i},  \\
\nonumber
R_{i} + R_{j,C}  &  \leqslant  &  I(\overrightarrow{Y}_i; U,U_i, U_j,V_i, V_j, X_i) \\
\nonumber
&=& I(\overrightarrow{Y}_i; U, U_j,V_j, X_i) \\
\label{EqRateRegion1}
&\triangleq&  \theta_{2,i}, \\
\nonumber
R_{j,C2}  &  \leqslant  & I(\overrightarrow{Y}_i; V_j | U, U_i, U_j, V_i, X_i)\\
\nonumber
&=& I(\overrightarrow{Y}_i; V_j | U, U_j, X_i) \\
\label{EqRateRegion2}
&\triangleq&  \theta_{3,i}, \\
\nonumber
R_{i,P}    &  \leqslant  &  I(\overrightarrow{Y}_i; X_i | U, U_i, U_j,V_i, V_j) \\
\label{EqRateRegion3}
&\triangleq& \theta_{4,i}, \\
\nonumber
R_{i,P}+R_{j,C2}  &  \leqslant  & I(\overrightarrow{Y}_i; V_j, X_i | U, U_i, U_j, V_i) \\
\label{EqRateRegion4}
&\triangleq& \theta_{5,i}, \\
\nonumber
R_{i,C2}+R_{i,P}  &  \leqslant  & I(\overrightarrow{Y}_i; V_i,X_i | U, U_i, U_j, V_j) \\
\nonumber
&=& I(\overrightarrow{Y}_i; X_i | U, U_i, U_j, V_j) \\
\label{EqRateRegion5}
&\triangleq& \theta_{6,i},  \mbox{ and } \\
\nonumber
R_{i,C2}+R_{i,P}+R_{j,C2} &  \leqslant  & I(\overrightarrow{Y}_i; V_i, V_j, X_i | U, U_i, U_j) \\
\nonumber
&=& I(\overrightarrow{Y}_i; V_j, X_i | U, U_i, U_j) \\
\label{EqRateRegion6}
&\triangleq& \theta_{7,i}.
\end{IEEEeqnarray}
\end{subequations}
Taking into account that $R_i=R_{i,C1}+R_{i,C2}+R_{i,P}$, a Fourier-Motzkin elimination process in \eqref{EqRateRegion-z} yields:
\begin{subequations}
\label{EqRateRegion2b}
\begin{IEEEeqnarray}{rcl}
\label{EqRateRegion21}
R_{1}  &  \leqslant  & \min\left(\theta_{2,1},\theta_{6,1}+\theta_{1,2},\theta_{4,1}+\theta_{1,2}+\theta_{3,2}\right), \qquad \\ 
\label{EqRateRegion22}
R_{2}   & \leqslant & \min\left(\theta_{2,2},\theta_{1,1}+a_{6,2},\theta_{1,1}+\theta_{3,1}+\theta_{4,2}\right), \qquad \\
\nonumber
R_{1}+R_{2}  & \leqslant & \min(\theta_{2,1}+\theta_{4,2}, \theta_{2,1}+a_{6,2}, \theta_{4,1}+\theta_{2,2}, \\
\nonumber
& & \theta_{6,1}+\theta_{2,2}, \theta_{1,1}+\theta_{3,1}+\theta_{4,1}+\theta_{1,2}+\theta_{5,2},  \\
\nonumber
& & \theta_{1,1}\!+\theta_{7,1}\!+\theta_{1,2}\!+\theta_{5,2}, \theta_{1,1}\!+\theta_{4,1}\!+\theta_{1,2}\!+\theta_{7,2}, \\
\nonumber
& & \theta_{1,1}\!+\theta_{5,1}\!+\theta_{1,2}\!+\theta_{3,2}\!+\theta_{4,2}, \\
\nonumber
& & \theta_{1,1}\!+\theta_{5,1}\!+\theta_{1,2}\!+\theta_{5,2}, \theta_{1,1}\!+\theta_{7,1}\!+\theta_{1,2}\!+\theta_{4,2}), \\
\label{EqRateRegion23}\\
\nonumber
2R_{1}+R_{2}  & \leqslant & \min(\theta_{2,1}+\theta_{4,1}+\theta_{1,2}+\theta_{7,2}, \\
\nonumber
& & \theta_{1,1}+\theta_{4,1}+\theta_{7,1}+2\theta_{1,2}+\theta_{5,2}, \\
\label{EqRateRegion24}
& & \theta_{2,1}+\theta_{4,1}+\theta_{1,2}+\theta_{5,2}), \\
\nonumber
R_{1}+2R_{2}  & \leqslant & \min(\theta_{1,1}+\theta_{5,1}+\theta_{2,2}+\theta_{4,2}, \\
\nonumber
& & \theta_{1,1}+\theta_{7,1}+\theta_{2,2}+\theta_{4,2}, \\
\label{EqRateRegion25}
& & 2\theta_{1,1}+\theta_{5,1}+\theta_{1,2}+\theta_{4,2}+\theta_{7,2}),
\end{IEEEeqnarray}
\end{subequations}
where $\theta_{l,i}$ are defined in \eqref{EqRateRegion-z} with $(l,i) \in \lbrace1,  \ldots, 7 \rbrace \times \lbrace 1, 2 \rbrace$.

\subsection{An Achievable Region for the Two-user Linear Deterministic Interference Channel with Noisy Channel-Output Feedback} \label{AppAch-LD-IC-NF}

Following the discussion in Section \ref{SecCommentsAchievabilityLD}, consider the random variables $\bs{U}$, $\bs{U}_i$, $\bs{V}_i$, and $\bs{X}_{i,P}$ independent and uniformly distributed over the sets $\lbrace 0 \rbrace$, $\lbrace 0,1\rbrace^{\left(n_{ji}-\left(\max\left(\overrightarrow{n}_{jj},n_{ji}\right)-\overleftarrow{n}_{jj}\right)^+\right)^+}$, 
$\lbrace 0,1\rbrace^{\left(\min\left(n_{ji},\left(\max\left(\overrightarrow{n}_{jj},n_{ji}\right)-\overleftarrow{n}_{jj}\right)^+\right)\right)}$ and 
$\lbrace 0,1\rbrace^{\left(\overrightarrow{n}_{ii}-n_{ji}\right)^+}$, respectively, with $i \in \lbrace 1,2 \rbrace$ and $j \in \lbrace 1,2 \rbrace\setminus\lbrace i \rbrace$.

Let the channel input symbol of transmitter $i$ at each channel use be $\bs{X}_i = \left( \bs{U}_{i}^{\sf{T}}, \bs{V}_{i}^{\sf{T}}, \bs{X}_{i,P}^{\sf{T}},(0,\ldots,0)\right)^{\sf{T}} \in \lbrace 0,1 \rbrace^{q}$.
Considering these random variables, the following holds for the terms $\theta_{l,i}$, with $(l,i) \in \lbrace 1, \ldots, 7 \rbrace \times \lbrace 1, 2 \rbrace$, in \eqref{EqRateRegion-z}: 
\begin{subequations}
\label{EqRateRegionLDM}
\begin{IEEEeqnarray}{rcl}
\nonumber
\theta_{1,i}& = & I\Big( \overleftarrow{\bs{Y}}_i ; \bs{U}_j  | \bs{U}, \bs{X}_i  \Big) \\
\nonumber
&\stackrel{(a)}{=}& H\left( \overleftarrow{\bs{Y}}_i  | \bs{U}, \bs{X}_i \right) \\
\nonumber
&=& H\left(\bs{U}_j\right) \\
\label{EqRateRegion26}
&=& \left(n_{ij}-\left(\max\left(\overrightarrow{n}_{ii},n_{ij}\right)-\overleftarrow{n}_{ii}\right)^+\right)^+, \\
\nonumber
\theta_{2,i}& = &I\Big(\overrightarrow{\bs{Y}}_i; \bs{U}, \bs{U}_j,\bs{V}_j, \bs{X}_i\Big) \\
\nonumber
&\stackrel{(b)}{=}&  H\left(\overrightarrow{\bs{Y}}_i\right) \\
\label{EqRateRegion27}
&=&  \max\left(\overrightarrow{n}_{ii},n_{ij}\right), \\
\nonumber
\theta_{3,i}& = &I\Big(\overrightarrow{\bs{Y}}_i; \bs{V}_j | \bs{U}, \bs{U}_j, \bs{X}_i\Big)  \\
\nonumber
&\stackrel{(b)}{=}& H\left(\overrightarrow{\bs{Y}}_i | \bs{U}, \bs{U}_j, \bs{X}_i\right) \\
\nonumber
&=& H\left(\bs{V}_j\right) \\
\label{EqRateRegion28}
&=& \min \left(n_{ij},\left(\max\left(\overrightarrow{n}_{ii},n_{ij}\right)-\overleftarrow{n}_{ii}\right)^+\right), \\
\nonumber
\theta_{4,i}& = &I\Big(\overrightarrow{\bs{Y}}_i; \bs{X}_i | \bs{U}, \bs{U}_i, \bs{U}_j, \bs{V}_i, \bs{V}_j\Big) \\
\nonumber
\nonumber
&\stackrel{(b)}{=}&  H\left(\overrightarrow{\bs{Y}}_i | \bs{U}, \bs{U}_i, \bs{U}_j, \bs{V}_i, \bs{V}_j\right) \\
\nonumber
&=&  H\left(\bs{X}_{i,P}\right) \\
\label{EqRateRegion29}
&=& \left(\overrightarrow{n}_{ii}-n_{ji}\right)^+, \mbox{ and }
\end{IEEEeqnarray}
\begin{IEEEeqnarray}{rcl}
\nonumber
\theta_{5,i}& = &I\Big(\overrightarrow{\bs{Y}}_i; \bs{V}_j, \bs{X}_i | \bs{U}, \bs{U}_i, \bs{U}_j, \bs{V}_i\Big)  \\
\nonumber
 &\stackrel{(b)}{=}&  H\left(\overrightarrow{\bs{Y}}_i | \bs{U}, \bs{U}_i, \bs{U}_j, \bs{V}_i\right) \\
\nonumber
&=& \max\left(\dim \bs{X}_{i,P}, \dim \bs{V}_j \right) \\ 
\nonumber
&=&  \max \Bigg(\left(\overrightarrow{n}_{ii}-n_{ji}\right)^+, \\
\label{EqRateRegion30}
& & \min\left(n_{ij},\left(\max\left(\overrightarrow{n}_{ii},n_{ij}\right)-\overleftarrow{n}_{ii}\right)^+\right)\Bigg),
\end{IEEEeqnarray}
where
(a) follows from the fact that $H\left( \overleftarrow{\bs{Y}}_i  | \bs{U}, \bs{U}_j, \bs{X}_i  \right)=0$; and 
(b) follows from the fact that $H(\overrightarrow{\bs{Y}}_i|\bs{U}, \bs{U}_j,\bs{V}_j, \bs{X}_i)=0$.

For the calculation of the last two mutual information terms in inequalities \eqref{EqRateRegion5} and \eqref{EqRateRegion6}, special notation is used. Let the vector $\bs{V}_i$ be the concatenation 
of the vectors $\bs{X}_{i,HA}$ and $\bs{X}_{i,HB}$, i.e., $\bs{V}_i=\left(\bs{X}_{i,HA}^{\sf{T}}, \bs{X}_{i,HB}^{\sf{T}}\right)^{\sf{T}}$. The vector $\bs{X}_{i,HA}$ contains the entries of $\bs{V}_i$ that are available in both receivers. The vector  $\bs{X}_{i,HB}$ is the part of $\bs{V}_i$ that is exclusively available in receiver $j$ (see Figure~\ref{FigachievabilityMessagesLDICNFB}). 
Note also that the vectors $\bs{X}_{i,HA}$ and $\bs{X}_{i,HB}$ possess the following dimensions:
\begin{IEEEeqnarray}{lcl}
\nonumber
\dim \bs{X}_{i,HA} & = & \min\Big(n_{ji} \!,\left(\max\left(\overrightarrow{n}_{jj},n_{ji}\right)-\overleftarrow{n}_{jj}\right)^+\Big) \\
\nonumber
& & \!-\!\min\Big(\! \left( \! n_{ji}\!-\!\overrightarrow{n}_{ii}\right)^+\!,\! \left(\max\left(\overrightarrow{n}_{jj},n_{ji}\right)\!-\!\overleftarrow{n}_{jj}\right)^+ \!\Big) 
\end{IEEEeqnarray}
\begin{IEEEeqnarray}{lcl}
\nonumber
\dim \bs{X}_{i,HB} & = & \min \!\Big( \! \left(n_{ji}\!-\!\overrightarrow{n}_{ii}\right)^+,\left(\max\left(\overrightarrow{n}_{jj},n_{ji}\right)-\overleftarrow{n}_{jj}\right)^+ \! \Big).
\end{IEEEeqnarray}
Using this notation, the following holds:
\begin{IEEEeqnarray}{rcl}
\nonumber
\theta_{6,i}& = &I\Big(\overrightarrow{\bs{Y}}_i; \bs{X}_i | \bs{U}, \bs{U}_i, \bs{U}_j, \bs{V}_j\Big) \\
\nonumber
&\stackrel{(c)}{=}& H\left(\overrightarrow{\bs{Y}}_i | \bs{U}, \bs{U}_i, \bs{U}_j, \bs{V}_j\right)\\
\nonumber
&= & H\left(\bs{X}_{i,HA},\bs{X}_{i,P}\right)\\
\nonumber
&=& \dim \bs{X}_{i,HA}+ \dim \bs{X}_{i,P}\\
\nonumber
&=&  \min\left(n_{ji},\left(\max\left(\overrightarrow{n}_{jj},n_{ji}\right)-\overleftarrow{n}_{jj}\right)^+\right) \\
\nonumber
& & -\min\big(\left(n_{ji}-\overrightarrow{n}_{ii}\right)^+,\left(\max\left(\overrightarrow{n}_{jj},n_{ji}\right)-\overleftarrow{n}_{jj}\right)^+\big)\\
\label{EqRateRegion31}
& & +\left(\overrightarrow{n}_{ii}-n_{ji}\right)^+,  \textrm{ and} \\
\nonumber
\theta_{7,i}& = &I\Big(\overrightarrow{\bs{Y}}_i; \bs{V}_j, \bs{X}_i | \bs{U}, \bs{U}_i, \bs{U}_j\Big) \\
\nonumber
&=& I\left(\overrightarrow{\bs{Y}}_i; \bs{X}_i | \bs{U}, \bs{U}_i, \bs{U}_j\right)+I\left(\overrightarrow{\bs{Y}}_i; \bs{V}_j  | \bs{U}, \bs{U}_i, \bs{U}_j, \bs{X}_i\right) \\
\nonumber
&=& I\left(\overrightarrow{\bs{Y}}_i; \bs{X}_i | \bs{U}, \bs{U}_i, \bs{U}_j\right)+I\left(\overrightarrow{\bs{Y}}_i; \bs{V}_j  | \bs{U}, \bs{U}_j, \bs{X}_i\right) \\
\nonumber
&\stackrel{(c)}{=}& H\left(\overrightarrow{\bs{Y}}_i | \bs{U}, \bs{U}_i, \bs{U}_j\right) \\
\nonumber
&=& \max\left(H\left(\bs{V}_j\right), H\left(\bs{X}_{i,HA}\right)+H\left(\bs{X}_{i,P}\right)\right) \\
\nonumber
&=& \max\left(\dim \bs{V}_j, \dim \bs{X}_{i,HA}+\dim \bs{X}_{i,P}\right) \\ 
\nonumber
&=&  \max\Big(\min\big(n_{ij},\left(\max\left(\overrightarrow{n}_{ii},n_{ij}\right)-\overleftarrow{n}_{ii}\right)^+\big), \\
\nonumber
& & \min\big(n_{ji},\left(\max\left(\overrightarrow{n}_{jj},n_{ji}\right)-\overleftarrow{n}_{jj}\right)^+\big)\\
\nonumber
& & -\min\big(\left(n_{ji}-\overrightarrow{n}_{ii}\right)^+,\left(\max\left(\overrightarrow{n}_{jj},n_{ji}\right)-\overleftarrow{n}_{jj}\right)^+\big)\\
\label{EqRateRegion32}
& & +\left(\overrightarrow{n}_{ii}-n_{ji} \right)^+\Big),
\end{IEEEeqnarray}
\end{subequations}
where
(c) follows from the fact that $H(\overrightarrow{\bs{Y}}_i|\bs{U}, \bs{U}_j,\bs{V}_j, \bs{X}_i)=0$.

Plugging \eqref{EqRateRegionLDM} into  \eqref{EqRateRegion2b}  (after some algebraic manipulation) yields the system of inequalities in Theorem~\ref{TheoremANFBLDMCap}.  

The sum-rate bound in \eqref{EqRateRegion23} is simplified as follows:
\begin{IEEEeqnarray}{rcl}
\nonumber
R_{1}+R_{2}  &  \leqslant  &  \min \Big(  \theta_{2,1}+\theta_{4,2}, \theta_{4,1}+\theta_{2,2}, \\
\label{EqSumrateRegion2bsimplified}
& & \theta_{1,1}+\theta_{5,1}+\theta_{1,2}+\theta_{5,2} \Big). 
\end{IEEEeqnarray}
Note that this follows from the fact that 
\begin{IEEEeqnarray}{rcl}
\nonumber
& & \max\Big(\theta_{2,1}+\theta_{4,2}, \theta_{4,1}+\theta_{2,2}, \theta_{1,1}+\theta_{5,1}+\theta_{1,2}+\theta_{5,2}\Big) \leqslant \\
\nonumber
& & \min\Big(\theta_{2,1}+a_{6,2}, \theta_{6,1}+\theta_{2,2}, \theta_{1,1}+\theta_{3,1}+\theta_{4,1}+\theta_{1,2}+\theta_{5,2}, \\
\nonumber
& & \theta_{1,1}+\theta_{7,1}+\theta_{1,2}+\theta_{5,2}, \theta_{1,1}+\theta_{4,1}+\theta_{1,2}+\theta_{7,2},  \\
\nonumber
& & \theta_{1,1}+\theta_{5,1}+\theta_{1,2}+\theta_{3,2}+\theta_{4,2}, \theta_{1,1}+\theta_{7,1}+\theta_{1,2}+\theta_{4,2}\Big).\\
\end{IEEEeqnarray}

\subsection{An Achievable Region for the Two-user Gaussian Interference Channel with Noisy Channel-Output Feedback} \label{AppAch-GIC-NF}

Consider that transmitter $i$ uses the following channel input: 
\begin{equation}
\label{EqXia}
X_i=U+U_i+V_i+X_{i,P}, 
\end{equation}
where $U$, $U_1$, $U_2$, $V_1$, $V_2$, $X_{1,P}$, and $X_{2,P}$ in \eqref{EqXia} are mutually independent and distributed as follows:  
\begin{subequations}
\label{EqAchievGaussDef}
\begin{IEEEeqnarray}{rcl}
\label{EqU}
U&  \sim &\mathcal{N}\left(0,\rho \right), \\
\label{EqUi}
U_i&\sim&\mathcal{N}\left(0,\mu_i \lambda_{i,C}\right), \\
\label{EqVi}
V_i&\sim&\mathcal{N}\left(0,(1-\mu_i)\lambda_{i,C}\right), \\
\label{EqXip}
X_{i,P}&\sim&\mathcal{N}\left(0,\lambda_{i,P}\right),
\end{IEEEeqnarray}
\end{subequations}
with
\begin{subequations}
\label{EqAchievGaussDef2}
\begin{equation}
\label{Eqpwrallc}
\rho+\lambda_{i,C}+\lambda_{i,P} = 1 \mbox{ and}
\end{equation}
\begin{IEEEeqnarray}{rcl}
\label{Eqpwrp}
\lambda_{i,P}&  = &\min\left(\frac{1}{\INR_{ji}},1\right), 
\end{IEEEeqnarray}
\end{subequations}
where $\mu_i \in \left[0,1\right]$ and $\rho \in \left[0,\left(1-\max\left(\frac{1}{\INR_{12}},\frac{1}{\INR_{21}}\right) \right)^+\right]$. 

The parameters $\rho$, $\mu_i$, and $\lambda_{i,P}$ define a particular coding scheme for transmitter $i$.
The assignment in \eqref{Eqpwrp} is based on the intuition obtained from the linear deterministic model, in which the power of the signal $X_{i,P}$ from transmitter $i$ to receiver $j$ must be observed at the noise level.  
From \eqref{Eqsignalyif}, \eqref{Eqsignalyib}, and \eqref{EqXia}, the right-hand side of the inequalities in \eqref{EqRateRegion-z} is written in terms of $\overrightarrow{\SNR}_1$, $\overrightarrow{\SNR}_2$, $\INR_{12}$, $\INR_{21}$, $\overleftarrow{\SNR}_1$, $\overleftarrow{\SNR}_2$, $\rho$, $\mu_1$, and $\mu_2$ as follows:
\begin{subequations}
\label{EqIab}
\begin{IEEEeqnarray}{rcl}
\nonumber
\theta_{1,i} & = & I\left(\overleftarrow{Y}_i ; U_j  | U, X_i  \right) \\
\nonumber
&=& \frac{1}{2}\log \left(\frac{\overleftarrow{\SNR}_i\Big(b_{2,i}(\rho)+2\Big)+b_{1,i}(1)+1}{\overleftarrow{\SNR}_i\Big(\left(1\!-\!\mu_j\right)b_{2,i}(\rho)\!+\!2\Big)\!+\!b_{1,i}(1)\!+1}\right) \\
\label{EqIab8}
&=& a_{3,i}(\rho,\mu_j), \\
\nonumber
\theta_{2,i}& = & I\left(\overrightarrow{Y}_i; U, U_j,V_j, X_i \right) \\
\nonumber
&=&  \frac{1}{2}\log \Big(b_{1,i}(\rho)+1\Big)-\frac{1}{2} \\
\label{EqIab9}
&=& a_{2,i}(\rho),
\end{IEEEeqnarray}
\begin{IEEEeqnarray}{rcl}
\nonumber
\theta_{3,i}& = & I\left(\overrightarrow{Y}_i; V_j | U, U_j, X_i \right) \\
\nonumber
&=&  \frac{1}{2}\log \bigg(\Big(1-\mu_j\Big)b_{2,i}(\rho)+2\bigg)-\frac{1}{2} \\
\label{EqIab10}
&=& a_{4,i}(\rho,\mu_j), \\
\nonumber
\theta_{4,i}& = & I\left(\overrightarrow{Y}_i; X_i | U, U_i, U_j,V_i, V_j   \right) \\
\nonumber
&=&  \frac{1}{2}\log\left(\frac{\overrightarrow{\SNR}_{i}}{\INR_{ji}}+2\right)-\frac{1}{2} \\
\label{EqIab11}
&=& a_{1,i}, \\
\nonumber
\theta_{5,i}& = & I\left(\overrightarrow{Y}_i; V_j, X_i | U, U_i, U_j, V_i   \right) \\
\nonumber
& & = \frac{1}{2}\log \left(2+\frac{\overrightarrow{\SNR}_{i}}{\INR_{ji}}+\Big(1-\mu_j\Big)b_{2,i}(\rho)\right)-\frac{1}{2} \\
\label{EqIab12}
&=& a_{5,i}(\rho,\mu_j), \\
\nonumber
\theta_{6,i} & = & I\left(\overrightarrow{Y}_i; X_i | U, U_i, U_j, V_j   \right) \\
\nonumber
& & = \frac{1}{2}\log \left(\frac{\overrightarrow{\SNR}_{i}}{\INR_{ji}}\bigg(\Big(1-\mu_i\Big)b_{2,j}(\rho)+1\right)+2\bigg)-\frac{1}{2} \\
\label{EqIab13}
&=& a_{6,i}(\rho,\mu_i), \mbox{ and } \\
\nonumber
\theta_{7,i}& = & I\left( \overrightarrow{Y}_i; V_j, X_i | U, U_i, U_j  \right)  \\
\nonumber
&=& \! \frac{1}{2} \! \log\! \left(\!\frac{\!\overrightarrow{\SNR}_{i}\!}{\!\INR_{ji}\!}\!\bigg(\!\Big(1\!-\!\mu_i\Big)b_{2,j}(\rho)\!+\!1\bigg)\!+\!\Big(1\!-\!\mu_j \! \Big) \! b_{2,i}(\rho)\!+\!2\!\right)\\
\nonumber
& & \!-\frac{1}{2}\\
\label{EqIab14}
&=& a_{7,i}(\rho,\mu_1, \mu_2).
\end{IEEEeqnarray}
\end{subequations}
Finally, plugging \eqref{EqIab} into  \eqref{EqRateRegion2b}  (after some algebraic manipulation) yields the system of inequalities in Theorem~\ref{TheoremA-G-IC-NF}.  
The sum-rate bound in \eqref{EqRateRegion23} is simplified to give:
\begin{IEEEeqnarray}{rcl}
\nonumber
R_{1}+R_{2}   &  \leqslant &  \min \Big(a_{2,1}(\rho)+a_{1,2}, a_{1,1}+a_{2,2}(\rho), \\
\nonumber
& & \! a_{3,1}(\rho,\mu_2)\!+\! a_{1,1}\!+\!a_{3,2}(\rho,\mu_1)\!+\!a_{7,2}(\rho,\mu_1,\mu_2),   \\
\nonumber
& & \! a_{3,1}(\rho,\mu_2)\!+\!a_{5,1}(\rho,\mu_2)\!+\!a_{3,2}(\rho,\mu_1)\!+\!a_{5,2}(\rho,\mu_1), \\
\nonumber
& & \!a_{3,1}(\rho,\mu_2)\!+\!a_{7,1}(\rho,\mu_1,\mu_2)\!+\!a_{3,2}(\rho,\mu_1)\!+\!a_{1,2} \! \Big).\\
\label{EqRateRegion22simplified} 
\end{IEEEeqnarray}
Note that this follows from the fact that 

\begin{IEEEeqnarray}{rcl}
\nonumber
& & \max\Big( a_{2,1}(\rho)+a_{1,2}, a_{1,1}+a_{2,2}(\rho), \\
\nonumber
& & a_{3,1}(\rho,\mu_2)+a_{1,1}+a_{3,2}(\rho,\mu_1)+a_{7,2}(\rho,\mu_1,\mu_2), \\
\nonumber
& & a_{3,1}(\rho,\mu_2)+a_{5,1}(\rho,\mu_2)+a_{3,2}(\rho,\mu_1)+a_{5,2}(\rho,\mu_1), \\
\nonumber
& & a_{3,1}(\rho,\mu_2)+a_{7,1}(\rho,\mu_1,\mu_2)+a_{3,2}(\rho,\mu_1)+a_{1,2}\Big)   \\
\nonumber
\leqslant & & \min\Big(a_{2,1}+a_{6,2}(\rho,\mu_2),a_{6,1}(\rho,\mu_1)+a_{2,2}(\rho), \\
\nonumber
& &a_{3,1}(\rho,\mu_2)+a_{4,1}(\rho,\mu_2)+a_{1,1}+a_{3,2}(\rho,\mu_1)+a_{5,2}(\rho,\mu_1), \\
\nonumber
& & a_{3,1}(\rho,\mu_2)+a_{7,1}(\rho,\mu_1,\mu_2)+a_{3,2}(\rho,\mu_1)+a_{5,2}(\rho,\mu_1), \\
\nonumber
& & a_{3,1}(\rho,\mu_2)+a_{5,1}(\rho,\mu_2)+a_{3,2}(\rho,\mu_1)+\theta_{3,2}+a_{1,2}\Big). \\
\end{IEEEeqnarray}
Therefore, the inequalities in \eqref{EqRateRegion2b} simplify into \eqref{EqRa-G-IC-NF} and this completes the proof of Theorem~\ref{TheoremA-G-IC-NF}. 
 \section{Converse Proof for Theorem~\ref{TheoremANFBLDMCap}} \label{App-C-LD-IC-NF}

This appendix provides the second part of the proof of Theorem~\ref{TheoremANFBLDMCap}. 
The proof of inequalities \eqref{EqRiV2} and \eqref{EqRi+Rj-1-V2} is presented in \cite{Suh-TIT-2011}.
The rest of this appendix provides a proof of the inequalities \eqref{EqRi-2-V2}, \eqref{EqRi+Rj-2-V2} and \eqref{Eq2Ri+Rj-V2}.

\textbf{Notation.}  For all $i \in \lbrace 1,2 \rbrace$, the channel input $\bs{X}_{i,n}$ of the LD-IC-NF in \eqref{EqLDICsignals} for any channel use $n \in \lbrace 1, 2,  \ldots, N\rbrace$  is  a $q$-dimensional vector, with $q$ in \eqref{Eqq}, that is written as the concatenation of four vectors: $\bs{X}_{i,C,n}$, $\bs{X}_{i,P,n}$, $\bs{X}_{i,D,n}$, and $\bs{X}_{i,Q,n}$, i.e., $\bs{X}_{i,n} = \Big(\bs{X}_{i,C,n}^{\sfT},\bs{X}_{i,P,n}^{\sfT}, \bs{X}_{i,D,n}^{\sfT},\bs{X}_{i,Q,n}^{\sfT}\Big)^{\sfT}$, as shown in Figure~\ref{FigMessagesLDICNFB}. Note that this notation is independent of the feedback parameters $\overleftarrow{n}_{11}$ and $\overleftarrow{n}_{22}$, and it holds for all $n\in\{1, 2, \ldots,N\}$. More specifically, 
\begin{subequations}
$\bs{X}_{i,C,n}$ represents the bits of $\bs{X}_{i,n}$ that are observed by both receivers. Then, 
\begin{IEEEeqnarray}{rcl}
\label{EqdimXic}
\dim \bs{X}_{i,C,n} & =  & \min\left( \overrightarrow{n}_{ii}, n_{ji} \right), 
\end{IEEEeqnarray}
$\bs{X}_{i,P,n}$ represents the bits of $\bs{X}_{i,n}$ that are observed only at receiver $i$. Then, 
\begin{IEEEeqnarray}{rcl}
\label{EqdimXip}
\dim\bs{X}_{i,P,n} &  = & (\overrightarrow{n}_{ii} - n_{ji})^+,
\end{IEEEeqnarray}
$\bs{X}_{i,D,n}$ represents the bits of $\bs{X}_{i,n}$ that are observed only at receiver $j$. Then, 
\begin{IEEEeqnarray}{rcl}
\label{EqdimXid}
\dim \bs{X}_{i,D,n} & =  & (n_{ji} - \overrightarrow{n}_{ii})^+,\mbox{ and }
\end{IEEEeqnarray}
$\bs{X}_{i,Q,n} = \left(0,\ldots,0\right)^{\sfT}$  is included for dimensional matching of the model in \eqref{EqLDICsignalsc}. Then,
\begin{IEEEeqnarray}{rcl}
\label{EqdimXid}
\dim \bs{X}_{i,Q,n} &  =  & q - \max\left( \overrightarrow{n}_{ii}, n_{ji} \right).
\end{IEEEeqnarray}
The bits $\bs{X}_{i,Q,n}$ are fixed and thus do not carry information. Hence, the following holds:
\begin{IEEEeqnarray}{rcl}
\nonumber
H\left( \bs{X}_{i,n} \right) & =  &H\big( \bs{X}_{i,C,n}, \bs{X}_{i,P,n}, \bs{X}_{i,D,n}, \bs{X}_{i,Q,n} \big)\\
\nonumber
& = &H\big( \bs{X}_{i,C,n}, \bs{X}_{i,P,n}, \bs{X}_{i,D,n} \big)\\
\label{EqHXi}
& \leqslant & \dim\bs{X}_{i,C,n} + \dim\bs{X}_{i,P,n} + \dim\bs{X}_{i,D,n}. \qquad
\end{IEEEeqnarray}
Note that vectors $\bs{X}_{i,P,n}$ and $\bs{X}_{i,D,n}$ do not exist simultaneously. The former exists when $\overrightarrow{n}_{ii} > n_{ji}$, while the latter exists when $\overrightarrow{n}_{ii} < n_{ji}$. 
\end{subequations}
Let $\bs{X}_{i,D,n}$ be written in terms of $\bs{X}_{i,DF,n}$ and $\bs{X}_{i,DG,n}$, i.e., $\bs{X}_{i,D,n}=\Big(\bs{X}_{i,DF,n}^{\sfT},\bs{X}_{i,DG,n}^{\sfT}\Big)^{\sfT}$.
The vector $\bs{X}_{i,DF,n}$ represents the bits of $\bs{X}_{i,D,n}$ that are above the noise level in the feedback link from receiver $j$ to transmitter $j$;
and $\bs{X}_{i,DG,n}$ represents the bits of $\bs{X}_{i,D,n}$ that are below the noise level in the feedback link from receiver $j$ to transmitter $j$, as shown in Figure~\ref{FigMessagesLDICNFB}.  
The dimension of vectors $\bs{X}_{i,DF,n}$ and $\bs{X}_{i,DG,n}$ are given by
\begin{subequations}
\begin{IEEEeqnarray}{lcl}
\nonumber
\dim  \bs{X}_{i,DF,n} &  = &  \min\Bigg(\left(n_{ji}-\overrightarrow{n}_{ii}\right)^+,\Big(\overleftarrow{n}_{jj}-\overrightarrow{n}_{ii}\\
\nonumber
& & -\min\left(\left(\overrightarrow{n}_{jj}-n_{ji}\right)^+,n_{ij}\right)\\
\label{dimXiDF-1}
& & -\left(\left(\overrightarrow{n}_{jj}-n_{ij}\right)^+-n_{ji}\right)^+\Big)^+\Bigg) \mbox{ and } \qquad\\
\label{dimXiDF-2}
\dim \bs{X}_{i,DG,n} &  = &\dim \bs{X}_{i,D,n}  - \dim \bs{X}_{i,DF,n}.
\end{IEEEeqnarray}
\end{subequations}
Let $\bs{X}_{i,C,n}$ be written in terms of $\bs{X}_{i,CF,n}$ and $\bs{X}_{i,CG,n}$, i.e., ${\bs{X}_{i,C,n}=\Big(\bs{X}_{i,CF,n}^{\sfT}, \bs{X}_{i,CG,n}^{\sfT}\Big)^{\sfT}}$.
The vector $\bs{X}_{i,CF,n}$ represents the bits of $\bs{X}_{i,C,n}$ that are above the noise level in the feedback link from receiver $j$ to transmitter $j$; 
and $\bs{X}_{i,CG,n}$ represents the bits of $\bs{X}_{i,C,n}$ that are below the noise level in the feedback link from receiver $j$ to transmitter $j$, as shown in Figure~\ref{FigMessagesLDICNFB}.  
Define the dimension of the vector $\left(\bs{X}_{i,CF,n}^{\sfT}, \bs{X}_{i,DF,n}^{\sfT} \right)^{\sfT}$ as follows:

\begin{IEEEeqnarray}{lll}
\nonumber
\dim \left(\left(\bs{X}_{i,CF,n}^{\sfT}, \bs{X}_{i,DF,n}^{\sfT} \right)^{\sfT}\right) &  =  & \Big(\min\left(\overleftarrow{n}_{jj}, \max\left(\overrightarrow{n}_{jj},n_{ji}\right)\right)\\
\label{EqdimXicfjXidf}
& & -\left(\overrightarrow{n}_{jj}-n_{ji}\right)^+\Big)^+. \quad
\end{IEEEeqnarray}
The dimension of vectors $\bs{X}_{i,CF,n}$ and $\bs{X}_{i,CG,n}$ is obtained as follows:
\begin{subequations}
\begin{IEEEeqnarray}{lcl}
\nonumber
\dim \bs{X}_{i,CF,n} &  = &\dim \left(\left(\bs{X}_{i,CF,n}^{\sfT}, \bs{X}_{i,DF,n}^{\sfT} \right)^{\sfT}\right)  - \dim \bs{X}_{i,DF,n} \\
\label{dimXiCFjn} \\
\nonumber
\textrm{and }\\
\label{dimXiCGjn}
\dim \bs{X}_{i,CG,n} &  = &\dim \bs{X}_{i,C,n}  - \dim \bs{X}_{i,CF,n}.
\end{IEEEeqnarray}
\end{subequations}
The vector $\bs{X}_{i,U,n}$ is used to represent the bits of vector  $\bs{X}_{i,n}$ that are observed at receiver $j$ without producing any interference with bits in $\bs{X}_{j,P,n}$.  An example is shown in Figure~\ref{FigXiu}. 

Based on its definition, $\bs{X}_{i,U,n}$ consists of the top
\begin{IEEEeqnarray}{rcl}
\nonumber
\dim \bs{X}_{i,U,n} & = & \min\left(\overrightarrow{n}_{jj},n_{ij} \right)-\min\left(\left(\overrightarrow{n}_{jj}-n_{ji}\right)^+,n_{ij} \right)\\
\label{EqHXtopi2}
& & +\left(n_{ji}-\overrightarrow{n}_{jj}\right)^+
\end{IEEEeqnarray}
bits of $\bs{X}_{i,n}$. 

Finally, for all $i \in \lbrace 1, 2 \rbrace$, with $j \in \lbrace 1, 2 \rbrace \setminus \lbrace i \rbrace$, the channel output $\overrightarrow{\bs{Y}}_{i,n}$ is written as the concatenation of three vectors: $\overrightarrow{\bs{Y}}_{i,Q,n}$, $\overleftarrow{\bs{Y}}_{i,n}$, and $\overrightarrow{\bs{Y}}_{i,G,n}$, i.e., $\overrightarrow{\bs{Y}}_{i,n} = \Big(\overrightarrow{\bs{Y}}_{i,Q,n}^{\sfT}, \overleftarrow{\bs{Y}}_{i,n}^{\sfT}, \overrightarrow{\bs{Y}}_{i,G,n}^{\sfT}\Big)^{\sfT}$, as shown in Figure~\ref{FigMessagesLDICNFB}.  More specifically, the vector $\overleftarrow{\bs{Y}}_{i,n}$ contains the bits that are above the noise level in the feedback link from receiver $i$ to transmitter $i$. Then, 
\begin{subequations}
\begin{IEEEeqnarray}{rcl}
\label{EqdimYib}
\dim \overleftarrow{\bs{Y}}_{i,n}  & =  & \min\Big( \overleftarrow{n}_{ii}, \max\left(\overrightarrow{n}_{ii},n_{ij}\right)\Big).
\end{IEEEeqnarray}

The vector $\overrightarrow{\bs{Y}}_{i,G,n}$ contains the bits that are below the noise level in the feedback link from receiver $i$ to transmitter $i$. Then, 
\begin{IEEEeqnarray}{rcl}
\label{EqdimYig}
\dim \overrightarrow{\bs{Y}}_{i,G,n} &  =  & \Big(\max\left(\overrightarrow{n}_{ii},n_{ij}\right)-\overleftarrow{n}_{ii} \Big)^+.
\end{IEEEeqnarray}

The vector $\overrightarrow{\bs{Y}}_{i,Q,n}=\left(0, \ldots, 0 \right)$ is included for dimensional matching with the model in \eqref{EqLDICsignalsc}. Then, 

\begin{IEEEeqnarray}{rcl}
\nonumber
H\left( \overrightarrow{\bs{Y}}_{i,n} \right) &  =  &H\big(\overrightarrow{\bs{Y}}_{i,Q,n}, \overleftarrow{\bs{Y}}_{i,n}, \overrightarrow{\bs{Y}}_{i,G,n} \big)\\
\nonumber
& = &H\big(\overleftarrow{\bs{Y}}_{i,n}, \overrightarrow{\bs{Y}}_{i,G,n} \big)\\
\label{EqHYi}
& \leqslant & \dim \overleftarrow{\bs{Y}}_{i,n} + \dim \overrightarrow{\bs{Y}}_{i,G,n}.
\end{IEEEeqnarray}
\end{subequations}
Using this notation, the proof continues as follows. 

\textbf{Proof of \eqref{EqRi-2-V2}:}
First, consider $n_{ji} \leqslant \overrightarrow{n}_{ii}$, i.e., vector $\bs{X}_{i,P,n}$ exists and vector $\bs{X}_{i,D,n}$ does not exist.  From the assumption that the message indices $W_1$ and $W_2$  are i.i.d. following a uniform distribution over the sets $\mathcal{W}_1$ and $\mathcal{W}_2$, respectively, the following holds for any $k \in \lbrace 1, 2, \ldots, N \rbrace$:   
\begin{IEEEeqnarray}{rcl}
\nonumber
NR_i &  = & H\left(W_i\right)\\
\nonumber
&\stackrel{(a)}{=}& H\left(W_i|W_j\right)\\
\nonumber
&\stackrel{(b)}{\leqslant}& I\left(W_i;\overrightarrow{\bs{Y}}_i,\overleftarrow{\bs{Y}}_j|W_j\right)+N\delta(N)\\
\nonumber
&=& H\left(\overrightarrow{\bs{Y}}_i,\overleftarrow{\bs{Y}}_j|W_j\right)+N\delta(N) \\
\nonumber
&\stackrel{(c)}{=}& \sum_{n=1}^{N}H\!\Big(\!\overrightarrow{\bs{Y}}_{i,n},\overleftarrow{\bs{Y}}_{j,n}|W_j,\overrightarrow{\bs{Y}}_{i,(1:n-1)},\overleftarrow{\bs{Y}}_{j,(1:n-1)}, \bs{X}_{j,n}\Big)\\
\nonumber
& & +N\delta(N)\\
\nonumber
&\leqslant& \sum_{n=1}^{N}H\Big(\bs{X}_{i,n},\overleftarrow{\bs{Y}}_{j,n}|\bs{X}_{j,n}\Big)+N\delta(N)\\
\nonumber
&\leqslant& \sum_{n=1}^{N}H\left(\bs{X}_{i,n}\right)+N\delta(N)\\
\nonumber
&=& NH\left(\bs{X}_{i,k}\right)+N\delta(N), \\
\label{EqRi-42}
&\leqslant & N\left(\dim \bs{X}_{i,C,k} + \dim \bs{X}_{i,P,k}\right)+N\delta(N).
\end{IEEEeqnarray}
Second, consider the case in which ${n_{ji} > \overrightarrow{n}_{ii}}$. In this case the vector $\bs{X}_{i,P,n}$ does not exist and the vector $\bs{X}_{i,D,n}$ exists.  From the assumption that the message indices $W_1$ and $W_2$  are i.i.d. following a uniform distribution over the sets $\mathcal{W}_1$ and $\mathcal{W}_2$, respectively, the following holds for any $k \in \lbrace 1, 2, \ldots, N \rbrace$:  
\begin{IEEEeqnarray}{rcl}
\nonumber
NR_i &  = & H\left(W_i\right)\\
\nonumber
&\stackrel{(a)}{=}& H\left(W_i|W_j\right)\\
\nonumber
&\stackrel{(b)}{\leqslant}& I\left(W_i;\overrightarrow{\bs{Y}}_i,\overleftarrow{\bs{Y}}_j|W_j\right)+N\delta(N) \\
\nonumber
&=& H\left(\overrightarrow{\bs{Y}}_i,\overleftarrow{\bs{Y}}_j|W_j\right)+N\delta(N)\\
\nonumber
&\stackrel{(c)}{=}& \sum_{n=1}^{N}\! H \! \Big(\overrightarrow{\bs{Y}}_{i,n},\overleftarrow{\bs{Y}}_{j,n}|W_j,\overrightarrow{\bs{Y}}_{i,(1:n-1)},\overleftarrow{\bs{Y}}_{j,(1:n-1)}, \bs{X}_{j,n}\Big)\\
\nonumber
& & +N\delta(N) \\
\nonumber
&\leqslant& \sum_{n=1}^{N}H\Big(\bs{X}_{i,C,n}, \bs{X}_{i,CF,n}, \bs{X}_{i,DF,n}\Big)+N\delta(N) \\
\nonumber
&=& \sum_{n=1}^{N}H\Big(\bs{X}_{i,C,n}, \bs{X}_{i,DF,n}\Big)+N\delta(N)\\
\nonumber
&=& NH\Big(\bs{X}_{i,C,k}, \bs{X}_{i,DF,k}\Big)+N\delta(N) \\
\label{EqRi-23}
&\leqslant & N\left(\dim \bs{X}_{i,C,k} + \dim \bs{X}_{i,DF,k}\right)+N\delta(N).
\end{IEEEeqnarray}

The inequalities $(a)$, $(b)$, and $(c)$ in \eqref{EqRi-42} and \eqref{EqRi-23} are justified in accordance with:
(a) follows from the fact that $W_1$ and $W_2$ are independent; 
(b) follows from considering enhanced receivers (see Figure \ref{Fig:G-IC-NF-Conv}(a) for the case $i=1$) and Fano's inequality; and 
(c) follows from the fact that $\bs{X}_{j,n}=f_j^{(n)}\left(W_j,\overleftarrow{\bs{Y}}_{j,(1:n-1)}\right)$.

Then,  \eqref{EqRi-42} and \eqref{EqRi-23} are expressed as one inequality in the asymptotic regime, as follows:
\begin{IEEEeqnarray}{rcl}
\label{EqRi-44}
R_i &  \leqslant &\dim \bs{X}_{i,C,k}+\dim \bs{X}_{i,P,k}+\dim \bs{X}_{i,DF,k},
\end{IEEEeqnarray}
which holds for any $k \in \{1, 2, \ldots,N\}$.

Plugging \eqref{EqdimXic}, \eqref{EqdimXip}, and \eqref{dimXiDF-1} in \eqref{EqRi-44}, and after some algebraic manipulation, the following holds: 
\begin{figure*}[t]
 \centerline{\epsfig{figure=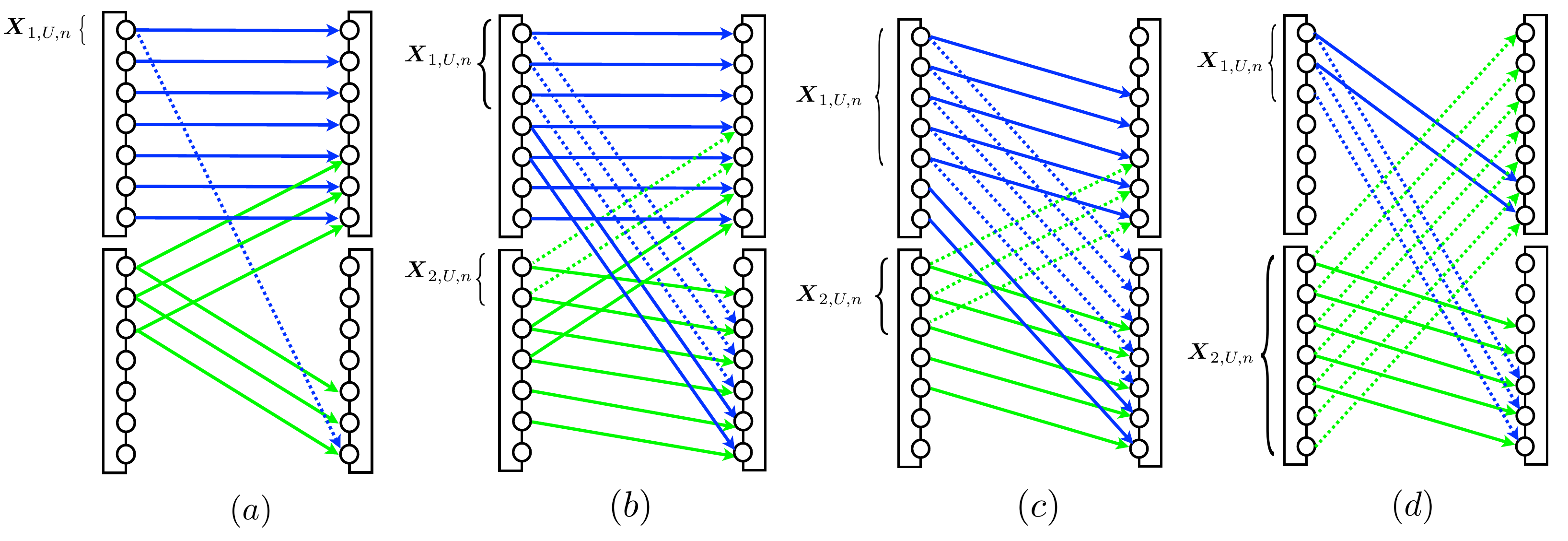,width=0.9\textwidth}}
 \caption{Vector $\bs{X}_{i,U,n}$ in different combination of interference regimes.}
\label{FigXiu}
\end{figure*}
\begin{IEEEeqnarray}{rcl}
\nonumber
R_i &  \leqslant & \min \! \Big( \! \max \! \left(\overrightarrow{n}_{ii},n_{ji}\right),\max\Big(\overrightarrow{n}_{ii},\overleftarrow{n}_{jj} \! - \! \left(\overrightarrow{n}_{jj} \! - \! n_{ji}\right)^+\Big)\Big).
\end{IEEEeqnarray}
This completes the proof of \eqref{EqRi-2-V2}. 

\textbf{Proof of \eqref{EqRi+Rj-2-V2}:}  From the assumption that the message indices $W_1$ and $W_2$  are i.i.d. following a uniform distribution over the sets $\mathcal{W}_1$ and $\mathcal{W}_2$, respectively, the following holds for any $k \in \lbrace 1, 2, \ldots, N \rbrace$: 
\begin{IEEEeqnarray}{lcl}
\nonumber
N&\big(\!R_1& \! + \! R_2 \! \big)\! = H\left(W_1\right)+H\left(W_2\right)\\
\nonumber
&\stackrel{(a)}{\leqslant}& I\left(W_1;\overrightarrow{\bs{Y}}_1,\overleftarrow{\bs{Y}}_1\right)+I\left(W_2;\overrightarrow{\bs{Y}}_2,\overleftarrow{\bs{Y}}_2\right)+N\delta(N) \\
\nonumber
&\leqslant&H \! \left(\!\overrightarrow{\bs{Y}}_1 \!\right)-H \! \left(\!\overleftarrow{\bs{Y}}_1|W_1 \!\right)-H \! \left(\! \bs{X}_{2,C}|W_1,\overleftarrow{\bs{Y}}_1,\bs{X}_1 \! \right)\\
\nonumber
& & +H\left(\overrightarrow{\bs{Y}}_2\right)-H\left(\overleftarrow{\bs{Y}}_2|W_2\right)\\
\nonumber
& & -H\left(\bs{X}_{1,C}|W_2,\overleftarrow{\bs{Y}}_2,\bs{X}_2\right)+N\delta(N) \\
\nonumber
&=&H\left(\overrightarrow{\bs{Y}}_1\right)-H\left(\overleftarrow{\bs{Y}}_1|W_1\right)-H\left(\bs{X}_{2,C},\bs{X}_{1,U}|W_1,\overleftarrow{\bs{Y}}_1,\bs{X}_1\right)\\
\nonumber
& & +H\left(\overrightarrow{\bs{Y}}_2\right)-H\left(\overleftarrow{\bs{Y}}_2|W_2\right) \\
\nonumber
& & -H\left(\bs{X}_{1,C},\bs{X}_{2,U}|W_2,\overleftarrow{\bs{Y}}_2,\bs{X}_2\right)+N\delta(N)\\
\nonumber
&=&H\left(\overrightarrow{\bs{Y}}_1\right)\!+\!\Big[I\left(\bs{X}_{2,C},\bs{X}_{1,U};W_1,\overleftarrow{\bs{Y}}_1\right) \!- \! H\left(\bs{X}_{2,C},\bs{X}_{1,U}\right)\Big]\\
\nonumber
& & +H\left(\overrightarrow{\bs{Y}}_2\right)\!+\!\left[I \!\left(\! \bs{X}_{1,C},\bs{X}_{2,U};W_2,\overleftarrow{\bs{Y}}_2\right)\!-\!H\left(\bs{X}_{1,C},\bs{X}_{2,U}\right)\right] \\
\nonumber
& & -H\left(\overleftarrow{\bs{Y}}_1|W_1\right)-H\left(\overleftarrow{\bs{Y}}_2|W_2\right)+N\delta(N)\\
\nonumber
&\stackrel{(b)}{=}&H\left(\overrightarrow{\bs{Y}}_1|\bs{X}_{1,C},\bs{X}_{2,U}\right)-H\left(\bs{X}_{1,C},\bs{X}_{2,U}|\overrightarrow{\bs{Y}}_1\right)\\
\nonumber
& & +H\left(\overrightarrow{\bs{Y}}_2|\bs{X}_{2,C},\bs{X}_{1,U}\right)-H\left(\bs{X}_{2,C},\bs{X}_{1,U}|\overrightarrow{\bs{Y}}_2\right)\\
\nonumber
& & +I\left(\bs{X}_{2,C},\bs{X}_{1,U};W_1,\overleftarrow{\bs{Y}}_1\right)+I\left(\bs{X}_{1,C},\bs{X}_{2,U};W_2,\overleftarrow{\bs{Y}}_2\right)\\
\nonumber
& & -H\left(\overleftarrow{\bs{Y}}_1|W_1\right)-H\left(\overleftarrow{\bs{Y}}_2|W_2\right)+N\delta(N) \\
\nonumber
& \leqslant & H\left(\overrightarrow{\bs{Y}}_1|\bs{X}_{1,C},\bs{X}_{2,U}\right)+H\left(\overrightarrow{\bs{Y}}_2|\bs{X}_{2,C},\bs{X}_{1,U}\right)\\
\nonumber
& & +I\left(\bs{X}_{2,C},\bs{X}_{1,U};W_1,\overleftarrow{\bs{Y}}_1\right)+I\left(\bs{X}_{1,C},\bs{X}_{2,U};W_2,\overleftarrow{\bs{Y}}_2\right)\\
\nonumber
& & -H\left(\overleftarrow{\bs{Y}}_1|W_1\right)-H\left(\overleftarrow{\bs{Y}}_2|W_2\right)+N\delta(N) \\
\nonumber
&\leqslant& H\left(\overrightarrow{\bs{Y}}_1|\bs{X}_{1,C},\bs{X}_{2,U}\right)+H\left(\overrightarrow{\bs{Y}}_2|\bs{X}_{2,C},\bs{X}_{1,U}\right)\\
\nonumber
& & +I\left(\bs{X}_{2,C},\bs{X}_{1,U},W_2,\overleftarrow{\bs{Y}}_2;W_1,\overleftarrow{\bs{Y}}_1\right)\\
\nonumber
& & +I\left(\bs{X}_{1,C},\bs{X}_{2,U},W_1,\overleftarrow{\bs{Y}}_1;W_2,\overleftarrow{\bs{Y}}_2\right)-H\left(\overleftarrow{\bs{Y}}_1|W_1\right)\\
\nonumber
& & -H\left(\overleftarrow{\bs{Y}}_2|W_2\right)+N\delta(N) \\
\nonumber
&=& H\left(\overrightarrow{\bs{Y}}_1|\bs{X}_{1,C},\bs{X}_{2,U}\right)+H\left(\overrightarrow{\bs{Y}}_2|\bs{X}_{2,C},\bs{X}_{1,U}\right)\\
\nonumber
& & +I\left(W_2;W_1,\overleftarrow{\bs{Y}}_1\right)+I\left(\bs{X}_{2,C},\bs{X}_{1,U},\overleftarrow{\bs{Y}}_2;W_1, \overleftarrow{\bs{Y}}_1|W_2\right) \\
\nonumber
& & +I\left(W_1;W_2,\overleftarrow{\bs{Y}}_2\right)+I\left(\bs{X}_{1,C},\bs{X}_{2,U},\overleftarrow{\bs{Y}}_1;W_2,\overleftarrow{\bs{Y}}_2|W_1\right)\\
\nonumber
& & -H\left(\overleftarrow{\bs{Y}}_1|W_1\right)-H\left(\overleftarrow{\bs{Y}}_2|W_2\right)+N\delta(N) 
\end{IEEEeqnarray}
\begin{IEEEeqnarray}{rcl}
\nonumber
&\stackrel{(c)}{=}& H\left(\overrightarrow{\bs{Y}}_1|\bs{X}_{1,C},\bs{X}_{2,U}\right)\!+\!H\left(\overrightarrow{\bs{Y}}_2|\bs{X}_{2,C},\bs{X}_{1,U}\right)\!+\!H\left(W_1\right)\\
\nonumber
& & +H\left(\overleftarrow{\bs{Y}}_1|W_1\right)-H\left(W_1|W_2\right)-H\left(\overleftarrow{\bs{Y}}_1|W_2,W_1\right) \\
\nonumber
& & +H\left(\bs{X}_{2,C},\bs{X}_{1,U},\overleftarrow{\bs{Y}}_2|W_2\right)+H\left(W_2\right)+H\left(\overleftarrow{\bs{Y}}_2|W_2\right)\\
\nonumber
& & -H\left(W_2|W_1\right)-H\left(\overleftarrow{\bs{Y}}_2|W_1,W_2\right) \\
\nonumber
& & +H \! \left( \! \bs{X}_{1,C},\bs{X}_{2,U},\overleftarrow{\bs{Y}}_1|W_1\right)\! - \! H \! \left( \! \overleftarrow{\bs{Y}}_1|W_1\right) \! - \! H \! \left(\! \overleftarrow{\bs{Y}}_2|W_2 \! \right)\\
\nonumber
& & +N\delta(N)\\
\nonumber
&\leqslant& H\left(\overrightarrow{\bs{Y}}_1|\bs{X}_{1,C},\bs{X}_{2,U}\right)+H\left(\overrightarrow{\bs{Y}}_2|\bs{X}_{2,C},\bs{X}_{1,U}\right)\\
\nonumber
& & +H\left(\bs{X}_{2,C},\bs{X}_{1,U},\overleftarrow{\bs{Y}}_2|W_2\right)\\
\nonumber
& & +H\left(\bs{X}_{1,C},\bs{X}_{2,U},\overleftarrow{\bs{Y}}_1|W_1\right)+N\delta(N)\\
\nonumber
&=& \sum_{n=1}^{N}\Big[H\left(\overrightarrow{\bs{Y}}_{1,n}|\bs{X}_{1,C},\bs{X}_{2,U},\overrightarrow{\bs{Y}}_{1,(1:n-1)}\right) + \\
\nonumber
& & H\left(\overrightarrow{\bs{Y}}_{2,n}|\bs{X}_{2,C},\bs{X}_{1,U},\overrightarrow{\bs{Y}}_{2,(1:n-1)}\right) + \\
\nonumber
& & H\Big(\!\!\bs{X}_{2,C,n},\bs{X}_{1,U,n},\!\overleftarrow{\bs{Y}}_{2,n}|\!W_2,\!\bs{X}_{2,C,(1:n-1)},\! \bs{X}_{1,U,(1:n-1)},\\
\nonumber
& &  \overleftarrow{\bs{Y}}_{2,(1:n-1)}\!\Big)\!+\! H\Big(\bs{X}_{1,C,n},\!\bs{X}_{2,U,n},\!\overleftarrow{\bs{Y}}_{1,n}|\!W_1,\! \bs{X}_{1,C,(1:n-1)},  \\
\nonumber
& & \bs{X}_{2,U,(1:n-1)}, \overleftarrow{\bs{Y}}_{1,(1:n-1)}\Big)\Big]+N\delta(N) \\
\nonumber
&\stackrel{(d)}{=}& \sum_{n=1}^{N}\Big[H\left(\overrightarrow{\bs{Y}}_{1,n}|\bs{X}_{1,C},\bs{X}_{2,U},\overrightarrow{\bs{Y}}_{1,(1:n-1)}\right) + \\
\nonumber
& & H\left(\overrightarrow{\bs{Y}}_{2,n}|\bs{X}_{2,C},\bs{X}_{1,U},\overrightarrow{\bs{Y}}_{2,(1:n-1)}\right) + \\
\nonumber
& & H\Big(\!\bs{X}_{2,C,n},\!\bs{X}_{1,U,n},\!\overleftarrow{\bs{Y}}_{2,n}|\!W_2,\!\bs{X}_{2,C,(1:n-1)},\!\bs{X}_{1,U,(1:n-1)}, \\
\nonumber
& & \overleftarrow{\bs{Y}}_{2,(1:n-1)},\bs{X}_{2,(1:n)}\Big) +H\Big(\bs{X}_{1,C,n},\bs{X}_{2,U,n},\overleftarrow{\bs{Y}}_{1,n}|W_1, \\
\nonumber
& & \bs{X}_{1,C,(1:n-1)}, \bs{X}_{2,U,(1:n-1)},\overleftarrow{\bs{Y}}_{1,(1:n-1)},\bs{X}_{1,(1:n)}\Big)\Big]\\
\nonumber
& & +N\delta(N) \\
\nonumber
&\stackrel{(e)}{\leqslant}& \! \sum_{n=1}^{N} \! \Big[ \! H \! \left( \! \overrightarrow{\bs{Y}}_{1,n}|\bs{X}_{1,C,n},\bs{X}_{2,U,n} \! \right) \!\! + \!\! H \! \left( \! \overrightarrow{\bs{Y}}_{2,n}|\bs{X}_{2,C,n},\bs{X}_{1,U,n} \! \right)\\
\nonumber
& &  +H\left(\bs{X}_{1,U,n},\overleftarrow{\bs{Y}}_{2,n}|\bs{X}_{2,n}\right)+H\left(\bs{X}_{2,U,n},\overleftarrow{\bs{Y}}_{1,n}|\bs{X}_{1,n}\right)\Big]\\
\nonumber
& & +N\delta(N)\\
\nonumber
&\leqslant& \! \sum_{n=1}^{N} \! \Big[ \! H\left(\bs{X}_{1,P,n}\right)\! + \! H\left(\bs{X}_{2,P,n}\right) \! + \! H\left(\bs{X}_{1,U,n},\overleftarrow{\bs{Y}}_{2,n}|\bs{X}_{2,n}\! \right)
\end{IEEEeqnarray}
\begin{IEEEeqnarray}{rcl}
\nonumber
& & +H\left(\bs{X}_{2,U,n},\overleftarrow{\bs{Y}}_{1,n}|\bs{X}_{1,n}\right)\Big]+N\delta(N)\\
\nonumber
&\stackrel{(e)}{\leqslant}& N\Big[H\left(\bs{X}_{1,P,k}\right)+H\left(\bs{X}_{2,P,k}\right)+H\left(\bs{X}_{1,U,k}\right)\\
\nonumber
& & +H\left(\overleftarrow{\bs{Y}}_{2,k}|\bs{X}_{2,k},\bs{X}_{1,U,k}\right)+H\left(\bs{X}_{2,U,k}\right) \\
\nonumber
& & +H\left(\overleftarrow{\bs{Y}}_{1,k}|\bs{X}_{1,k},\bs{X}_{2,U,k}\right)\Big]+N\delta(N), \\
\nonumber
&=& N\Big[H\left(\bs{X}_{1,P,k}\right)+H\left(\bs{X}_{2,P,k}\right)+H\left(\bs{X}_{1,U,k}\right)\\
\nonumber
& & +H\left(\bs{X}_{1,CF,k},\bs{X}_{1,DF,k}|\bs{X}_{2,k},\bs{X}_{1,U,k}\right)+H\left(\bs{X}_{2,U,k}\right)\\
\nonumber
& & +H\left(\bs{X}_{2,CF,k},\bs{X}_{2,DF,k}|\bs{X}_{1,k},\bs{X}_{2,U,k}\right)\Big]+N\delta(N)\\
\nonumber
&\leqslant& N\Big[H\left(\bs{X}_{1,P,k}\right)+H\left(\bs{X}_{2,P,k}\right)+H\left(\bs{X}_{1,U,k}\right)\\
\nonumber
& & +H\left(\bs{X}_{1,CF,k},\bs{X}_{1,DF,k}|\bs{X}_{1,U,k}\right)+H\left(\bs{X}_{2,U,k}\right)\\
\nonumber
& & +H\left(\bs{X}_{2,CF,k},\bs{X}_{2,DF,k}|\bs{X}_{2,U,k}\right)\Big]+N\delta(N) \\
\nonumber
&  \leqslant & N \Big[ \dim \bs{X}_{1,P,k}+\dim \bs{X}_{2,P,k}+\dim \bs{X}_{1,U,k}\\
\nonumber
& & \!+\!\Big( \dim \! \left( \bs{X}_{1,CF,k}\!,\!\bs{X}_{1,DF,k} \right) \!-\! \dim \! \bs{X}_{1,U,k} \Big)^+\!+\!\dim \! \bs{X}_{2,U,k}\\
\nonumber
& & +\Big(\dim \left(\bs{X}_{2,CF,k},\bs{X}_{2,DF,k}\right)- \dim \bs{X}_{2,U,k}\Big)^+ +\delta(N)\Big]. \\
\label{EqR1+R2-32c}
\end{IEEEeqnarray}
where, 
(a) follows from considering enhanced receivers (see Figure \ref{Fig:G-IC-NF-Conv}(b)) and Fano's inequality; 
(b) follows from the fact that $H(Y)-H(X)=H(Y|X)-H(X|Y)$; 
(c) follows from the fact that $H\Big(\bs{X}_{i,C}$, $\bs{X}_{j,U}$, $\overleftarrow{\bs{Y}}_i|W_i$, $W_j$, $\overleftarrow{\bs{Y}}_j\Big)=0$; 
(d) follows from the fact that $\bs{X}_{i,n}=f_i^{(n)}\left(W_i,\overleftarrow{\bs{Y}}_{i,(1:n-1)}\right)$; and 
(e) follows from the fact that conditioning does not increase entropy.
Plugging \eqref{EqdimXip}, \eqref{EqdimXicfjXidf}, and \eqref{EqHXtopi2} in \eqref{EqR1+R2-32c} and after some algebraic manipulation, the following holds in the asymptotic regime:  
\begin{IEEEeqnarray}{rcl}
\nonumber
&R_1&+R_2   \leqslant   \max\Big(\left(\overrightarrow{n}_{11}-{n}_{12} \right)^+, n_{21}, \\
\nonumber
& & \overrightarrow{n}_{11}-\left(\max\left(\overrightarrow{n}_{11},n_{12}\right)-\overleftarrow{n}_{11}\right)^+\Big) + \\
\nonumber
& &\max\!\Bigg(\!\!\left(\overrightarrow{n}_{22}-{n}_{21} \right)^+\!\!, \!n_{12},\!\!\overrightarrow{n}_{22}\!-\!\left(\max\left(\overrightarrow{n}_{22},n_{21}\right)\!-\!\overleftarrow{n}_{22}\right)^+\!\!\!\Bigg).
\end{IEEEeqnarray}
This completes the proof of \eqref{EqRi+Rj-2-V2}. 

\textbf{Proof of \eqref{Eq2Ri+Rj-V2}:} From the assumption that the message indices $W_1$ and $W_2$  are i.i.d. following a uniform distribution over the sets $\mathcal{W}_1$ and $\mathcal{W}_2$ respectively, it follows that for all $i \in \lbrace1,2\rbrace$, with $j \in \lbrace1,2 \rbrace \setminus\lbrace i \rbrace$ and for all $k \in \lbrace 1, 2, \ldots, N \rbrace$: 
\begin{IEEEeqnarray}{rll}
\nonumber
&N\big(&2 R_i + R_j\big) = 2H\left(W_i\right)+H\left(W_j\right)\\
\nonumber
&\stackrel{(a)}{  \leqslant }& I\left(W_i;\overrightarrow{\bs{Y}}_i,\overleftarrow{\bs{Y}}_i\right)+I\left(W_i;\overrightarrow{\bs{Y}}_i,\overleftarrow{\bs{Y}}_j|W_j\right)\\
\nonumber
& & +I\left(W_j;\overrightarrow{\bs{Y}}_j,\overleftarrow{\bs{Y}}_j\right)+N\delta(N) \\
\nonumber
&\stackrel{(b)}{=}& H\left(\overrightarrow{\bs{Y}}_i\right) \! - \! H\left(\overleftarrow{\bs{Y}}_i|W_i\right) \! - \! H\left(\overrightarrow{\bs{Y}}_i|W_i,\overleftarrow{\bs{Y}}_i\right) \! \\
\nonumber
& & + \! H\left(\overrightarrow{\bs{Y}}_i|W_j,\overleftarrow{\bs{Y}}_j\right)+H\left(\overrightarrow{\bs{Y}}_j\right)-H\left(\overrightarrow{\bs{Y}}_j|W_j,\overleftarrow{\bs{Y}}_j\right)\\
\nonumber
& & +N\delta(N) \\
\nonumber
&=& H\left(\overrightarrow{\bs{Y}}_i\right)-H\left(\overleftarrow{\bs{Y}}_i|W_i\right)-H\left(\bs{X}_{j,C},\bs{X}_{j,D}|W_i,\overleftarrow{\bs{Y}}_i\right)\\
\nonumber
& & +H\left(\overrightarrow{\bs{Y}}_i|W_j,\overleftarrow{\bs{Y}}_j\right)+H\left(\overrightarrow{\bs{Y}}_j\right)\\
\nonumber
& & -H\left(\bs{X}_{i,C}, \bs{X}_{i,D}|W_j,\overleftarrow{\bs{Y}}_j\right)+N\delta(N) \\
\nonumber
&\leqslant& H\left(\overrightarrow{\bs{Y}}_i\right)-H\left(\overleftarrow{\bs{Y}}_i|W_i\right)-H\left(\bs{X}_{j,C},\bs{X}_{i,U}|W_i,\overleftarrow{\bs{Y}}_i\right)\\
\nonumber
& & +H\left(\overrightarrow{\bs{Y}}_i|W_j,\overleftarrow{\bs{Y}}_j\right)+H\left(\overrightarrow{\bs{Y}}_j\right)\\
\nonumber
& & -H\left(\bs{X}_{i,C}|W_j,\overleftarrow{\bs{Y}}_j\right)+N\delta(N)\\
\nonumber
&\leqslant& H\left(\overrightarrow{\bs{Y}}_i\right)-H\left(\overleftarrow{\bs{Y}}_i|W_i\right)+\big[I\left(\bs{X}_{j,C},\bs{X}_{i,U};W_i,\overleftarrow{\bs{Y}}_i\right)\\
\nonumber
& & -H\left(\bs{X}_{j,C},\bs{X}_{i,U}\right)  \big]+H\left(\overrightarrow{\bs{Y}}_i,\bs{X}_{i,C}|W_j,\overleftarrow{\bs{Y}}_j\right)\\
\nonumber
& & +H\left(\overrightarrow{\bs{Y}}_j\right)-H\left(\bs{X}_{i,C}|W_j,\overleftarrow{\bs{Y}}_j\right)+N\delta(N) \\
\nonumber
&=& H\left(\overrightarrow{\bs{Y}}_i\right)-H\left(\overleftarrow{\bs{Y}}_i|W_i\right)+\Big[I\left(\bs{X}_{j,C},\bs{X}_{i,U};W_i,\overleftarrow{\bs{Y}}_i\right)\\ 
\nonumber
& & -H\left(\bs{X}_{j,C},\bs{X}_{i,U}\right)  \Big]+H\left(\overrightarrow{\bs{Y}}_i|W_j,\overleftarrow{\bs{Y}}_j,\bs{X}_{i,C}\right)\\
\nonumber
& & +H\left(\overrightarrow{\bs{Y}}_j\right)+N\delta(N) \\
\nonumber
&\leqslant& H\left(\overrightarrow{\bs{Y}}_i\right)-H\left(\overleftarrow{\bs{Y}}_i|W_i\right)+\Big[I\left(\bs{X}_{j,C},\bs{X}_{i,U};W_i,\overleftarrow{\bs{Y}}_i\right)\\
\nonumber
& & -H\left(\bs{X}_{j,C},\bs{X}_{i,U}\right)  \Big]+H\left(\overrightarrow{\bs{Y}}_i|W_j,\overleftarrow{\bs{Y}}_j,\bs{X}_{i,C}\right)\\
\nonumber
& & +H\left(\overrightarrow{\bs{Y}}_j,\bs{X}_{j,C},\bs{X}_{i,U}\right)+N\delta(N) \\
\nonumber
&\stackrel{(c)}{=}& H\left(\overrightarrow{\bs{Y}}_i\right)-H\left(\overleftarrow{\bs{Y}}_i|W_i\right)+I\left(\bs{X}_{j,C},\bs{X}_{i,U};W_i,\overleftarrow{\bs{Y}}_i\right)\\
\nonumber
& & +H\left(\overrightarrow{\bs{Y}}_i|W_j,\overleftarrow{\bs{Y}}_j,\bs{X}_{i,C}\right)+H\left(\overrightarrow{\bs{Y}}_j|\bs{X}_{j,C},\bs{X}_{i,U}\right)\\
\nonumber
& & +N\delta(N) \\
\nonumber
&\leqslant& H\left(\overrightarrow{\bs{Y}}_i\right)-H\left(\overleftarrow{\bs{Y}}_i|W_i\right)\\
\nonumber
& & +I\left(\bs{X}_{j,C},\bs{X}_{i,U},W_j,\overleftarrow{\bs{Y}}_j;W_i,\overleftarrow{\bs{Y}}_i\right)\\
\nonumber
& & +H\left(\overrightarrow{\bs{Y}}_i|W_j,\overleftarrow{\bs{Y}}_j,\bs{X}_{i,C}\right)+H\left(\overrightarrow{\bs{Y}}_j|\bs{X}_{j,C},\bs{X}_{i,U}\right)\\
\nonumber
& & +N\delta(N) \\
\nonumber
&  \stackrel{(d) }{=}& H\left(\overrightarrow{\bs{Y}}_i\right)-H\left(\overleftarrow{\bs{Y}}_i|W_j,W_i,\right)\\
\nonumber
& & +H \! \left(\bs{X}_{j,C},\bs{X}_{i,U},\overleftarrow{\bs{Y}}_j|W_j\right) \! + \! H\left(\! \overrightarrow{\bs{Y}}_i|W_j,\overleftarrow{\bs{Y}}_j,\bs{X}_{i,C}\right)\\
\nonumber
& & +H\left(\overrightarrow{\bs{Y}}_j|\bs{X}_{j,C},\bs{X}_{i,U}\right)+N\delta(N) \\
\nonumber
&\leqslant& H\left(\overrightarrow{\bs{Y}}_i\right)+H\left(\bs{X}_{j,C},\bs{X}_{i,U},\overleftarrow{\bs{Y}}_j|W_j\right)\\
\nonumber
& & +H\left(\overrightarrow{\bs{Y}}_i|W_j,\overleftarrow{\bs{Y}}_j,\bs{X}_{i,C}\right)+H\left(\overrightarrow{\bs{Y}}_j|\bs{X}_{j,C},\bs{X}_{i,U}\right)\\
\nonumber
& & +N\delta(N) 
\end{IEEEeqnarray}
\begin{IEEEeqnarray}{rcl}
\nonumber
&\leqslant& \sum_{n=1}^N \Big[H\left(\overrightarrow{\bs{Y}}_{i,n}\right)+H\Big(\bs{X}_{j,C,n},\bs{X}_{i,U,n},\overleftarrow{\bs{Y}}_{j,n}|W_j,\\
\nonumber
& & \bs{X}_{j,C,(1:n-1)},\bs{X}_{i,U,(1:n-1)},\overleftarrow{\bs{Y}}_{j,(1:n-1)}\Big)\\
\nonumber
& & +H\left(\overrightarrow{\bs{Y}}_{i,n}|W_j,\overleftarrow{\bs{Y}}_{j},\bs{X}_{i,C},\overrightarrow{\bs{Y}}_{i,(1:n-1)}\right)\\
\nonumber
& & +H\left(\overrightarrow{\bs{Y}}_{j,n}|\bs{X}_{j,C},\bs{X}_{i,U},\overrightarrow{\bs{Y}}_{j,(1:n-1)}\right)\Big] +N\delta(N) \\
\nonumber
&=& \sum_{n=1}^N \Big[H\left(\overrightarrow{\bs{Y}}_{i,n}\right)+H\Big(\bs{X}_{j,C,n},\bs{X}_{i,U,n},\overleftarrow{\bs{Y}}_{j,n}|W_j, \\
\nonumber
& & \bs{X}_{j,C,(1:n-1)},\bs{X}_{i,U,(1:n-1)},\overleftarrow{\bs{Y}}_{j,(1:n-1)},\bs{X}_{j,(1:n)}\Big)\\
\nonumber
& & +H\left(\overrightarrow{\bs{Y}}_{i,n}|W_j,\overleftarrow{\bs{Y}}_{j},\bs{X}_{i,C},\overrightarrow{\bs{Y}}_{i,(1:n-1)},\bs{X}_{j,(1:n)}\right)\\
\nonumber
& & +H\left(\overrightarrow{\bs{Y}}_{j,n}|\bs{X}_{j,C},\bs{X}_{i,U},\overrightarrow{\bs{Y}}_{j,(1:n-1)}\right)\Big]+N\delta(N) \\
\nonumber
&\leqslant& \sum_{n=1}^N \Big[H\left(\overrightarrow{\bs{Y}}_{i,n}\right)+H\left(\bs{X}_{i,U,n}|\bs{X}_{j,n}\right)\\
\nonumber
& & +H\left(\overleftarrow{\bs{Y}}_{j,n}|\bs{X}_{j,n},\bs{X}_{i,U,n}\right)+H\left(\overrightarrow{\bs{Y}}_{i,n}|\bs{X}_{i,C,n},\bs{X}_{j,n}\right)\\
\nonumber
& &  +H\left(\overrightarrow{\bs{Y}}_{j,n}|\bs{X}_{j,C,n},\bs{X}_{i,U,n}\right)\Big]+N\delta(N) \\
\nonumber
&\leqslant& N\Big[H\left(\overrightarrow{\bs{Y}}_{i,k}\right)+H\left(\bs{X}_{i,U,k}\right)+H\left(\overleftarrow{\bs{Y}}_{j,k}|\bs{X}_{j,k},\bs{X}_{i,U,k}\right)\\
\nonumber
& & +H\left(\bs{X}_{i,P,k}\right)+H\left(\bs{X}_{j,P,k}\right)\Big]+N\delta(N) \\
\nonumber
&=& N \! \Big[ \!H\left(\overrightarrow{\bs{Y}}_{i,k}\right)\!+\!H\left(\bs{X}_{i,U,k}\right)\!+\!H\left(\bs{X}_{i,CF,k}\!,\!\bs{X}_{i,DF,k} |\bs{X}_{i,U,k}\right)\\
\nonumber
& & +H\left(\bs{X}_{i,P,k}\right)+H\left(\bs{X}_{j,P,k}\right)\Big]+N\delta(N) \\
\nonumber
&\leqslant& N\Big[\dim \overleftarrow{\bs{Y}}_{i,k}+\dim \overrightarrow{\bs{Y}}_{i,G,k}+\dim \bs{X}_{i,U,k}\\
\nonumber
& & +\left(\dim\left(\bs{X}_{i,CF,k},\bs{X}_{i,DF,k}\right) - \dim \bs{X}_{i,U,k}\right)^++\dim \bs{X}_{i,P,k}\\
\label{Eq2Ri+Rjc38}
& &+\dim\bs{X}_{j,P,k}\big]+N\delta(N),
\end{IEEEeqnarray}
where, 
(a) follows from considering enhanced receivers (see Figure \ref{Fig:G-IC-NF-Conv}(c) for the case $i=1$) and Fano's inequality; 
(b) follows from the fact that $H\left(\overrightarrow{\bs{Y}}_i,\overleftarrow{\bs{Y}}_j|W_i,W_j\right)=0$; 
(c) follows from the fact that $H(Y|X)=H(X,Y)-H(X)$; and
(d) follows from the fact that $H\Big(\bs{X}_{j,C}$, $\bs{X}_{i,U}$, $\overleftarrow{\bs{Y}}_j|W_j$, $W_i$, $\overleftarrow{\bs{Y}}_i\Big)=0$.

Plugging \eqref{EqdimXip}, \eqref{EqdimXicfjXidf}, \eqref{EqHXtopi2}, \eqref{EqdimYib}, and \eqref{EqdimYig} in \eqref{Eq2Ri+Rjc38} and after some algebraic manipulation, the following holds in the asymptotic regime: 
\begin{IEEEeqnarray}{rcl}
\nonumber
&2 &R_i+R_j \leqslant  \max\left(\overrightarrow{n}_{ii},{n}_{ji} \right)+\left(\overrightarrow{n}_{ii}-{n}_{ij} \right)^+\\
\nonumber
& +& \max\Bigg(\!\!\left(\overrightarrow{n}_{jj}-{n}_{ji} \right)^+\!\!\!, n_{ij},\! \overrightarrow{n}_{jj} \!-\!\Big(\!\max\left(\overrightarrow{n}_{jj},{n}_{ji} \!\!\right)
\label{Eq2Ri+Rjc39}
 -\overleftarrow{n}_{jj}\Big)^+\!\!\Bigg). \qquad
\end{IEEEeqnarray}

This completes the proof of \eqref{Eq2Ri+Rj-V2}. 

 \section{Proof of Theorem~\ref{TheoremC-G-IC-NF}} \label{App-C-G-IC-NF}

The outer bounds \eqref{EqRic-12-G-IC-NF} and \eqref{EqR1+R2c-12-G-IC-NF} correspond to the outer bounds of the case of perfect channel-output feedback derived in \cite{Suh-TIT-2011}. 
The bounds \eqref{EqRic-3-G-IC-NF}, \eqref{EqR1+R2c-3g-G-IC-NF} and \eqref{Eq2Ri+Rjc-g-G-IC-NF} correspond to new outer bounds.
Before presenting the proof, consider the parameter $h_{ji,U}$, with $i \in \lbrace 1, 2 \rbrace$ and $j \in \lbrace 1, 2 \rbrace\setminus\lbrace i \rbrace$, defined as follows:
\begin{equation}
h_{ji,U}=\left\{
    \begin{array}{ll}
      0 & \textrm{if } (S_{1,i} \lor S_{2,i} \lor S_{3,i}) \\
    \sqrt{\frac{\INR_{ij}\INR_{ji}}{\overrightarrow{\SNR}_{j}}} & \textrm{if } (S_{4,i} \lor S_{5,i}),                                                                                                
  \end{array} \right.
\label{Eqdefhjiu}
\end{equation} 
where, the events $S_{1,i}$, $S_{2,i}$, $S_{3,i}$, $S_{4,i}$, and $S_{5,i}$ are defined in \eqref{EqSi}. 
Consider also the following signals:
\begin{IEEEeqnarray}{rcl}
\label{EqsignalXiCG}
X_{i,C,n}& = & \sqrt{\INR_{ji}}X_{i,n}+\overrightarrow{Z}_{j,n} \mbox{ and }\\
\label{EqsignalXiUG}
X_{i,U,n}& = &h_{ji,U}X_{i,n}+\overrightarrow{Z}_{j,n},
\end{IEEEeqnarray} 
where,  $X_{i,n}$ and $\overrightarrow{Z}_{j,n}$ are the channel input of transmitter $i$ and the noise observed at receiver $j$ during a given channel use $n \in \lbrace1, 2,  \ldots, N \rbrace$, as described by \eqref{Eqsignalyif}. 

\textbf{Proof of \eqref{EqRic-3-G-IC-NF}}: From the assumption that the message indices $W_1$ and $W_2$ are i.i.d. following a uniform distribution over the sets $\mathcal{W}_1$ and $\mathcal{W}_2$, respectively, the following holds for any $k \in \lbrace 1, 2, \ldots, N \rbrace$: 
\begin{IEEEeqnarray}{rcl}
\nonumber
NR_i \! & \!=\! & \! H\left(W_i\right)\\
\nonumber
&=& H\left(W_i|W_j\right)\\
\nonumber
&\stackrel{(a)}{\leqslant}& I\left(W_i;\overrightarrow{\bs{Y}}_i, \overleftarrow{\bs{Y}}_j |W_j\right)+N\delta(N)\\
\nonumber
&\leqslant& \sum_{n=1}^{N} \! \Big[ \! h\Big(\overrightarrow{Y}_{i,n}, \! \overleftarrow{Y}_{j,n} |W_j, \! \overrightarrow{\bs{Y}}_{i,(1:n-1)}, \! \overleftarrow{\bs{Y}}_{j,(1:n-1)}, \! \bs{X}_{j,(1:n)}\Big)\\
\nonumber
& & -h\left(\overrightarrow{Z}_{i,n} \right)-h\left(\overleftarrow{Z}_{j,n} \right)\Big]+N\delta(N)\\
\nonumber
&\leqslant& \sum_{n=1}^{N}\Big[h\left(\overrightarrow{Y}_{i,n},\overleftarrow{Y}_{j,n} | X_{j,n}\right)-h\left(\overrightarrow{Z}_{i,n} \right)-h\left(\overleftarrow{Z}_{j,n} \right)\Big]\\
\nonumber
& & +N\delta(N)\\
\label{EqproofRi-1gc1}
&=& N\left[h\left(\overrightarrow{Y}_{i,k},\overleftarrow{Y}_{j,k} | X_{j,k}\right)-\log\left(2 \pi e\right)\right]+N\delta(N),
\end{IEEEeqnarray}
where 
(a) follows from considering enhanced receivers (see Figure \ref{Fig:G-IC-NF-Conv}(a) for the case $i=1$) and Fano's inequality. 

From \eqref{EqproofRi-1gc1}, the following holds in the asymptotic regime: 
\begin{IEEEeqnarray} {rcl}
\label{EqproofRi-1gc2}
\nonumber
R_i&  \leqslant & h\left(\overrightarrow{Y}_{i,k},\overleftarrow{Y}_{j,k} | X_{j,k}\right)-\log\left(2 \pi e\right) \\
\nonumber
&\leqslant& \frac{1}{2}\log\Big(b_{3,i}+1\Big)\\
\label{EqproofRi-1gc3}
& & +\frac{1}{2} \! \log \! \left( \! \frac{\Big(b_{3,i}+b_{4,j}(\rho)+1\Big)\overleftarrow{\SNR}_j}{\Big(b_{1,j}(\rho)+1\Big)\Big(b_{3,i}+\left(1-\rho^2\right)\Big)} \!+ \! 1 \! \right). \quad
\end{IEEEeqnarray}
This completes the proof of \eqref{EqRic-3-G-IC-NF}. 

\textbf{Proof of \eqref{EqR1+R2c-3g-G-IC-NF}}: 

From the assumption that the message indices $W_1$ and $W_2$  are i.i.d. following a uniform distribution over the sets $\mathcal{W}_1$ and $\mathcal{W}_2$ respectively, the following holds for any $k \in \lbrace 1, 2, \ldots, N \rbrace$:
\begin{IEEEeqnarray}{rcl}
\nonumber
N \!&\! \Big(\!& \! R_1+R_2\Big) = H\left(W_1\right)+H\left(W_2\right)\\
\nonumber
&\stackrel{(a)}{ \leqslant }& I\left(W_1;\overrightarrow{\bs{Y}}_1,\overleftarrow{\bs{Y}}_1\right)+I\left(W_2;\overrightarrow{\bs{Y}}_2,\overleftarrow{\bs{Y}}_2\right)+N\delta(N)\\
\nonumber
&=& h\left(\overrightarrow{\bs{Y}}_1\right)+h\left(\overleftarrow{\bs{Z}}_1|\overrightarrow{\bs{Y}}_1\right)-h\left(\overleftarrow{\bs{Y}}_1|W_1\right)\\
\nonumber
& & -h\left(\overrightarrow{\bs{Y}}_1|W_1,\overleftarrow{\bs{Y}}_1, \bs{X}_1\right)+h\left(\overrightarrow{\bs{Y}}_2\right)+h\left(\overleftarrow{\bs{Z}}_2|\overrightarrow{\bs{Y}}_2\right) \\
\nonumber
& & -h\left(\overleftarrow{\bs{Y}}_2|W_2\right)-h\left(\overrightarrow{\bs{Y}}_2|W_2,\overleftarrow{\bs{Y}}_2, \bs{X}_2\right)+N\delta(N) \\
\nonumber
&\leqslant& h\left(\overrightarrow{\bs{Y}}_1\right)+h\left(\overleftarrow{\bs{Z}}_1\right)-h\left(\overleftarrow{\bs{Y}}_1|W_1\right)\\
\nonumber
& & -h\left(\bs{X}_{2,C}|W_1,\overleftarrow{\bs{Y}}_1, \bs{X}_1\right)+h\left(\overrightarrow{\bs{Y}}_2\right)+h\left(\overleftarrow{\bs{Z}}_2\right)\\
\nonumber
& & -h\left(\overleftarrow{\bs{Y}}_2|W_2\right)-h\left(\bs{X}_{1,C}|W_2,\overleftarrow{\bs{Y}}_2, \bs{X}_2\right)+N\delta(N)\\
\nonumber
&=& h\left(\overrightarrow{\bs{Y}}_1\right)-h\left(\overleftarrow{\bs{Y}}_1|W_1\right)-h\left(\bs{X}_{2,C},\overrightarrow{\bs{Z}}_{2} |W_1,\overleftarrow{\bs{Y}}_1, \bs{X}_1\right)\\
\nonumber
& & +h\left(\overrightarrow{\bs{Z}}_{2} |W_1,\overleftarrow{\bs{Y}}_1, \bs{X}_1, \bs{X}_{2,C}\right)+h\left(\overrightarrow{\bs{Y}}_2\right)-h\left(\overleftarrow{\bs{Y}}_2|W_2\right)\\
\nonumber
& & -h\left(\bs{X}_{1,C}, \overrightarrow{\bs{Z}}_{1}|W_2,\overleftarrow{\bs{Y}}_2, \bs{X}_2\right)\\
\nonumber
& & +h\left(\overrightarrow{\bs{Z}}_{1}|W_2,\overleftarrow{\bs{Y}}_2, \bs{X}_2, \bs{X}_{1,C}\right)+N\log\left(2\pi e\right)+N\delta(N)\\
\nonumber
&=& h\left(\overrightarrow{\bs{Y}}_1\right)-h\left(\overleftarrow{\bs{Y}}_1|W_1\right)-h\left(\bs{X}_{2,C},\bs{X}_{1,U} |W_1,\overleftarrow{\bs{Y}}_1, \bs{X}_1\right)\\
\nonumber
& & +h\left(\overrightarrow{\bs{Z}}_{2} |W_1,\overleftarrow{\bs{Y}}_1, \bs{X}_1, \bs{X}_{2,C}\right)+h\left(\overrightarrow{\bs{Y}}_2\right)-h\left(\overleftarrow{\bs{Y}}_2|W_2\right)\\
\nonumber
& & -h\left(\bs{X}_{1,C}, \bs{X}_{2,U}|W_2,\overleftarrow{\bs{Y}}_2, \bs{X}_2\right)\\
\nonumber
& & +h\left(\overrightarrow{\bs{Z}}_{1}|W_2,\overleftarrow{\bs{Y}}_2, \bs{X}_2, \bs{X}_{1,C}\right)+N\log\left(2\pi e\right)+N\delta(N) \\
\nonumber
&=& h\left(\overrightarrow{\bs{Y}}_1\right) \! - \! h\left(\overleftarrow{\bs{Y}}_1|W_1\right) \!+ \! \Big[I\left(\bs{X}_{2,C},\bs{X}_{1,U} ; W_1,\overleftarrow{\bs{Y}}_1\right)\\
\nonumber
& & -h\left(\bs{X}_{2,C},\bs{X}_{1,U}\right)\Big]+h\left(\overrightarrow{\bs{Y}}_2\right)-h\left(\overleftarrow{\bs{Y}}_2|W_2\right)\\
\nonumber
& & +\left[I\left(\bs{X}_{1,C}, \bs{X}_{2,U};W_2,\overleftarrow{\bs{Y}}_2\right)-h\left(\bs{X}_{1,C}, \bs{X}_{2,U}\right)\right]\\
\nonumber
& & +h\left(\overrightarrow{\bs{Z}}_{1}|W_2,\overleftarrow{\bs{Y}}_2, \bs{X}_2, \bs{X}_{1,C}\right)\\
\nonumber
& & +h\left(\overrightarrow{\bs{Z}}_{2} |W_1,\overleftarrow{\bs{Y}}_1, \bs{X}_1, \bs{X}_{2,C}\right)+N\log\left(2\pi e\right)+N\delta(N)\\
\nonumber
&\leqslant& h\left(\overrightarrow{\bs{Y}}_1\right) \! - \! h\left(\overleftarrow{\bs{Y}}_1|W_1\right) \! + \! \Big[I\left(\bs{X}_{2,C},\bs{X}_{1,U} ; W_1,\overleftarrow{\bs{Y}}_1\right)\\
\nonumber
& & -h\left(\bs{X}_{2,C},\bs{X}_{1,U}\right)\Big]+h\left(\overrightarrow{\bs{Y}}_2\right)-h\left(\overleftarrow{\bs{Y}}_2|W_2\right)\\
\nonumber
& & +\left[I\left(\bs{X}_{1,C}, \bs{X}_{2,U};W_2,\overleftarrow{\bs{Y}}_2\right)-h\left(\bs{X}_{1,C}, \bs{X}_{2,U}\right)\right] \\
\nonumber
& & + \!   \Big[h\left(\!\bs{X}_{2,C},\bs{X}_{1,U}| \overrightarrow{\bs{Y}}_{2}\right) \! - \! h\left(\bs{X}_{2,C},\bs{X}_{1,U}| \overrightarrow{\bs{Y}}_{2}, \bs{X}_{1}, \bs{X}_{2}\right) \!  \Big] \\
\nonumber
& &  + \!  \Big[h\left(\!\bs{X}_{1,C},\bs{X}_{2,U}| \overrightarrow{\bs{Y}}_{1}\right)  \! - \!  h\left(\bs{X}_{1,C},\bs{X}_{2,U}| \overrightarrow{\bs{Y}}_{1}, \bs{X}_{2}, \bs{X}_{1}\right) \! \Big]\\
\nonumber
& & +h\left(\overrightarrow{\bs{Z}}_{1}|W_2,\overleftarrow{\bs{Y}}_2, \bs{X}_2, \bs{X}_{1,C}\right) \\
\nonumber
& & +h\left(\overrightarrow{\bs{Z}}_{2} |W_1,\overleftarrow{\bs{Y}}_1, \bs{X}_1, \bs{X}_{2,C}\right)+N\log\left(2\pi e\right)+N\delta(N) \\
\nonumber
&\stackrel{(b)}{=}& h\left(\overrightarrow{\bs{Y}}_1|\bs{X}_{1,C},\bs{X}_{2,U}\right)-h\left(\overleftarrow{\bs{Y}}_1|W_1\right)\\
\nonumber
& & +I\left(\bs{X}_{2,C},\bs{X}_{1,U} ; W_1,\overleftarrow{\bs{Y}}_1\right)+h\left(\overrightarrow{\bs{Y}}_2|\bs{X}_{2,C},\bs{X}_{1,U}\right)\\
\nonumber
& & -h\left(\overleftarrow{\bs{Y}}_2|W_2\right)+I\left(\bs{X}_{1,C}, \bs{X}_{2,U};W_2,\overleftarrow{\bs{Y}}_2\right)\\
\nonumber
& & -h\left(\overrightarrow{\bs{Z}}_1,\overrightarrow{\bs{Z}}_2| \overrightarrow{\bs{Y}}_{2}, \bs{X}_{1}, \bs{X}_{2}\right)\!-\!h\left(\overrightarrow{\bs{Z}}_2,\overrightarrow{\bs{Z}}_1| \overrightarrow{\bs{Y}}_{1}, \bs{X}_{2}, \bs{X}_{1}\right)\\
\nonumber
& & +h\left(\overrightarrow{\bs{Z}}_{1}|W_2,\overleftarrow{\bs{Y}}_2, \bs{X}_2, \bs{X}_{1,C}\right)\\
\nonumber
& & +h\left(\overrightarrow{\bs{Z}}_{2} |W_1,\overleftarrow{\bs{Y}}_1, \bs{X}_1, \bs{X}_{2,C}\right)\\
\nonumber
& & +N\log\left(2\pi e\right)+N\delta(N)
\end{IEEEeqnarray}
\begin{IEEEeqnarray}{rcl}
\nonumber
&\stackrel{(c)}{\leqslant}& h\left(\overrightarrow{\bs{Y}}_1|\bs{X}_{1,C},\bs{X}_{2,U}\right)-h\left(\overleftarrow{\bs{Y}}_1|W_1\right)\\
\nonumber
& & +I\left(\bs{X}_{2,C},\bs{X}_{1,U}; W_1,\overleftarrow{\bs{Y}}_1\right)+h\left(\overrightarrow{\bs{Y}}_2|\bs{X}_{2,C},\bs{X}_{1,U}\right)\\
\nonumber
& & -h\left(\overleftarrow{\bs{Y}}_2|W_2\right)+I\left(\bs{X}_{1,C}, \bs{X}_{2,U};W_2,\overleftarrow{\bs{Y}}_2\right)+N\log\left(2\pi e\right)\\
\nonumber
& & +N\delta(N) \\
\nonumber
&  \leqslant & h\left(\overrightarrow{\bs{Y}}_1|\bs{X}_{1,C},\bs{X}_{2,U}\right)-h\left(\overleftarrow{\bs{Y}}_1|W_1\right)\\
\nonumber
& & +I\left(\bs{X}_{2,C},\bs{X}_{1,U}, W_2,\overleftarrow{\bs{Y}}_2; W_1,\overleftarrow{\bs{Y}}_1\right)\!\\
\nonumber
& & +\!h\left(\overrightarrow{\bs{Y}}_2|\bs{X}_{2,C},\bs{X}_{1,U}\right)-h\left(\overleftarrow{\bs{Y}}_2|W_2\right)\\
\nonumber
& & +I\left(\bs{X}_{1,C}, \bs{X}_{2,U}, W_1,\overleftarrow{\bs{Y}}_1;W_2,\overleftarrow{\bs{Y}}_2\right)+N\log\left(2\pi e\right)\\
\nonumber
& & +N\delta(N) \\
\nonumber
&\stackrel{(d)}{\leqslant}& \sum_{n=1}^{N}\Big[h\left(\overrightarrow{Y}_{1,n}|\bs{X}_{1,C},\bs{X}_{2,U}, \overrightarrow{\bs{Y}}_{1,(1:n-1)}\right)\\
\nonumber
& & +h\left(X_{1,U,n}|X_{2,C,n} \right)+h\left(\overleftarrow{Y}_{2,n}|X_{2,n},X_{1,U,n}\right)\!\\
\nonumber
& & +\!h\left(\overrightarrow{Y}_{2,n}|\bs{X}_{2,C},\bs{X}_{1,U} \overrightarrow{\bs{Y}}_{2,(1:n-1)}\right)+h\left(X_{2,U,n}|X_{1,C,n} \right)\\
\nonumber
& & +h\left(\overleftarrow{Y}_{1,n}|X_{1,n},X_{2,U,n}\right)-3\log\left(2\pi e\right)\Big]+N\log\left(2\pi e\right)\\
\nonumber
& & +N\delta(N)\\
\nonumber
&\leqslant& \sum_{n=1}^{N}\Big[h\left(\overrightarrow{Y}_{1,n}|X_{1,C,n},X_{2,U,n}\right)+h\left(X_{1,U,n}|X_{2,C,n} \right)\\
\nonumber
& & +h\left(\overleftarrow{Y}_{2,n}|X_{2,n},X_{1,U,n}\right)+h\left(\overrightarrow{Y}_{2,n}|X_{2,C,n},X_{1,U,n}\right)\\
\nonumber
& & +h\left(X_{2,U,n}|X_{1,C,n} \right)+h\left(\overleftarrow{Y}_{1,n}|X_{1,n},X_{2,U,n}\right)\\
\nonumber
& & -3\log\left(2\pi e\right)\Big]+N\log\left(2\pi e\right)+N\delta(N)\\
\nonumber
&=& N\Big[h\left(\overrightarrow{Y}_{1,k}|X_{1,C,k},X_{2,U,k}\right)+h\left(X_{1,U,k}|X_{2,C,k} \right)\\
\nonumber
& & +h\left(\overleftarrow{Y}_{2,k}|X_{2,k},X_{1,U,k}\right)+h\left(\overrightarrow{Y}_{2,k}|X_{2,C,k},X_{1,U,k}\right)\\
\nonumber
& & +h\left(X_{2,U,k}|X_{1,C,k} \right)+h\left(\overleftarrow{Y}_{1,k}|X_{1,k},X_{2,U,k}\right)\\
\label{EqproofR1+R2gc1}
& & -3\log\left(2 \pi e\right)\Big]+N\log\left(2\pi e\right)+N\delta(N), 
\end{IEEEeqnarray}
where
(a) follows from considering enhanced receivers (see Figure \ref{Fig:G-IC-NF-Conv}(b)) and Fano's inequality; 
(b) follows from the fact that $h\left(\overrightarrow{\bs{Y}}_i\right)-h\left(\bs{X}_{i,C},\bs{X}_{j,U} \right)+h\left(\bs{X}_{i,C},\bs{X}_{j,U}|\overrightarrow{\bs{Y}}_i \right)=h\left(\overrightarrow{\bs{Y}}_i|\bs{X}_{i,C},\bs{X}_{j,U}\right)$; 
(c) follows from the fact that $h\left(\overrightarrow{\bs{Z}}_{i}|W_j,\overleftarrow{\bs{Y}}_j, \bs{X}_j, \bs{X}_{i,C}\right) -h\left(\overrightarrow{\bs{Z}}_i,\overrightarrow{\bs{Z}}_j| \overrightarrow{\bs{Y}}_{j}, \bs{X}_{i}, \bs{X}_{j}\right) \leqslant 0$; and 
(d) follows from Lemma~\ref{Lemmahelpsumandweightedrates} at the end of this appendix.

From \eqref{EqproofR1+R2gc1},  the following holds in the asymptotic regime for any $k \in \lbrace 1, 2, \ldots, N \rbrace$:  
\begin{IEEEeqnarray} {rcl}
\nonumber
\! R_1 \!+ \!R_2  &   \leqslant  & h\left(\overrightarrow{Y}_{1,k}|X_{1,C,k},X_{2,U,k}\right)+h\left(X_{1,U,k}|X_{2,C,k} \right)\\
\nonumber
& & +h\left(\overleftarrow{Y}_{2,k}|X_{2,k},X_{1,U,k}\right)+h\left(\overrightarrow{Y}_{2,k}|X_{2,C,k},X_{1,U,k}\right)\\
\nonumber
& & +h\left(X_{2,U,k}|X_{1,C,k} \right)+h\left(\overleftarrow{Y}_{1,k}|X_{1,k},X_{2,U,k}\right)\\
\nonumber
& & -2\log\left(2 \pi e\right)\\
\nonumber
&\leqslant&\frac{1}{2}\log\left(\det \left( \textrm{Var}\left( \overrightarrow{Y}_{1,k},X_{1,C,k},X_{2,U,k} \right) \right) \right)\\
\nonumber
& & -\frac{1}{2} \log \left( \INR_{12}+1\right) -\frac{1}{2}\log\left(\INR_{21}+1\right)
\end{IEEEeqnarray}
\begin{IEEEeqnarray} {rcl}
\nonumber
& & +\frac{1}{2}\log\left(\det\left(\textrm{Var}\left(\overleftarrow{Y}_{2,k},X_{2,k},X_{1,U,k}\right)\right)\right)\\
\nonumber
& & -\frac{1}{2}\log\left(\det\left(\textrm{Var}\left(X_{2,k},X_{1,U,k}\right)\right)\right)\\
\nonumber
& & +\frac{1}{2}\log\left(\det\left(\textrm{Var}\left(\overrightarrow{Y}_{2,k},X_{2,C,k},X_{1,U,k}\right)\right)\right)\\
\nonumber
& & +\frac{1}{2}\log\left(\det\left(\textrm{Var}\left(\overleftarrow{Y}_{1,k},X_{1,k},X_{2,U,k}\right)\right)\right)\\
\label{EqproofR1+R2gc3}
& & -\frac{1}{2}\log\left(\det\left(\textrm{Var}\left(X_{1,k},X_{2,U,k}\right)\right)\right)+\log\left(2\pi e\right),
\end{IEEEeqnarray}
where, for all $i \in \lbrace 1, 2 \rbrace$, with $j \in {\lbrace 1, 2 \rbrace\setminus\lbrace i \rbrace}$ the following holds for any $k \in \lbrace 1, 2, \ldots, N \rbrace$:
\begin{subequations}
\label{EqVarAux1}
\begin{IEEEeqnarray} {lcl}
\nonumber
 \det \Big(& & \textrm{Var}\left(\overrightarrow{Y}_{j,k},X_{j,C,k},X_{i,U,k}\right)\Big)  =  \overrightarrow{\SNR}_{j}+\INR_{ji}+h_{ji,U}^2\\
\nonumber
& & -2h_{ji,U}\sqrt{\INR_{ji}}+\left(1-\rho^2\right)\Bigg(\INR_{ij}\INR_{ji}\\
\nonumber
& & +h_{ji,U}^2\left(\overrightarrow{\SNR}_{j}+\INR_{ij} \right)-2h_{ji,U}\INR_{ij}\sqrt{\INR_{ji}} \Bigg)\\
\label{EqvarYjXjcgXiug}
& & +2\rho\sqrt{\overrightarrow{\SNR}_{j}}\left(\sqrt{\INR_{ji}}-h_{ji,U}\right),\\
\nonumber
\det \Big(& & \textrm{Var}\left(\overleftarrow{Y}_{j,k},X_{j,k},X_{i,U,k}\right)\Big) = 1+h_{ji,U}^2\left(1-\rho^2\right)\\
\nonumber
& & +\frac{\overleftarrow{\SNR}_j\left(1-\rho^2\right)\left(h_{ji,U}^2-2h_{ji,U}\sqrt{\INR_{ji}}+\INR_{ji} \right )}{\left(\overrightarrow{\SNR}_j + 2\rho\sqrt{\overrightarrow{\SNR}_j\INR_{ji}}+\INR_{ji}+1\right)}, \\
\label{EqvarYjXjXiug} \\
\nonumber
\mbox{and } \\
\label{EqvarXjXiug}
\det \Big( & & \textrm{Var}\left(X_{j,k},X_{i,U,k}\right)\Big) = 1+\left(1-\rho^2 \right)h_{ji,U}^2. 
\end{IEEEeqnarray}
\end{subequations}
The expressions in \eqref{EqVarAux1} depend on $S_{1,i}$, $S_{2,i}$, $S_{3,i}$, $S_{4,i}$, and $S_{5,i}$ via the parameter $h_{ji,U}$ in  \eqref{Eqdefhjiu}.
Hence, the following cases are identified:  \newline
\textbf{Case 1:  $(S_{1,2} \lor S_{2,2} \lor S_{5,2})\land(S_{1,1} \lor S_{2,1} \lor S_{5,1})$}.  From \eqref{Eqdefhjiu}, it follows that $h_{12,U}=0$ and $h_{21,U}=0$. Therefore, plugging \eqref{EqVarAux1} into \eqref{EqproofR1+R2gc3} yields \eqref{Eqconv61}. \newline
\textbf{Case 2: $(S_{1,2} \lor S_{2,2} \lor S_{5,2})\land(S_{3,1} \lor S_{4,1})$}.  From \eqref{Eqdefhjiu}, it follows that $h_{12,U}=0$ and ${h_{21,U}= \sqrt{\frac{\INR_{12}\INR_{21}}{\overrightarrow{\SNR}_{2}}}}$. Therefore, plugging \eqref{EqVarAux1} into \eqref{EqproofR1+R2gc3} yields \eqref{Eqconv62}. \newline
\textbf{Case 3: $(S_{3,2} \lor S_{4,2})\land(S_{1,1} \lor S_{2,1} \lor S_{5,1})$}.  From \eqref{Eqdefhjiu}, it follows that $h_{12,U}=\sqrt{\frac{\INR_{12}\INR_{21}}{\overrightarrow{\SNR}_{1}}}$ and $h_{21,U}=0$. Therefore, plugging \eqref{EqVarAux1} into \eqref{EqproofR1+R2gc3} yields \eqref{Eqconv63}. \newline
\textbf {Case 4: $(S_{3,2} \lor S_{4,2})\land(S_{3,1} \lor S_{4,1})$}. From \eqref{Eqdefhjiu}, it follows that $h_{12,U}= \sqrt{\frac{\INR_{12}\INR_{21}}{\overrightarrow{\SNR}_{1}}}$ and $h_{21,U}=\sqrt{\frac{\INR_{12}\INR_{21}}{\overrightarrow{\SNR}_{2}}}$. Therefore, plugging \eqref{EqVarAux1} into \eqref{EqproofR1+R2gc3} yields \eqref{Eqconv64}. \newline
This completes the proof of \eqref{EqR1+R2c-3g-G-IC-NF}. 

\textbf{Proof of \eqref{Eq2Ri+Rjc-g-G-IC-NF}}:  From the assumption that the message indices $W_1$ and $W_2$  are i.i.d. following a uniform distribution over the sets $\mathcal{W}_1$ and $\mathcal{W}_2$, respectively, for all $i \in \lbrace1,2\rbrace$, with $j \in \lbrace1,2 \rbrace \setminus\lbrace i \rbrace$, the following holds for any $k \in \lbrace 1, 2, \ldots, N \rbrace$:
\begin{IEEEeqnarray}{rcl}
\nonumber
&N \Big(&2 R_i+R_j\Big) = 2H\left(W_i\right)+H\left(W_j\right)\\
\nonumber
&\stackrel{(a)}{ = }& H\left(W_i\right)+H\left(W_i|W_j\right)+H\left(W_j\right)\\
\nonumber
&\stackrel{(b)}{\leqslant}& \! I \! \left(W_i;\overrightarrow{\bs{Y}}_i,\overleftarrow{\bs{Y}}_i\right)\! + \! I\left(W_i;\overrightarrow{\bs{Y}}_i,\overleftarrow{\bs{Y}}_j|W_j\right)\! + \! I\left(W_j;\overrightarrow{\bs{Y}}_j,\overleftarrow{\bs{Y}}_j\right)\\
\nonumber
& & +N\delta(N)\\
\nonumber
&\leqslant& h\left(\overrightarrow{\bs{Y}}_i\right)+h\left(\overleftarrow{\bs{Z}}_i\right)-h\left(\overleftarrow{\bs{Y}}_i|W_i\right)-h\left(\overrightarrow{\bs{Y}}_i|W_i,\overleftarrow{\bs{Y}}_i\right)\\
\nonumber
& & +h\left(\overleftarrow{\bs{Y}}_j|W_j\right)-h\left(\overleftarrow{\bs{Y}}_j|W_i,W_j\right)+I\left(W_i;\overrightarrow{\bs{Y}}_i|W_j,\overleftarrow{\bs{Y}}_j\right)\\
\nonumber
& & +h\left(\overrightarrow{\bs{Y}}_j\right)+h\left(\overleftarrow{\bs{Z}}_j\right)-h\left(\overleftarrow{\bs{Y}}_j|W_j\right)-h\left(\overrightarrow{\bs{Y}}_j|W_j,\overleftarrow{\bs{Y}}_j\right)\\
\nonumber
& & +N\delta(N)\\
\nonumber
&=& h\left(\overrightarrow{\bs{Y}}_i\right)-h\left(\overleftarrow{\bs{Y}}_i|W_i\right)-h\left(\overrightarrow{\bs{Y}}_i|W_i,\overleftarrow{\bs{Y}}_i, \bs{X}_i\right)\\
\nonumber
& & -h\left(\overleftarrow{\bs{Y}}_j|W_i,W_j\right)+I\left(W_i;\overrightarrow{\bs{Y}}_i|W_j,\overleftarrow{\bs{Y}}_j\right)+h\left(\overrightarrow{\bs{Y}}_j\right)\\
\nonumber
& & -h\left(\overrightarrow{\bs{Y}}_j|W_j,\overleftarrow{\bs{Y}}_j, \bs{X}_j\right)+N\log\left(2\pi e\right)+N\delta(N) \\
\nonumber
&\leqslant& h\left(\overrightarrow{\bs{Y}}_i\right)-h\left(\overleftarrow{\bs{Y}}_i|W_i\right)-h\left(\overrightarrow{\bs{Y}}_i|W_i,\overleftarrow{\bs{Y}}_i, \bs{X}_i\right)\\
\nonumber
& & +I\left(W_i;\overrightarrow{\bs{Y}}_i|W_j,\overleftarrow{\bs{Y}}_j\right)+h\left(\overrightarrow{\bs{Y}}_j\right)-h\left(\overrightarrow{\bs{Y}}_j|W_j,\overleftarrow{\bs{Y}}_j, \bs{X}_j\right)\\
\nonumber
& & +N\log\left(2\pi e\right)+N\delta(N)\\ 
\nonumber
&\stackrel{(c)}{=}& h\left(\overrightarrow{\bs{Y}}_i\right)-h\left(\overleftarrow{\bs{Y}}_i|W_i\right)-h\left(\bs{X}_{j,C}|W_i,\overleftarrow{\bs{Y}}_i, \bs{X}_i\right)\\
\nonumber
& & \! +I\left(W_i;\overrightarrow{\bs{Y}}_i|W_j,\overleftarrow{\bs{Y}}_j\right)\!+\!h\left(\overrightarrow{\bs{Y}}_j\right)\!-\!h\left(\bs{X}_{i,C}|W_j,\overleftarrow{\bs{Y}}_j, \bs{X}_j\right)\\
\nonumber
& & +N\log\left(2\pi e\right)+N\delta(N)\\
\nonumber
&=& h\left(\overrightarrow{\bs{Y}}_i\right)-h\left(\overleftarrow{\bs{Y}}_i|W_i\right)-h\left(\bs{X}_{j,C}, \overrightarrow{\bs{Z}}_j|W_i,\overleftarrow{\bs{Y}}_i, \bs{X}_i\right) \\
\nonumber
& & +h\left(\overrightarrow{\bs{Z}}_j|W_i,\overleftarrow{\bs{Y}}_i, \bs{X}_i,\bs{X}_{j,C}\right)+I\left(W_i;\overrightarrow{\bs{Y}}_i|W_j,\overleftarrow{\bs{Y}}_j\right)\\
\nonumber
& & +h\left(\overrightarrow{\bs{Y}}_j\right)-h\left(\bs{X}_{i,C}|W_j,\overleftarrow{\bs{Y}}_j, \bs{X}_j\right)+N\log\left(2\pi e\right)\\
\nonumber
& & +N\delta(N)\\
\nonumber
&\stackrel{(d)}{=}& h\left(\overrightarrow{\bs{Y}}_i\right)-h\left(\overleftarrow{\bs{Y}}_i|W_i\right)-h\left(\bs{X}_{j,C}, \bs{X}_{i,U}|W_i,\overleftarrow{\bs{Y}}_i, \bs{X}_i\right)\\
\nonumber
& & +h\left(\overrightarrow{\bs{Z}}_j|W_i,\overleftarrow{\bs{Y}}_i, \bs{X}_i,\bs{X}_{j,C}\right)+I\left(W_i;\overrightarrow{\bs{Y}}_i|W_j,\overleftarrow{\bs{Y}}_j\right)\\
\nonumber
& & +h\left(\overrightarrow{\bs{Y}}_j\right)-h\left(\bs{X}_{i,C}|W_j,\overleftarrow{\bs{Y}}_j, \bs{X}_j\right)+N\log\left(2\pi e\right)\\
\nonumber
& & +N\delta(N) \\
\nonumber
&\leqslant& h\left(\overrightarrow{\bs{Y}}_i\right)-h\left(\overleftarrow{\bs{Y}}_i|W_i\right)-h\left(\bs{X}_{j,C}, \bs{X}_{i,U}|W_i,\overleftarrow{\bs{Y}}_i\right)\\
\nonumber
& & \!+\!h\left(\overrightarrow{\bs{Z}}_j|W_i,\overleftarrow{\bs{Y}}_i, \bs{X}_i,\bs{X}_{j,C}\right)\!+\!I\left(W_i;\overrightarrow{\bs{Y}}_i, \bs{X}_{i,C}|W_j,\overleftarrow{\bs{Y}}_j\right) \\
\nonumber
& & +h\left(\overrightarrow{\bs{Y}}_j\right)-h\left(\bs{X}_{i,C}|W_j,\overleftarrow{\bs{Y}}_j\right)+N\log\left(2\pi e\right)+N\delta(N) \\
\nonumber
& = & h\left(\overrightarrow{\bs{Y}}_i\right)-h\left(\overleftarrow{\bs{Y}}_i|W_i\right)-h\left(\bs{X}_{j,C}, \bs{X}_{i,U}|W_i,\overleftarrow{\bs{Y}}_i\right)\\
\nonumber
& & +h\left(\overrightarrow{\bs{Z}}_j|W_i,\overleftarrow{\bs{Y}}_i, \bs{X}_i,\bs{X}_{j,C}\right)+h\left(\overrightarrow{\bs{Y}}_i|W_j,\overleftarrow{\bs{Y}}_j, \bs{X}_{i,C}\right)\\
\nonumber
& & -h\left(\overrightarrow{\bs{Y}}_i, \bs{X}_{i,C}|W_i, W_j,\overleftarrow{\bs{Y}}_j\right)+h\left(\overrightarrow{\bs{Y}}_j\right)+N\log\left(2\pi e\right)\\
\nonumber
& & +N\delta(N) \\
\nonumber
&\stackrel{(e)}{\leqslant}& h\left(\overrightarrow{\bs{Y}}_i\right)-h\left(\overleftarrow{\bs{Y}}_i|W_i\right)-h\left(\bs{X}_{j,C}, \bs{X}_{i,U}|W_i,\overleftarrow{\bs{Y}}_i\right)\\
\nonumber
& & +h\left(\overrightarrow{\bs{Z}}_j|W_i,\overleftarrow{\bs{Y}}_i, \bs{X}_i,\bs{X}_{j,C}\right)+h\left(\overrightarrow{\bs{Y}}_i|W_j,\overleftarrow{\bs{Y}}_j, \bs{X}_{i,C}\right)\\
\nonumber
& & -h\left(\overrightarrow{\bs{Y}}_i, \bs{X}_{i,C}|W_i, W_j,\overleftarrow{\bs{Y}}_j, \bs{X}_{i}, \bs{X}_{j}\right)+h\left(\overrightarrow{\bs{Y}}_j\right)\\
\nonumber
& & +N\log\left(2\pi e\right)+N\delta(N) \\
\nonumber
&=& h\left(\overrightarrow{\bs{Y}}_i\right)-h\left(\overleftarrow{\bs{Y}}_i|W_i\right)-h\left(\bs{X}_{j,C}, \bs{X}_{i,U}|W_i,\overleftarrow{\bs{Y}}_i\right)\\
\nonumber
& & +h\left(\overrightarrow{\bs{Z}}_j|W_i,\overleftarrow{\bs{Y}}_i, \bs{X}_i,\bs{X}_{j,C}\right)+h\left(\overrightarrow{\bs{Y}}_i|W_j,\overleftarrow{\bs{Y}}_j, \bs{X}_{i,C}\right) 
\end{IEEEeqnarray}
\begin{IEEEeqnarray}{rcl}
\nonumber
& & -h\left(\overrightarrow{\bs{Z}}_i, \overrightarrow{\bs{Z}}_j|W_i, W_j,\overleftarrow{\bs{Y}}_j, \bs{X}_{i}, \bs{X}_{j}\right)+h\left(\overrightarrow{\bs{Y}}_j\right)\\
\nonumber
& & +N\log\left(2\pi e\right)+N\delta(N) \\
\nonumber
&\stackrel{(f)}{\leqslant}& h\left(\overrightarrow{\bs{Y}}_i\right)-h\left(\overleftarrow{\bs{Y}}_i|W_i\right)-h\left(\bs{X}_{j,C}, \bs{X}_{i,U}|W_i,\overleftarrow{\bs{Y}}_i\right)\\
\nonumber
& & +h\left(\overrightarrow{\bs{Y}}_i|W_j,\overleftarrow{\bs{Y}}_j, \bs{X}_{i,C}\right)+h\left(\overrightarrow{\bs{Y}}_j\right)+N\log\left(2\pi e\right)\\
\nonumber
& & +N\delta(N)\\
\nonumber
&\leqslant& h\left(\overrightarrow{\bs{Y}}_i\right)-h\left(\overleftarrow{\bs{Y}}_i|W_i\right)+I\left(\bs{X}_{j,C}, \bs{X}_{i,U};W_i,\overleftarrow{\bs{Y}}_i\right)\\
\nonumber
& & -h\left(\bs{X}_{j,C}, \bs{X}_{i,U}\right)+h\left(\overrightarrow{\bs{Y}}_i|W_j,\overleftarrow{\bs{Y}}_j, \bs{X}_{i,C}\right)+h\left(\overrightarrow{\bs{Y}}_j\right)\\
\nonumber
& & +h\left(\bs{X}_{j,C}, \bs{X}_{i,U}|\overrightarrow{\bs{Y}}_j\right)+N\log\left(2\pi e\right)+N\delta(N) \\
\nonumber
&\stackrel{(g)}{=}& h\left(\overrightarrow{\bs{Y}}_i\right)-h\left(\overleftarrow{\bs{Y}}_i|W_i\right)+I\left(\bs{X}_{j,C}, \bs{X}_{i,U};W_i,\overleftarrow{\bs{Y}}_i\right)\\
\nonumber
& & +h\left(\overrightarrow{\bs{Y}}_i|W_j,\overleftarrow{\bs{Y}}_j, \bs{X}_{i,C}\right)+h\left(\overrightarrow{\bs{Y}}_j|\bs{X}_{j,C}, \bs{X}_{i,U}\right)\\
\nonumber
& & +N\log\left(2\pi e\right)+N\delta(N)\\
\nonumber
&\leqslant& \! h\left(\overrightarrow{\bs{Y}}_i\right)\!-\! h\left(\overleftarrow{\bs{Y}}_i|W_i\right)\!+\! I\left(\bs{X}_{j,C}, \bs{X}_{i,U}, W_j,\overleftarrow{\bs{Y}}_j;W_i,\overleftarrow{\bs{Y}}_i\right)\\
\nonumber
& & +h\left(\overrightarrow{\bs{Y}}_i|W_j,\overleftarrow{\bs{Y}}_j, \bs{X}_{i,C}\right)+h\left(\overrightarrow{\bs{Y}}_j|\bs{X}_{j,C}, \bs{X}_{i,U}\right)\\
\nonumber
& & +N\log\left(2\pi e\right)+N\delta(N)\\
\nonumber
&\stackrel{(h)}{\leqslant}& \! h\left(\overrightarrow{\bs{Y}}_i\right)\!+\!\sum_{n=1}^{N}\Big[h\left(X_{i,U,n}|X_{j,C,n} \right)\!+\!h\left(\overleftarrow{\bs{Y}}_{j,n}|X_{j,n},X_{i,U,n}\right)\\
\nonumber
& & -\frac{3}{2}\log\left(2\pi e\right) \Big]+h\left(\overrightarrow{\bs{Y}}_i|W_j,\overleftarrow{\bs{Y}}_j, \bs{X}_{i,C}\right)\\
\nonumber
& & +h\left(\overrightarrow{\bs{Y}}_j|\bs{X}_{j,C}, \bs{X}_{i,U}\right)+N\log\left(2\pi e\right)+N\delta(N)\\
\nonumber
&\stackrel{(i)}{\leqslant}& \! h\left(\overrightarrow{\bs{Y}}_i\right)\!+\! \sum_{n=1}^{N}\Big[h\left(X_{i,U,n}|X_{j,C,n} \right)\!+\! h\left(\overleftarrow{\bs{Y}}_{j,n}|X_{j,n},X_{i,U,n}\right)\\
\nonumber
& & -\frac{3}{2}\log\left(2\pi e\right) \Big]+h\left(\overrightarrow{\bs{Y}}_i|\bs{X}_{i,C}, \bs{X}_j \right)\\
\nonumber
& & +h\left(\overrightarrow{\bs{Y}}_j|\bs{X}_{j,C}, \bs{X}_{i,U}\right)+N\log\left(2\pi e\right)+N\delta(N) \\
\nonumber
&\leqslant& \! \sum_{n=1}^{N} \! \Big[h\left(\overrightarrow{Y}_{i,n}\right)\! + \! h\left(X_{i,U,n}|X_{j,C,n} \right)\! + \! h\left(\overleftarrow{Y}_{j,n}|X_{j,n},X_{i,U,n}\right)\\
\nonumber
& & -\frac{3}{2}\log\left(2\pi e\right)+h\left(\overrightarrow{Y}_{i,n}|X_{i,C,n}, X_{j,n} \right)\\
\nonumber
& & +h\left(\overrightarrow{Y}_{j,n}|X_{j,C,n}, X_{i,U,n}\right)\Big]+N\log\left(2\pi e\right)+N\delta(N) \\
\nonumber
&=& N\Big[h\left(\overrightarrow{Y}_{i,k}\right)+h\left(X_{i,U,k}|X_{j,C,k} \right)+h\left(\overleftarrow{Y}_{j,k}|X_{j,k},X_{i,U,j}\right)\\
\nonumber
& & -\frac{5}{2}\log\left(2 \pi e\right)+h\left(\overrightarrow{Y}_{i,k}|X_{i,C,k}, X_{j,k} \right)\\
\label{Eqproof2Ri+Rjgc1}
& & +h\left(\overrightarrow{Y}_{j,k}|X_{j,C,k}, X_{i,U,k}\right)+2\log\left(2\pi e\right)+\delta(N)\Big], 
\end{IEEEeqnarray}
where,
(a) follows from the fact that $W_1$ and $W_2$ are mutually independent; 
(b) follows from considering enhanced receivers (see Figure \ref{Fig:G-IC-NF-Conv}(c) for the case $i=1$) and Fano's inequality; 
(c) follows from \eqref{Eqsignalyif} and \eqref{EqsignalXiCG}; 
(d) follows from \eqref{EqsignalXiUG}; 
(e) follows from \eqref{Eqencod} and the fact that conditioning does not increase entropy; 
(f) follows from the fact that $h\left(\overrightarrow{\bs{Z}}_{j}|W_j,\overleftarrow{\bs{Y}}_i, \bs{X}_i, \bs{X}_{j,C}\right)-h\left(\overrightarrow{\bs{Z}}_i,\overrightarrow{\bs{Z}}_j| W_i, W_j, \overleftarrow{\bs{Y}}_{j}, \bs{X}_{i}, \bs{X}_{j}\right) \leqslant 0$;
(g) follows from the fact that $h\left(\overrightarrow{\bs{Y}}_j\right)-h\left(\bs{X}_{j,C},\bs{X}_{i,U} \right)+h\left(\bs{X}_{j,C},\bs{X}_{i,U}|\overrightarrow{\bs{Y}}_j \right)=h\left(\overrightarrow{\bs{Y}}_j|\bs{X}_{j,C},\bs{X}_{i,U}\right)$; 
(h) follows from Lemma~\ref{Lemmahelpsumandweightedrates} at the end of this appendix; and
(i) follows from the fact that conditioning does not increase entropy.

From \eqref{Eqproof2Ri+Rjgc1},  the following holds in the asymptotic regime for any $k \in \lbrace 1, 2, \ldots N \rbrace$:  
\begin{IEEEeqnarray} {rcl}
\nonumber
2R_i\!&+&\!R_j \!   \! \leqslant \!  \! h\left(\!\overrightarrow{Y}_{i,k}\right)\!+\!h\left(\!X_{i,U,k}|X_{j,C,k} \right)\!+\!h\left(\!\overleftarrow{Y}_{j,k}|X_{j,k}, \! X_{i,U,k} \! \right) \! \\
\nonumber
& & + \! h\left(\overrightarrow{Y}_{i,k}|X_{i,C,k}, X_{j,k} \right) \! + \! h\left(\overrightarrow{Y}_{j,k}|X_{j,C,k}, X_{i,U,k}\right)\\
\nonumber
& &  \! - \! \frac{1}{2}\log\left(2 \pi e\right)\\
\nonumber
&\leqslant& \frac{1}{2}\log\left(\overrightarrow{\SNR}_{i}+2\rho\sqrt{\overrightarrow{\SNR}_{i}\INR_{ij}}+\INR_{ij}+1\right)\\
\nonumber
& & \!-\!\frac{1}{2}\log\left(\INR_{ij} \!+ \! 1 \! \right)\!+\!\frac{1}{2}\! \log\! \left(\! \det \! \left(\! \textrm{Var} \! \left(\! \overleftarrow{Y}_{j,k},X_{j,k},X_{i,U,k}\! \right)\! \right)\! \right)\\
\nonumber
& & -\frac{1}{2}\log\left(\det\left(\textrm{Var}\left(X_{j,k},X_{i,U,k}\right)\right)\right)\\
\nonumber
& & +\frac{1}{2}\log\left(1+\left(1-\rho^2\right)\left(\overrightarrow{\SNR}_{i}+\INR_{ji}\right)\right)\\
\nonumber
& & -\frac{1}{2}\log\left(1+\left(1-\rho^2 \right)\INR_{ji}\right)\\
\nonumber
& & \! +  \frac{1}{2} \! \log \! \left( \! \det \! \left( \! \textrm{Var}\left(\overrightarrow{Y}_{j,k},X_{j,C,k}, X_{i,U,k}\right)\right)\right) \! + \! 2\log\left(2 \pi e\right). \\
\label{Eqproof2Ri+Rjgc3}
\end{IEEEeqnarray}

The outer bound  on  \eqref{Eqproof2Ri+Rjgc3} depends on $S_{1,i}$, $S_{2,i}$, $S_{3,i}$, $S_{4,i}$, and $S_{5,i}$ via the parameter $h_{ji,U}$ in  \eqref{Eqdefhjiu}. Hence, as in the previous part, the following cases are identified:

\textbf{Case 1: $(S_{1,i} \lor S_{2,i} \lor S_{5,i})$}. From \eqref{Eqdefhjiu}, it follows that $h_{ji,U}=0$. Therefore, plugging \eqref{EqVarAux1} into \eqref{Eqproof2Ri+Rjgc3} yields \eqref{Eqconv7i1}.

\textbf{Case 2: $(S_{3,i} \lor S_{4,i})$}. From \eqref{Eqdefhjiu}, it follows that $h_{ji,U}=\sqrt{\frac{\INR_{ij}\INR_{ji}}{\overrightarrow{\SNR}_{j}}}$. Therefore, plugging \eqref{EqVarAux1} into \eqref{Eqproof2Ri+Rjgc3} yields \eqref{Eqconv7i2}.

This completes the proof of \eqref{Eq2Ri+Rjc-g-G-IC-NF} and the proof of Theorem~\ref{TheoremC-G-IC-NF}. 

\begin{lemma} \label{Lemmahelpsumandweightedrates} \emph{
For all $i \in \lbrace 1, 2 \rbrace$, with $j \in \lbrace 1, 2 \rbrace\setminus\lbrace i \rbrace$,  the following holds: 
\begin{IEEEeqnarray}{rcl}
\nonumber
 I&\Big(&\bs{X}_{i,C}, \bs{X}_{j,U}, \overleftarrow{\bs{Y}}_i, W_i; \overleftarrow{\bs{Y}}_j, W_j\Big)  \leqslant  h\left(\overleftarrow{\bs{Y}}_j| W_j \right) \\
 \nonumber
 & & +\sum_{n=1}^{N}\Big[h\left(X_{j,U,n}|X_{i,C,n} \right)+h\left(\overleftarrow{Y}_{i,n}|X_{i,n},X_{j,U,n}\right)\\
 \label{Eqhelpc1}
 & & -\frac{3}{2}\log\left(2\pi e\right) \Big]. 
\end{IEEEeqnarray}
}
\end{lemma}

\begin{IEEEproof}
Lemma~\ref{Lemmahelpsumandweightedrates} is proved as follows:
\begin{IEEEeqnarray}{rcl}
\nonumber
& & I\left(\bs{X}_{i,C}, \bs{X}_{j,U}, \overleftarrow{\bs{Y}}_i, W_i; \overleftarrow{\bs{Y}}_j, W_j\right)  \\
\nonumber
&  = & I\left( W_i; \overleftarrow{\bs{Y}}_j, W_j\right) + I\left(\bs{X}_{i,C}, \bs{X}_{j,U}, \overleftarrow{\bs{Y}}_i; \overleftarrow{\bs{Y}}_j, W_j| W_i\right) \\
\nonumber
&=& h\left(\overleftarrow{\bs{Y}}_j, W_j\right)-h\left(\overleftarrow{\bs{Y}}_j, W_j| W_i\right) + h\left(\bs{X}_{i,C}, \bs{X}_{j,U}, \overleftarrow{\bs{Y}}_i| W_i\right)\\
\nonumber
& & -h\left(\bs{X}_{i,C}, \bs{X}_{j,U}, \overleftarrow{\bs{Y}}_i | W_i, W_j, \overleftarrow{\bs{Y}}_j\right) \\
\nonumber
&=& h\left(\overleftarrow{\bs{Y}}_j|W_j\right)-h\left(\overleftarrow{\bs{Y}}_j| W_i, W_j\right) + h\left(\bs{X}_{i,C}, \bs{X}_{j,U}, \overleftarrow{\bs{Y}}_i| W_i\right)\\
\nonumber
& & -h\left(\bs{X}_{i,C}, \bs{X}_{j,U}, \overleftarrow{\bs{Y}}_i | W_i, W_j, \overleftarrow{\bs{Y}}_j \right)\\
\nonumber
&=& h\left(\overleftarrow{\bs{Y}}_j|W_j\right) + h\left(\bs{X}_{i,C}, \bs{X}_{j,U}, \overleftarrow{\bs{Y}}_i| W_i\right)\\
\nonumber
& & -h\left(\bs{X}_{i,C}, \bs{X}_{j,U}, \overleftarrow{\bs{Y}}_i, \overleftarrow{\bs{Y}}_j | W_i, W_j\right) \\
\nonumber
&=& \!h \!\left(\! \overleftarrow{\bs{Y}}_j|W_j \! \right) \! + \!  \sum_{n=1}^{N} \! \Bigg[ \! h\Big( \! X_{i,C,n}, X_{j,U,n}, \overleftarrow{Y}_{i,n}| W_i, \bs{X}_{i,C,(\!1:n-1)}, 
\end{IEEEeqnarray}
\begin{IEEEeqnarray}{rcl} 
\nonumber
\nonumber
& & \bs{X}_{j,U,(1:n-1)}, \overleftarrow{\bs{Y}}_{i,(1:n-1)} \! , \bs{X}_{i,(1:n)}\Big)\\
\nonumber
& & -h\Big(X_{i,C,n}, X_{j,U,n}, \overleftarrow{Y}_{i,n}, \overleftarrow{Y}_{j,n} | W_i, W_j,  \bs{X}_{i,C,(1:n-1)},  \\
\nonumber
& & \bs{X}_{j,U,(1:n-1)}, \overleftarrow{\bs{Y}}_{i,(1:n-1)}, \overleftarrow{\bs{Y}}_{j,(1:n-1)}, \bs{X}_{i,(1:n)}, \bs{X}_{j,(1:n)}  \Big)\Bigg]\\
\nonumber
&\leqslant& h\left(\overleftarrow{\bs{Y}}_j|W_j\right) + \sum_{n=1}^{N}\Bigg[h\Big(X_{i,C,n}, X_{j,U,n}, \overleftarrow{Y}_{i,n}| X_{i,n}\Big)\\
\nonumber
& &  -h\Big(\overrightarrow{Z}_{j,n}, \overrightarrow{Z}_{i,n}, \overleftarrow{Y}_{i,n}, \overleftarrow{Y}_{j,n} | W_i, W_j,  \bs{X}_{i,C,(1:n-1)}, \\
\nonumber
& & \bs{X}_{j,U,(1:n-1)}, \overleftarrow{\bs{Y}}_{i,(1:n-1)}, \overleftarrow{\bs{Y}}_{j,(1:n-1)}, \bs{X}_{i,(1:n)}, \bs{X}_{j,(1:n)}  \Big)\Bigg] \\
\nonumber
&  = & \! h\left(\overleftarrow{\bs{Y}}_j|W_j\right) \! + \!  \sum_{n=1}^{N}\Bigg[h\Big(X_{i,C,n}| X_{i,n}\Big)\\
\nonumber
& & \!+\!h\Big(\!X_{j,U,n}| X_{i,n}, X_{i,C,n} \! \Big)\! + \!h\Big(\!\overleftarrow{Y}_{i,n}| X_{i,n}, X_{i,C,n}, X_{j,U,n} \! \Big)\\
\nonumber
& & -h\left(\overrightarrow{Z}_{j,n}\right)-h\left(\overrightarrow{Z}_{i,n}\right)-h\Big(\overleftarrow{Y}_{i,n}, \overleftarrow{Y}_{j,n} | W_i, W_j, \\
\nonumber
& & \bs{X}_{i,C,(1:n-1)}, \bs{X}_{j,U,(1:n-1)}, \overleftarrow{\bs{Y}}_{i,(1:n-1)}, \overleftarrow{\bs{Y}}_{j,(1:n-1)}, \\
\nonumber
& & \bs{X}_{i,(1:n)}, \bs{X}_{j,(1:n)}, \overrightarrow{Z}_{j,n}, \overrightarrow{Z}_{i,n}  \Big)\Bigg] \\
\nonumber
&\leqslant& h\left(\overleftarrow{\bs{Y}}_j|W_j\right) + \sum_{n=1}^{N}\Bigg[h\Big(\overrightarrow{Z}_{j,n}| X_{i,n}\Big)+h\Big(X_{j,U,n}| X_{i,C,n} \Big)\\
\nonumber
& & +h\Big(\overleftarrow{Y}_{i,n}| X_{i,n}, X_{j,U,n} \Big)-h\left(\overrightarrow{Z}_{j,n}\right)-h\left(\overrightarrow{Z}_{i,n}\right) \\
\nonumber
& &  -h\Big(\overleftarrow{Z}_{i,n}, \overleftarrow{Z}_{j,n} | W_i, W_j, \bs{X}_{i,C,(1:n-1)}, \bs{X}_{j,U,(1:n-1)},  \\
\nonumber
& & \overleftarrow{\bs{Y}}_{i,(1:n-1)}, \overleftarrow{\bs{Y}}_{j,(1:n-1)},  \bs{X}_{i,(1:n)}, \bs{X}_{j,(1:n)}, \overrightarrow{Z}_{j,n}, \overrightarrow{Z}_{i,n}  \Big)\Bigg] \\
\nonumber
&\stackrel{(a)}{=}&  h  \left( \overleftarrow{\bs{Y}}_j|W_j \right) + \sum_{n=1}^{N} \Bigg[ h  \Big(X_{j,U,n}| X_{i,C,n}  \Big)  \\
\nonumber
& & +  h  \Big( \overleftarrow{Y}_{i,n}| X_{i,n}, X_{j,U,n}  \Big) -h\left(\overrightarrow{Z}_{i,n}\right)-h\Big(\overleftarrow{Z}_{i,n}\Big)\\
\nonumber
& & -h\Big(\overleftarrow{Z}_{j,n}\Big)\Bigg] \\
\nonumber
&=& h\left(\overleftarrow{\bs{Y}}_j|W_j\right) + \sum_{n=1}^{N}\Bigg[h\left(X_{j,U,n}| X_{i,C,n} \right)\\
\nonumber
& & +h\left(\overleftarrow{Y}_{i,n}| X_{i,n}, X_{j,U,n} \right)-\frac{3}{2}\log\left(2\pi e\right)\Bigg],
\end{IEEEeqnarray}
where
(a) follows from the fact that $\overleftarrow{Z}_{i,n}$ and $\overleftarrow{Z}_{j,n}$ are independent of $W_i$, $W_j$, $\bs{X}_{i,C,(1:n-1)}$, $\bs{X}_{j,U,(1:n-1)}$, $\overleftarrow{\bs{Y}}_{i,(1:n-1)}$, $\overleftarrow{\bs{Y}}_{j,(1:n-1)}$,  $\bs{X}_{i,(1:n)}$, $\bs{X}_{j,(1:n)}$, $\overrightarrow{Z}_{j,n}$, and $\overrightarrow{Z}_{i,n}$.
This completes the proof of  Lemma~\ref{Lemmahelpsumandweightedrates}.  
\end{IEEEproof}

\section{Proof of Theorem~\ref{TheoremGAP-G-IC-NF}} \label{AppG-Gap}
The  gap between the sets $\cgicnof$ and $\agicnof$ (Definition~\ref{DefGap}), denoted by $\delta$,  is approximated as follows: 
\begin{IEEEeqnarray} {rcl}
\label{Eqdelta}
\delta& = & \max\left(\delta_{R_1}^{*},\delta_{R_2}^{*},\frac{\delta_{2R}^{*}}{2},\frac{\delta_{3R_1}^{*}}{3}, \frac{\delta_{3R_2}^{*}}{3}\right),
\end{IEEEeqnarray}      
where, 
\begin{subequations}
\label{EqDeltasStar}
\begin{IEEEeqnarray}{rcl}
\label{EqdeltaR1star}
\delta_{R_1}^{*} & = & \sup_{\rho' \in [0,1]} \left \lbrace \delta_{R_1}\left(\rho', \rho, \mu_1 \right) \right \rbrace, \\
\label{EqdeltaR2star}
\delta_{R_2}^{*} & = & \sup_{\rho' \in [0,1]} \left \lbrace \delta_{R_2}\left(\rho', \rho, \mu_2 \right) \right \rbrace, \\
\label{Eqdelta2Rstar}
\delta_{2R}^{*} & = & \sup_{\rho' \in [0,1]}  \left \lbrace \delta_{2R}  \left(\rho', \rho, \mu_1, \mu_2 \right) \right \rbrace,\\
\label{Eqdelta3R1star}
\delta_{3R_1}^{*} & = & \sup_{\rho' \in [0,1]} \left \lbrace \delta_{3R_1} \left(\rho', \rho, \mu_1, \mu_2 \right) \right \rbrace,\\
\label{Eqdelta3R2star}
\delta_{3R_2}^{*} & = & \sup_{\rho' \in [0,1]} \left \lbrace \delta_{3R_2} \left(\rho', \rho, \mu_1, \mu_2 \right) \right \rbrace,
\end{IEEEeqnarray}
\end{subequations}
for some $\rho, \mu_1, \mu_2$ chosen arbitrarily, with
\begin{subequations}
\label{EqDeltas}
\begin{IEEEeqnarray} {rcl}
\nonumber
\delta_{R_1}&\Big(&\rho', \rho, \mu_1 \Big) = \min\Big(\kappa_{1,1}(\rho'),\kappa_{2,1}(\rho'),\kappa_{3,1}(\rho')\Big)\\
\nonumber
& & -\min\Big(a_{2,1}(\rho), \! a_{6,1}(\rho,\mu_1) \! + \!  a_{3,2}(\rho,\mu_1), \\
\label{EqdeltaR1}
& & a_{1,1} \! + \! a_{3,2}(\rho,\mu_1) \! + \! a_{4,2}(\rho,\mu_1)\Big), \\
\nonumber
\delta_{R_2}&\Big(&\rho', \rho, \mu_2 \Big) =\min\Big(\kappa_{1,2}(\rho'),\kappa_{2,2}(\rho'),\kappa_{3,2}(\rho')\Big)\\
\nonumber
& & -\min\Big(a_{2,2}(\rho), \!  a_{3,1}(\rho,\mu_2) \! + \!  a_{6,2}(\rho,\mu_2), \\
\label{EqdeltaR2}
& & a_{3,1}(\rho,\mu_2) \! + \! a_{4,1}(\rho,\mu_2) \! + \! a_{1,2}\Big), \\
\nonumber
\delta_{2R}&\Big(&\rho', \rho, \mu_1, \mu_2 \Big)=\min\Big(\kappa_{4}(\rho'),\kappa_{5}(\rho'),\kappa_{6}(\rho')\Big)\\
\nonumber
& & -\min\Big(a_{2,1}(\rho)+a_{1,2}, a_{1,1}+a_{2,2}(\rho), \\
\nonumber
& & a_{3,1}(\rho,\mu_2)+a_{1,1}+a_{3,2}(\rho,\mu_1)+a_{7,2}(\rho,\mu_1,\mu_2), \\
\nonumber
& & a_{3,1}(\rho,\mu_2)+a_{5,1}(\rho,\mu_2)+a_{3,2}(\rho,\mu_1)+a_{5,2}(\rho,\mu_1),  \\
\nonumber
&  & a_{3,1}(\rho,\mu_2)+a_{7,1}(\rho,\mu_1,\mu_2)+a_{3,2}(\rho,\mu_1)+a_{1,2} \Big), \\
\label{Eqdelta2R} \\
\nonumber
\delta_{3R_1}&\Big(&\rho', \rho, \mu_1, \mu_2 \Big)=\kappa_{7,1}(\rho')\\
\nonumber
& & -\min\Big(a_{2,1}(\rho)+a_{1,1}+a_{3,2}(\rho,\mu_1)+a_{7,2}(\rho,\mu_1,\mu_2), \\
\nonumber
& & a_{3,1}(\rho,\mu_2)+a_{1,1}+a_{7,1}(\rho,\mu_1,\mu_2)+2a_{3,2}(\rho,\mu_1)\\
\nonumber
& & +a_{5,2}(\rho,\mu_1), a_{2,1}(\rho)+a_{1,1}+a_{3,2}(\rho,\mu_1) \\
\label{Eqdelta3R1} 
& &+a_{5,2}(\rho,\mu_1)\Big),
\end{IEEEeqnarray} 
\begin{IEEEeqnarray} {rcl}
\nonumber
\delta_{3R_2}&\Big(&\rho', \rho, \mu_1, \mu_2 \Big) =  \kappa_{7,2}(\rho')\\
\nonumber
& & -\min\Big(a_{3,1}(\rho,\mu_2)+a_{5,1}(\rho,\mu_2)+a_{2,2}(\rho)+a_{1,2}, \\
\nonumber
& & a_{3,1}(\rho,\mu_2)+a_{7,1}(\rho,\mu_1,\mu_2)+a_{2,2}(\rho)+a_{1,2}, \\
\nonumber
& & 2a_{3,1}(\rho,\mu_2)+a_{5,1}(\rho,\mu_2)+a_{3,2}(\rho,\mu_1)+a_{1,2}\\
\label{Eqdelta3R2}
& & +a_{7,2}(\rho,\mu_1,\mu_2)\Big).
\end{IEEEeqnarray} 
\end{subequations}

Note that $\delta_{R_1}^*$ and $\delta_{R_2}^*$ represent the gap between an achievable individual rate bound (determined by $\rho, \mu_1, \mu_2$) and the active converse individual rate bound; $\delta_{2R}^*$ represents the gap between an achievable sum-rate bound (determined by $\rho, \mu_1, \mu_2$) and the active converse sum-rate bound; $\delta_{3R_1}^*$ and $\delta_{3R_2}^*$ represent the gap between an achievable weighted sum-rate bound (determined by $\rho, \mu_1, \mu_2$) and the active converse weighted sum-rate bound. 

It is important to highlight that, as suggested in \cite{Suh-TIT-2011}, \cite{Etkin-TIT-2008}, and \cite{SyQuoc-TIT-2015}, the gap between $\agicnof$ and $\cgicnof$ can be calculated more precisely. However, the approximation in \eqref{Eqdelta} eases the calculations at the expense of less precision. 
A key point in order to find the gap between the achievable region and the converse region is to choose convenient values of $\rho$, $\mu_1$, and $\mu_2$, such that the optimization problem in \eqref{EqDeltas} becomes an optimization only on $\rho' \in [0,1]$. The following describes all the key cases.
 
\underline{\textbf{Case 1}}: $\INR_{12} > \overrightarrow{\SNR}_1$ and $\INR_{21}>\overrightarrow{\SNR}_2$.  This case corresponds to the scenario in which both transmitter-receiver pairs are in high interference regime (HIR). Three subcases follow considering the SNR in the feedback links.

\underline{Case 1.1}: $\overleftarrow{\SNR}_2 \leqslant \overrightarrow{\SNR}_1$ and $\overleftarrow{\SNR}_1 \leqslant \overrightarrow{\SNR}_2$. The choice is: $\rho=0$, $\mu_1=0$, and $\mu_2=0$.

\underline{Case 1.2}: $\overleftarrow{\SNR}_2 > \overrightarrow{\SNR}_1$ and $\overleftarrow{\SNR}_1 > \overrightarrow{\SNR}_2$. The choice is: $\rho=0$, $\mu_1=1$, and $\mu_2=1$.

\underline{Case 1.3}: $\overleftarrow{\SNR}_2 \leqslant \overrightarrow{\SNR}_1$ and $\overleftarrow{\SNR}_1 > \overrightarrow{\SNR}_2$. The choice is: $\rho=0$, $\mu_1=0$, and $\mu_2=1$.

\underline{\textbf{Case 2}}: $\INR_{12} \leqslant \overrightarrow{\SNR}_1$ and $\INR_{21} \leqslant \overrightarrow{\SNR}_2$. This case corresponds to the scenario in which both transmitter-receiver pairs are in low interference regime (LIR). There are twelve subcases that must be studied separately.

In the following four subcases, the choice is: $\rho=0$, $\mu_1=0$, and $\mu_2=0$.

\underline{Case 2.1}: $\overleftarrow{\SNR}_1 \leqslant  \INR_{21}$, $\overleftarrow{\SNR}_2 \leqslant \INR_{12}$, $\INR_{12}\INR_{21} >  \overrightarrow{\SNR}_1$ and $\INR_{12}\INR_{21} >  \overrightarrow{\SNR}_2$.

\underline{Case 2.2}: $\overleftarrow{\SNR}_1 \leqslant  \INR_{21}$, $\overleftarrow{\SNR}_2\INR_{21} \leqslant \overrightarrow{\SNR}_2$, $\INR_{12}\INR_{21} >  \overrightarrow{\SNR}_1$ and $\INR_{12}\INR_{21} <  \overrightarrow{\SNR}_2$.

\underline{Case 2.3}: $\overleftarrow{\SNR}_1\INR_{12} \leqslant \overrightarrow{\SNR}_1$, $\overleftarrow{\SNR}_2 \leqslant \INR_{12}$, $\INR_{12}\INR_{21} <  \overrightarrow{\SNR}_1$ and $\INR_{12}\INR_{21} >  \overrightarrow{\SNR}_2$.

\underline{Case 2.4}: $\overleftarrow{\SNR}_1\INR_{12} \leqslant \overrightarrow{\SNR}_1$, $\overleftarrow{\SNR}_2\INR_{21} \leqslant \overrightarrow{\SNR}_2$, $\INR_{12}\INR_{21} <  \overrightarrow{\SNR}_1$ and $\INR_{12}\INR_{21} < \overrightarrow{\SNR}_2$.

In the following four subcases, the choice is: \hspace{4mm} $\rho=0$, \hspace{2mm} $\mu_1=\frac{\INR_{21}^2\overleftarrow{\SNR}_2}{\left(\INR_{21}-1\right)\left(\INR_{21}\overleftarrow{\SNR}_2+\overrightarrow{\SNR}_2\right)}$, \hspace{5mm} and \hspace{5mm} ${\mu_2=\frac{\INR_{12}^2\overleftarrow{\SNR}_1}{\left(\INR_{12}-1\right)\left(\INR_{12}\overleftarrow{\SNR}_1+\overrightarrow{\SNR}_1\right)}}$.

\underline{Case 2.5}: $\overleftarrow{\SNR}_1 >  \INR_{21}$, $\overleftarrow{\SNR}_2 > \INR_{12}$, $\INR_{12}\INR_{21} >  \overrightarrow{\SNR}_1$ and $\INR_{12}\INR_{21} >  \overrightarrow{\SNR}_2$.

\underline{Case 2.6}: $\overleftarrow{\SNR}_1 >  \INR_{21}$, $\overleftarrow{\SNR}_2\INR_{21} > \overrightarrow{\SNR}_2$, $\INR_{12}\INR_{21} >  \overrightarrow{\SNR}_1$ and $\INR_{12}\INR_{21} <  \overrightarrow{\SNR}_2$.

\underline{Case 2.7}: $\overleftarrow{\SNR}_1\INR_{12} > \overrightarrow{\SNR}_1$, $\overleftarrow{\SNR}_2 > \INR_{12}$, $\INR_{12}\INR_{21} <  \overrightarrow{\SNR}_1$ and $\INR_{12}\INR_{21} >  \overrightarrow{\SNR}_2$.

\underline{Case 2.8}: $\overleftarrow{\SNR}_1\INR_{12} > \overrightarrow{\SNR}_1$, $\overleftarrow{\SNR}_2\INR_{21} > \overrightarrow{\SNR}_2$, $\INR_{12}\INR_{21} <  \overrightarrow{\SNR}_1$ and $\INR_{12}\INR_{21} < \overrightarrow{\SNR}_2$.

In the following four subcases, the choice is: $\rho=0$, $\mu_1=0$, and $\mu_2=\frac{\INR_{12}^2\overleftarrow{\SNR}_1}{\left(\INR_{12}-1\right)\left(\INR_{12}\overleftarrow{\SNR}_1+\overrightarrow{\SNR}_1\right)}$.

\underline{Case 2.9}: $\overleftarrow{\SNR}_1 > \INR_{21}$, $\overleftarrow{\SNR}_2 \leqslant \INR_{12}$, $\INR_{12}\INR_{21} >  \overrightarrow{\SNR}_1$ and $\INR_{12}\INR_{21} >  \overrightarrow{\SNR}_2$.

\underline{Case 2.10}: $\overleftarrow{\SNR}_1 >  \INR_{21}$, $\overleftarrow{\SNR}_2\INR_{21} \leqslant \overrightarrow{\SNR}_2$, $\INR_{12}\INR_{21} >  \overrightarrow{\SNR}_1$ and $\INR_{12}\INR_{21} <  \overrightarrow{\SNR}_2$.

\underline{Case 2.11}: $\overleftarrow{\SNR}_1\INR_{12} > \overrightarrow{\SNR}_1$, $\overleftarrow{\SNR}_2 \leqslant \INR_{12}$, $\INR_{12}\INR_{21} <  \overrightarrow{\SNR}_1$ and $\INR_{12}\INR_{21} >  \overrightarrow{\SNR}_2$.

\underline{Case 2.12}: $\overleftarrow{\SNR}_1\INR_{12} > \overrightarrow{\SNR}_1$, $\overleftarrow{\SNR}_2\INR_{21} \leqslant \overrightarrow{\SNR}_2$, $\INR_{12}\INR_{21} <  \overrightarrow{\SNR}_1$ and $\INR_{12}\INR_{21} < \overrightarrow{\SNR}_2$.

\textbf{\underline{Case 3}}: $\INR_{12}>\overrightarrow{\SNR}_1$ and $\INR_{21}\leqslant\overrightarrow{\SNR}_2$. This case corresponds to the scenario in which  transmitter-receiver pair $1$ is in HIR and transmitter-receiver pair $2$ is in LIR. There are four subcases that must be studied separately.

In the following two subcases, the choice is: $\rho=0$, $\mu_1=0$, and $\mu_2=0$.

\underline{Case 3.1}: $\overleftarrow{\SNR}_2 \leqslant \INR_{12}$ and $\INR_{12}\INR_{21} >  \overrightarrow{\SNR}_2$.

\underline{Case 3.2}: $\overleftarrow{\SNR}_2\INR_{21} \leqslant  \overrightarrow{\SNR}_2$ and $\INR_{12}\INR_{21} <  \overrightarrow{\SNR}_2$.

In the following two subcases, the choice is: $\rho=0$, $\mu_1=1$, and $\mu_2=0$.

\underline{Case 3.3}: $\overleftarrow{\SNR}_2 > \INR_{12}$ and $\INR_{12}\INR_{21} >  \overrightarrow{\SNR}_2$.

\underline{Case 3.4}: $\overleftarrow{\SNR}_2\INR_{21} >  \overrightarrow{\SNR}_2$ and $\INR_{12}\INR_{21} <  \overrightarrow{\SNR}_2$.

The following is the calculation of the gap $\delta$ in Case~$1.1$. 
\begin{enumerate}

\item \underline {Calculation of $\delta_{R_1}^*$.}

From \eqref{EqdeltaR1} and considering the choice of values: $\rho=0$, $\mu_1=0$ and $\mu_2=0$, it follows that
\begin{IEEEeqnarray} {rcl}
\nonumber
\delta_{R_1}^*&  \leqslant & \sup_{\rho' \in [0,1]} \left \lbrace \min\Big(\kappa_{1,1}(\rho'),\kappa_{2,1}(\rho'),\kappa_{3,1}(\rho')\Big) \right \rbrace \\
\label{EqdeltaR1proofgap}
& & -\min\Big(a_{6,1}(0,0), a_{1,1}+a_{4,2}(0,0)\Big). 
\end{IEEEeqnarray} 
Note that in this case:
\begin{subequations}
\label{EqprofGapcHIRbb}
\begin{IEEEeqnarray}{rcl}
\nonumber
\kappa_{1,1}&(\rho') & =   \frac{1}{2}\log \Big(b_{1,1}(\rho')+1\Big) \\
\nonumber
&\stackrel{(a)}{\leqslant} &   \frac{1}{2}\log \! \left( \! \overrightarrow{\SNR}_{1} \! + \! 2\sqrt{\overrightarrow{\SNR}_{1}\INR_{12}} \! + \! \INR_{12} \! + \! 1 \!\right) \\
\nonumber
&  \stackrel{(b)}{\leqslant} &  \frac{1}{2}\log \left(2\overrightarrow{\SNR}_{1}+2\INR_{12}+1\right) \\
\label{EqprofGapcHIRbb1}
&  \leqslant &  \frac{1}{2}\log \left(\overrightarrow{\SNR}_{1}+\INR_{12}+1\right)+\frac{1}{2}, \\
\nonumber
\kappa_{2,1}&(\rho')  &  =   \frac{1}{2} \log \left(1+b_{4,1}(\rho')+b_{5,2}(\rho')\right) \\
\label{EqprofGapcHIRbb2}
& \leqslant & \frac{1}{2} \log \left(\overrightarrow{\SNR}_{1}+\INR_{21}+1\right), \\
\nonumber
\kappa_{3,1}&(\rho') &  =   \frac{1}{2}\log\Big(b_{4,1}(\rho')+1\Big)\\
\nonumber
& & +\frac{1}{2}\log \! \left(\frac{\overleftarrow{\SNR}_2\left(b_{4,1}(\rho')\!+\!b_{5,2}(\rho')\!+\!1\right)}{\left(b_{1,2}(1) \! + \! 1\right)\left(b_{4,1}(\rho') \! + \! 1\right)} \! + \! 1\right) \\
\nonumber
&\stackrel{(c)}{  \leqslant } & \frac{1}{2}\log\left(\overrightarrow{\SNR}_1+1\right) \\
\nonumber
& & \!+ \! \frac{1}{2} \! \log \! \left( \! \frac{\overleftarrow{\SNR}_2 \! \left(\! \overrightarrow{\SNR}_1\!+\!\INR_{21}\!+\!1\right)}{\left(\overrightarrow{\SNR}_{2} \! +\! \INR_{21} \! + \! 1\right) \! \left( \! \overrightarrow{\SNR}_1 \! + \! 1\right)} \! + \! 1 \! \right)\\
\nonumber
&=&\! \frac{1}{2} \! \log \! \left( \! \frac{\overleftarrow{\SNR}_2 \! \left( \! \overrightarrow{\SNR}_1 \! + \! \INR_{21} \! + \! 1\right)}{\overrightarrow{\SNR}_{2}+\INR_{21}+1} \! + \! \overrightarrow{\SNR}_1 \! + \!1 \! \right) \! , \\
\label{EqprofGapcHIRbb3}
\end{IEEEeqnarray}
\end{subequations}
where
(a) follows from the fact that $0\leqslant\rho'\leqslant1$; 
(b) follows from the fact that 
\begin{equation}
\label{Eqgapb}
\left(\sqrt{\overrightarrow{\SNR}_{1}}-\sqrt{\INR_{12}}\right)^2 \geqslant 0; 
\end{equation}
and
(c) follows from the fact that $\kappa_{3,1}(\rho')$ is a monotonically decreasing function of $\rho'$. 

Note also that the achievable bound $a_{1,1}+a_{4,2}(0,0)$ is lower bounded as follows:
\begin{IEEEeqnarray}{rcl}
\nonumber
a_{1,1}  &+&  a_{4,2}(0,0)  =  \frac{1}{2}  \log  \left(  \frac{\overrightarrow{\SNR}_{1}}{\INR_{21}}  +  2\right)  \\
\nonumber
& & +  \frac{1}{2}\log\Big(\INR_{21}  +  1  \Big) -  1 \\
\nonumber
&\geqslant& \frac{1}{2}  \log  \left(  \frac{\overrightarrow{\SNR}_{1}}{\INR_{21}}  +  2\right)  +  \frac{1}{2}  \log  \Big(  \INR_{21}  \Big)  -  1 \\
\nonumber
&=& \frac{1}{2}\log\left(\overrightarrow{\SNR}_{1}+2\INR_{21}\right)-1 \\
\nonumber
&=& \frac{1}{2}\log\left(\overrightarrow{\SNR}_{1}+\INR_{21}+\INR_{21}\right)-1 \\
\label{Eqgapv1}
&\geqslant& \frac{1}{2}\log\left(\overrightarrow{\SNR}_{1}+\INR_{21}+1\right)-1.
\end{IEEEeqnarray}
From \eqref{EqdeltaR1proofgap}, \eqref{EqprofGapcHIRbb} and \eqref{Eqgapv1}, assuming that ${a_{1,1}+a_{4,2}(0,0) < a_{6,1}(0,0)}$, it follows that
\begin{IEEEeqnarray} {rcl}
\nonumber
\delta_{R_1}^*&  \leqslant & \sup_{\rho' \in [0,1]} \left \lbrace \min\Big(\kappa_{1,1}(\rho'),\kappa_{2,1}(\rho'),\kappa_{3,1}(\rho')\Big) \right \rbrace  \\
\nonumber
& & -  \Big(  a_{1,1} +  a_{4,2}(0,0)  \Big) \\
\nonumber
&\leqslant& \sup_{\rho' \in [0,1]} \kappa_{2,1}(\rho')-\Big(a_{1,1}+a_{4,2}(0,0)\Big)\\
\label{EqdeltaR11}
&\leqslant& 1. 
\end{IEEEeqnarray} 
Now, assuming that $a_{6,1}(0,0) < a_{1,1}+a_{4,2}(0,0)$, the following holds: 
\begin{IEEEeqnarray} {rcl}
\nonumber
\delta_{R_1}^*&  \leqslant & \sup_{\rho' \in [0,1]} \left \lbrace \min\Big(\kappa_{1,1}(\rho'),\kappa_{2,1}(\rho'),\kappa_{3,1}(\rho')\Big) \right \rbrace\\
\label{EqdeltaR12p}
& & -a_{6,1}(0,0). 
\end{IEEEeqnarray} 

To calculate an upper bound for \eqref{EqdeltaR12p}, the following cases are considered: \newline
\underline{Case 1.1.1}: $\overrightarrow{\SNR}_{1} \geqslant \INR_{21} \land \overrightarrow{\SNR}_{2} < \INR_{12}$,\newline
\underline{Case 1.1.2}: $\overrightarrow{\SNR}_{1} < \INR_{21} \land \overrightarrow{\SNR}_{2} \geqslant \INR_{12}$, and \newline
\underline{Case 1.1.3}: $\overrightarrow{\SNR}_{1} < \INR_{21} \land \overrightarrow{\SNR}_{2} < \INR_{12}$. 

In Case~$1.1.1$,  from \eqref{EqprofGapcHIRbb} and \eqref{EqdeltaR12p}, it follows that
\begin{IEEEeqnarray} {rcl}
\nonumber
\delta_{R_1}^*&  \leqslant & \sup_{\rho' \in [0,1]} \kappa_{2,1}(\rho') -a_{6,1}(0,0) \\
\nonumber
&\leqslant& \frac{1}{2} \! \log \! \left(\overrightarrow{\SNR}_{1}\!+\!\INR_{21}\!+\!1\right)\!-\!\frac{1}{2}\log \! \left(\overrightarrow{\SNR}_{1}\!+\!2\right)\\
\nonumber
& & +\frac{1}{2}\\
\nonumber
& \leqslant & \frac{1}{2} \! \log \left(\overrightarrow{\SNR}_{1} \! + \! \overrightarrow{\SNR}_{1}\! + \! 1\right)\!-\!\frac{1}{2}\log\left(\overrightarrow{\SNR}_{1}\!+\!2\right)\\
\nonumber
& & +\frac{1}{2}\\
\label{EqdeltaR12}
&\leqslant& 1. 
\end{IEEEeqnarray} 

In Case~$1.1.2$, from \eqref{EqprofGapcHIRbb} and \eqref{EqdeltaR12p}, it follows that
\begin{IEEEeqnarray} {rcl}
\nonumber
\delta_{R_1}^*&  \! \leqslant \! & \sup_{\rho' \in [0,1]} \kappa_{3,1}(\rho')-a_{6,1}(0,0)\\
\nonumber
& \leqslant & \! \frac{1}{2} \! \log \! \left(\!\frac{\overleftarrow{\SNR}_2 \!\left(\!\overrightarrow{\SNR}_1\!+\!\INR_{21}\!+\!1\right)}{\!\overrightarrow{\SNR}_{2}\!+\!\INR_{21}\!+\!1}\!+\!\overrightarrow{\SNR}_1\!+\!1\right)\\
\nonumber
& & -\frac{1}{2}\!\log \! \left( \! \overrightarrow{\SNR}_{1}\!+\!2 \!\right)\!+\!\frac{1}{2} \\
\nonumber
& \leqslant & \frac{1}{2}\log\left(\overleftarrow{\SNR}_2+\overrightarrow{\SNR}_1+1\right)\\
\nonumber
& & -\frac{1}{2}\log\left(\overrightarrow{\SNR}_{1}+2\right)+\frac{1}{2} \\
\nonumber
& \leqslant & \frac{1}{2}\log\left(\overrightarrow{\SNR}_1+\overrightarrow{\SNR}_1+1\right)\\
\nonumber
& & -\frac{1}{2}\log\left(\overrightarrow{\SNR}_{1}+2\right)+\frac{1}{2}\\
\label{EqdeltaR13}
&\leqslant& 1. 
\end{IEEEeqnarray} 
In Case~$1.1.3$ two additional cases are considered: \newline
\underline{Case 1.1.3.1}: $\overrightarrow{\SNR}_{1} \geqslant \overrightarrow{\SNR}_{2}$ and \newline
\underline{Case 1.1.3.2}: $\overrightarrow{\SNR}_{1} < \overrightarrow{\SNR}_{2}$. \newline
In  Case~$1.1.3.1$, from \eqref{EqprofGapcHIRbb} and \eqref{EqdeltaR12p}, it follows that
\begin{IEEEeqnarray} {rcl}
\nonumber
\delta_{R_1}^*&  \leqslant & \sup_{\rho' \in [0,1]} \kappa_{3,1}(\rho') -a_{6,1}(0,0)\\
\nonumber
& \leqslant & \frac{1}{2}\log\left(\frac{\overleftarrow{\SNR}_2\left(\overrightarrow{\SNR}_1\!+\!\INR_{21}\!+\!1\right)}{\overrightarrow{\SNR}_{2}+\INR_{21}+1}\!+\!\overrightarrow{\SNR}_1\!+\!1\right)\\
\nonumber
& & -\frac{1}{2}\log\left(\overrightarrow{\SNR}_{1}+2\right)+\frac{1}{2} \\
\nonumber
& = & \frac{1}{2}\log\left(\overrightarrow{\SNR}_1+1\right)\\
\nonumber
& & +\frac{1}{2}\log\left(\frac{\overleftarrow{\SNR}_2\left(\overrightarrow{\SNR}_1+\INR_{21}+1\right)}{\left(\overrightarrow{\SNR}_{2}\!+\!\INR_{21}\!+\!1\right)\left(\overrightarrow{\SNR}_1\!+\!1\right)}\!+\!1\right)\\
\nonumber
& & -\frac{1}{2}\log\left(\overrightarrow{\SNR}_{1}+2\right)+\frac{1}{2} \\
\nonumber
& \leqslant & \frac{1}{2}\log\left(\frac{\overrightarrow{\SNR}_1\left(\INR_{21}+\INR_{21}+\INR_{21}\right)}{\INR_{21}\overrightarrow{\SNR}_1}+1\right)\\
\nonumber
& & +\frac{1}{2} \\
\label{EqdeltaR14}
&=& \frac{3}{2}. 
\end{IEEEeqnarray} 

In  Case~$1.1.3.2$, from \eqref{EqprofGapcHIRbb} and \eqref{EqdeltaR12p}, it follows that
\begin{IEEEeqnarray} {rcl}
\nonumber
\delta_{R_1}^*&  \leqslant & \sup_{\rho' \in [0,1]} \kappa_{3,1}(\rho')-a_{6,1}(0,0)\\
\nonumber
& \leqslant & \frac{1}{2}\log\left(\frac{\overleftarrow{\SNR}_2\left(\overrightarrow{\SNR}_1\!+\!\INR_{21}\!+\!1\right)}{\overrightarrow{\SNR}_{2}+\INR_{21}+1}\!+\!\overrightarrow{\SNR}_1\!+\!1\right)\\
\nonumber
& & -\frac{1}{2}\log\left(\overrightarrow{\SNR}_{1}+2\right)+\frac{1}{2} \\
\nonumber
& \leqslant & \frac{1}{2}\log\left(\overleftarrow{\SNR}_2+\overrightarrow{\SNR}_1+1\right)-\frac{1}{2}\log\left(\overrightarrow{\SNR}_{1}+2\right)\\
\nonumber
& & +\frac{1}{2}
\end{IEEEeqnarray}
\begin{IEEEeqnarray}{rcl}
\nonumber
& \leqslant & \frac{1}{2}\log\left(\overrightarrow{\SNR}_1+\overrightarrow{\SNR}_1+1\right)-\frac{1}{2}\log\left(\overrightarrow{\SNR}_{1}+2\right)\\
\nonumber
& & +\frac{1}{2} \\
\label{EqdeltaR15}
& \leqslant & 1. 
\end{IEEEeqnarray} 
Then, from \eqref{EqdeltaR11}, \eqref{EqdeltaR12}, \eqref{EqdeltaR13}, \eqref{EqdeltaR14}, and \eqref{EqdeltaR15}, it follows that in Case~$1.1$: 
\begin{IEEEeqnarray}{rcl}
\label{EqGapR1}
\delta_{R_1}^* &  \leqslant & \frac{3}{2}. 
\end{IEEEeqnarray} 

The same procedure holds to calculate $\delta_{R_2}^*$ and it yields:
\begin{IEEEeqnarray}{rcl}
\label{EqGapR2}
\delta_{R_2}^* &  \leqslant & \frac{3}{2}. 
\end{IEEEeqnarray} 

\item \underline {Calculation of $\delta_{2R}^*$}. From \eqref{Eqdelta2R} and considering the values $\rho=0$, $\mu_1=0$, and $\mu_2=0$,  it follows that
\begin{IEEEeqnarray} {rcl}
\nonumber
\delta_{2R}^*&  \leqslant & \sup_{\rho' \in [0,1]} \left \lbrace \min\Big(\kappa_{4}(\rho'),\kappa_{5}(\rho'),\kappa_{6}(\rho')\Big) \right \rbrace\\
\nonumber
& & -\min\Big(a_{2,1}(0)+a_{1,2}, a_{1,1}+a_{2,2}(0), \\
\nonumber
& & a_{5,1}(0,0)+a_{5,2}(0,0)\Big)\\
\nonumber
&\leqslant&\sup_{\rho' \in [0,1]} \left \lbrace \min\Big(\kappa_{4}(\rho'),\kappa_{5}(\rho') \Big) \right \rbrace\\
\nonumber
& & -\min\Big(a_{2,1}(0)+a_{1,2}, a_{1,1}+a_{2,2}(0), \\
\label{Eqdelta2Rproofgap}
& & a_{5,1}(0,0)+a_{5,2}(0,0)\Big).
\end{IEEEeqnarray} 

Note that 
\begin{subequations}
\label{EqprofGapcHIRsumratebb}
\begin{IEEEeqnarray}{rcl}
\nonumber
\kappa_{4}(\rho')  &  =  & \! \frac{1}{2} \! \log \!  \left( \! 1 \! + \! \frac{b_{4,1}(\rho')}{1+b_{5,2}(\rho')}\right)\!+\!\frac{1}{2}\!\log \! \bigg(\! b_{1,2}(\rho')+1\!\bigg) \\
\nonumber
& \leqslant & \frac{1}{2}\log \left(1+\frac{b_{4,1}(\rho')}{b_{5,2}(\rho')}\right)+\frac{1}{2}\log \bigg(b_{1,2}(\rho')+1\bigg) \\
\nonumber
& = & \frac{1}{2}\log \left(1+\frac{\overrightarrow{\SNR}_{1}}{\INR_{21}}\right)+\frac{1}{2}\log \bigg(b_{1,2}(\rho')+1\bigg) \\
\nonumber
& \stackrel{(h)}{\leqslant} & \frac{1}{2}\log \left(1+\frac{\overrightarrow{\SNR}_{1}}{\INR_{21}}\right)\\
\nonumber
& & +\frac{1}{2}\log \left(2\overrightarrow{\SNR}_{2}+2\INR_{21}+1\right), \\
\nonumber
& \leqslant & \frac{1}{2}\log \left(1+\frac{\overrightarrow{\SNR}_{1}}{\INR_{21}}\right)\\
\nonumber
& & +\frac{1}{2}\log \left(\overrightarrow{\SNR}_{2}+\INR_{21}+1\right)+\frac{1}{2} \\
\nonumber
& \leqslant & \frac{1}{2}\log \left(2+\frac{\overrightarrow{\SNR}_{1}}{\INR_{21}}\right)\\
\label{EqprofGapcHIRsumratebb1}
& & +\frac{1}{2}\log \left(\overrightarrow{\SNR}_{2}+\INR_{21}+1\right)+\frac{1}{2}, 
\end{IEEEeqnarray}
and
\begin{IEEEeqnarray}{rcl}
\nonumber
\kappa_{5}(\rho') &  =  & \frac{1}{2}\log \left(1 \! + \! \frac{b_{4,2}(\rho')}{1+b_{5,1}(\rho')}\right) \! + \! \frac{1}{2}\log \bigg(b_{1,1}(\rho') \! + \! 1\bigg) \\
\nonumber
& \leqslant & \frac{1}{2}\log \left(1+\frac{b_{4,2}(\rho')}{b_{5,1}(\rho')}\right)+\frac{1}{2}\log \bigg(b_{1,1}(\rho')+1\bigg) \\
\nonumber
& = & \frac{1}{2}\log \left(1+\frac{\overrightarrow{\SNR}_{2}}{\INR_{12}}\right)+\frac{1}{2}\log \bigg(b_{1,1}(\rho')+1\bigg) \\
\nonumber
&\stackrel{(i)}{\leqslant}& \frac{1}{2}\log \left(1+\frac{\overrightarrow{\SNR}_{2}}{\INR_{12}}\right)\\
\nonumber
& & +\frac{1}{2}\log \left(2\overrightarrow{\SNR}_{1}+2\INR_{12}+1\right) \\
\nonumber
&\leqslant& \frac{1}{2}\log \left(1+\frac{\overrightarrow{\SNR}_{2}}{\INR_{12}}\right)\\
\nonumber
& & +\frac{1}{2}\log \left(\overrightarrow{\SNR}_{1}+\INR_{12}+1\right)+\frac{1}{2} \\
\nonumber
&\leqslant& \frac{1}{2}\log \left(2+\frac{\overrightarrow{\SNR}_{2}}{\INR_{12}}\right)\\
\label{EqprofGapcHIRsumratebb2}
& & +\frac{1}{2}\log \left(\overrightarrow{\SNR}_{1}+\INR_{12}+1\right)+\frac{1}{2}, 
\end{IEEEeqnarray}
\end{subequations}
where
(h)  follows from the fact that 
\begin{equation}
\label{Eqgapeqh}
\left(\sqrt{\overrightarrow{\SNR}_{2}}-\sqrt{\INR_{21}}\right)^2 \geqslant 0; 
\end{equation}
and (i)  follows from the fact that 
\begin{equation}
\label{Eqgapeqi}
\left(\sqrt{\overrightarrow{\SNR}_{1}}-\sqrt{\INR_{12}}\right)^2 \geqslant 0. 
\end{equation}
From \eqref{Eqdelta2Rproofgap} and  \eqref{EqprofGapcHIRsumratebb}, assuming that $a_{2,1}(0)+a_{1,2} < \min\Big(a_{1,1}+a_{2,2}(0), a_{5,1}(0,0)+a_{5,2}(0,0)\Big)$, it follows that
\begin{IEEEeqnarray}{rcl}
\nonumber
\delta_{2R}^*&  \leqslant & \sup_{\rho' \in [0,1]} \! \left \lbrace \! \min \! \Big( \! \kappa_{4}(\rho'),\kappa_{5}(\rho') \! \Big) \right \rbrace-\Big(\! a_{2,1}(0)+a_{1,2} \! \Big) \\
\nonumber
&\leqslant& \sup_{\rho' \in [0,1]} \kappa_{5}(\rho') -\Big(a_{2,1}(0)+a_{1,2}\Big) \\
\nonumber
&\leqslant& \frac{1}{2}\log \left(2+\frac{\overrightarrow{\SNR}_{2}}{\INR_{12}}\right)\\
\nonumber
& & +\frac{1}{2}\log \left(\overrightarrow{\SNR}_{1}+\INR_{12}+1\right)+\frac{1}{2}\\
\nonumber
& & - \frac{1}{2}\log\left(\overrightarrow{\SNR}_{1}+\INR_{12}+1\right)\\
\nonumber
& & -\frac{1}{2}\log\left(\frac{\overrightarrow{\SNR}_{2}}{\INR_{12}}+2\right)+1 \\
\label{Eqgapsumrate1}
&=& \frac{3}{2}. 
\end{IEEEeqnarray} 
From \eqref{Eqdelta2Rproofgap} and  \eqref{EqprofGapcHIRsumratebb}, assuming that $a_{1,1}+a_{2,2}(0) < \min\Big(a_{2,1}(0)+a_{1,2}, a_{5,1}(0,0)+a_{5,2}(0,0)\Big)$, it follows that
\begin{IEEEeqnarray}{rcl}
\nonumber
\delta_{2R}^*&  \! \leqslant \!  & \! \sup_{\rho' \in [0,1]} \! \left \lbrace \! \min \! \Big(\! \kappa_{4}(\rho'),\kappa_{5}(\rho') \! \Big) \right \rbrace-\Big(\! a_{1,1}+a_{2,2}(0)\! \Big) \\
\nonumber
&\leqslant& \sup_{\rho' \in [0,1]} \kappa_{4}(\rho')-\Big(a_{1,1}+a_{2,2}(0)\Big) \\
\nonumber
&\leqslant& \frac{1}{2}\log \left(2+\frac{\overrightarrow{\SNR}_{1}}{\INR_{21}}\right)\\
\nonumber
& & +\frac{1}{2}\log \left(\overrightarrow{\SNR}_{2}+\INR_{21}+1\right)+\frac{1}{2}\\
\nonumber
& & - \frac{1}{2}\log\left(\overrightarrow{\SNR}_{2}+\INR_{21}+1\right)\\
\nonumber
& & -\frac{1}{2}\log\left(\frac{\overrightarrow{\SNR}_{1}}{\INR_{21}}+2\right)+1 \\
\label{Eqgapsumrate2}
&=& \frac{3}{2}. 
\end{IEEEeqnarray} 
Now,  assume that $a_{5,1}(0,0)+a_{5,2}(0,0) < \min(a_{2,1}(0)+a_{1,2}, a_{1,1}+a_{2,2}(0))$. In this case, the following holds: 
\begin{IEEEeqnarray}{rcl}
\nonumber
\delta_{2R}^*& \leqslant &  \sup_{\rho' \in [0,1]} \left \lbrace \min\Big(\kappa_{4}(\rho'),\kappa_{5}(\rho') \Big) \right \rbrace \\
\label{Eqgapsumrate3p}
& & -  \Big(  a_{5,1}(0,0)  +  a_{5,2}(0,0)  \Big). 
\end{IEEEeqnarray}
To calculate \eqref{Eqgapsumrate3p}, the cases $1.1.1$, $1.1.2$, and $1.1.3$ defined above are analyzed hereunder. 

In Case~$1.1.1$, $a_{5,1}(0,0)+a_{5,2}(0,0)$ is lower bounded as follows:
\begin{IEEEeqnarray}{rcl}
\nonumber
a_{5,1}(0,0) &+& a_{5,2}(0,0) =  \frac{1}{2} \! \log \! \left( \! \frac{\overrightarrow{\SNR}_{1}}{\INR_{21}} \! + \! \INR_{12} \! + \! 1 \! \right) \! \\
\nonumber
& & + \! \frac{1}{2} \! \log \! \left( \! \frac{\overrightarrow{\SNR}_{2}}{\INR_{12}} \! + \! \INR_{21} \! + \! 1\right) \! - \! 1\\
\label{Eqgapv1v21}
&\geqslant& \frac{1}{2}\log\left(\INR_{12}+1\right)-1.
\end{IEEEeqnarray}

From \eqref{EqprofGapcHIRsumratebb}, \eqref{Eqgapsumrate3p}, and \eqref{Eqgapv1v21}, it follows that
\begin{IEEEeqnarray}{rcl}
\nonumber
\delta_{2R}^*&  \leqslant &  \sup_{\rho' \in [0,1]} \left \lbrace \min\Big(\kappa_{4}(\rho'),\kappa_{5}(\rho') \Big) \right \rbrace \\
\nonumber
& & - \! \Big( \! a_{5,1}(0,0) \! + \! a_{5,2}(0,0) \! \Big)  \\
\nonumber
&\leqslant& \sup_{\rho' \in [0,1]} \kappa_{5}(\rho') - \Big(a_{5,1}(0,0)+a_{5,2}(0,0)\Big) \\
\nonumber
&\leqslant& \frac{1}{2} \! \log \! \left( \! 2 \! + \! \frac{\overrightarrow{\SNR}_{2}}{\INR_{12}}\right) \! + \! \frac{1}{2}\log \left(\overrightarrow{\SNR}_{1} \! + \! \INR_{12} \! + \! 1\right) \\
\nonumber
& & +\frac{1}{2}-\frac{1}{2}\log\left(\INR_{12}+1\right)+1 \\
\nonumber
& \leqslant & \frac{1}{2}\log \left(2+1\right)+\frac{1}{2}\log \left(\INR_{12}+\INR_{12}+1\right)\\
\nonumber
& & -\frac{1}{2}\log\left(\INR_{12}+1\right)+\frac{3}{2} \\
\label{Eqgapsumrate3}
& \leqslant & \frac{1}{2}\log \left(3\right)+2.
\end{IEEEeqnarray}

In Case~$1.1.2$, $a_{5,1}(0,0)+a_{5,2}(0,0)$ is lower bounded as follows:
\begin{IEEEeqnarray}{rcl}
\nonumber
a_{5,1}(0,0) \! &+& \! a_{5,2}(0,0) =  \frac{1}{2} \!  \log \! \left( \! \frac{\overrightarrow{\SNR}_{1}}{\INR_{21}} \! + \! \INR_{12} \! + \! 1 \! \right) \! \\
\nonumber
& & +  \frac{1}{2}  \log  \left(  \frac{\overrightarrow{\SNR}_{2}}{\INR_{12}}  +  \INR_{21}  +  1\right)  -  1  \\
\label{Eqgapv1v22}
& \geqslant & \frac{1}{2}\log\left(\INR_{21}+1\right)-1.
\end{IEEEeqnarray}
From \eqref{EqprofGapcHIRsumratebb}, \eqref{Eqgapsumrate3p}, and \eqref{Eqgapv1v22}, it follows that
\begin{IEEEeqnarray}{rcl}
\nonumber
\delta_{2R}^*&  \leqslant &  \sup_{\rho' \in [0,1]} \left \lbrace \min\Big(\kappa_{4}(\rho'),\kappa_{5}(\rho') \Big) \right \rbrace \\
\nonumber
& & - \! \Big( \! a_{5,1}(0,0) \! + \! a_{5,2}(0,0) \! \Big) \\
\nonumber
&\leqslant& \sup_{\rho' \in [0,1]} \kappa_{4}(\rho')-\Big(a_{5,1}(0,0)+a_{5,2}(0,0)\Big) \\
\nonumber
& \leqslant & \frac{1}{2} \! \log \! \left( \! 2 \! + \! \frac{\overrightarrow{\SNR}_{1}}{\INR_{21}}\right) \! + \! \frac{1}{2} \! \log \! \left(\! \overrightarrow{\SNR}_{2} \!+ \! \INR_{21} \!+ \! 1\right)\\
\nonumber
& & +\frac{1}{2}-\frac{1}{2}\log\left(\INR_{21}+1\right)+1 \\
\nonumber
& \leqslant & \frac{1}{2}\log \left(2+1\right)+\frac{1}{2}\log \left(\INR_{21}+\INR_{21}+1\right)\\
\nonumber
& & -\frac{1}{2}\log\left(\INR_{21}+1\right)+\frac{3}{2} \\
\label{Eqgapsumrate4}
& \leqslant & \frac{1}{2}\log \left(3\right)+2.
\end{IEEEeqnarray}

In Case~$1.1.3$, from  \eqref{EqprofGapcHIRsumratebb}, \eqref{Eqgapsumrate3p}, and \eqref{Eqgapv1v21}, it follows that
\begin{IEEEeqnarray}{rcl}
\nonumber
\delta_{2R}^*&  \leqslant & \sup_{\rho' \in [0,1]} \left \lbrace \min\Big(\kappa_{4}(\rho'),\kappa_{5}(\rho') \Big) \right \rbrace \\
\nonumber
& & -  \Big(  a_{5,1}(0,0)  +  a_{5,2}(0,0)  \Big) \\
\nonumber
&\leqslant& \sup_{\rho' \in [0,1]} \kappa_{5}(\rho')-\Big(a_{5,1}(0,0)+a_{5,2}(0,0)\Big) \\
\nonumber
&\leqslant& \frac{1}{2} \! \log \! \left( \! 2 \! + \! \frac{\overrightarrow{\SNR}_{2}}{\INR_{12}}\right) \! + \! \frac{1}{2}\log \left(\overrightarrow{\SNR}_{1} \! + \! \INR_{12} \! + \! 1\right) \\
\nonumber
& & +\frac{1}{2}-\frac{1}{2}\log\left(\INR_{12}+1\right)+1 \\
\nonumber
& \leqslant & \frac{1}{2}\log \left(2+1\right)+\frac{1}{2}\log \left(\INR_{12}+\INR_{12}+1\right) \\
\nonumber
& & -\frac{1}{2}\log\left(\INR_{12}+1\right)+\frac{3}{2}\\
\label{Eqgapsumrate5}
& \leqslant & \frac{1}{2}\log \left(3\right)+2. 
\end{IEEEeqnarray}
Then, from \eqref{Eqgapsumrate1}, \eqref{Eqgapsumrate2}, \eqref{Eqgapsumrate3}, \eqref{Eqgapsumrate4}, and \eqref{Eqgapsumrate5}, it follows that in Case~$1.1$: 
\begin{IEEEeqnarray}{rcl}
\label{EqGapR1R2}
\delta_{2R}* &  \leqslant & 2+\frac{1}{2}\log \left(3\right). 
\end{IEEEeqnarray} 

\item \underline {Calculation of $\delta_{3R_1}$}. From \eqref{Eqdelta3R1} and considering the choice $\rho=0$, $\mu_1=0$, and $\mu_2=0$, it follows that
\begin{IEEEeqnarray} {rcl}
\nonumber
\delta_{3R_1}^* & \leqslant& \! \sup_{\rho' \in [0,1]} \!  \kappa_{7,1}(\rho') \! - \! \Big(a_{1,1} \! + \! a_{7,1}(0,0,0) \! + \! a_{5,2}(0,0) \! \Big). \\
\label{Eqdelta3Rproofgap}
\end{IEEEeqnarray} 

The sum $a_{1,1}+a_{7,1}(0,0,0)+a_{5,2}(0,0)$ is lower bounded as follows:
\begin{IEEEeqnarray}{rcl}
\nonumber
a_{1,1}&+&a_{7,1}(0,0,0)+a_{5,2}(0,0)  =  \frac{1}{2}\log\left(\frac{\overrightarrow{\SNR}_{1}}{\INR_{21}}+2\right)\\
\nonumber
& & +\frac{1}{2}\log\left(\overrightarrow{\SNR}_{1}+\INR_{12}+1\right)\\
\nonumber
& & +\frac{1}{2}\log\left(\frac{\overrightarrow{\SNR}_{2}}{\INR_{12}}+\INR_{21}+1\right)-\frac{3}{2} \\
\nonumber
&\geqslant& \frac{1}{2} \! \log \! \left( \! \frac{\overrightarrow{\SNR}_{1}}{\INR_{21}} \! + \! 2\right) \! + \! \frac{1}{2}\log\left(\overrightarrow{\SNR}_{1} \! + \! \INR_{12} \! + \! 1\right)\\
\label{Eqgap2v1v2}
& & +\frac{1}{2}\log\left(\INR_{21}+1\right)-\frac{3}{2}.
\end{IEEEeqnarray}
When the term $\kappa_{7,1}(\rho')$ is active in the converse region, for a given $\rho' \in [0,1]$, it is upper bounded by the sum $\kappa_{1,1}(\rho')+\kappa_{4}(\rho')$, i.e., the sum of the single rate and sum-rate outer bounds respectively, and this is upper bounded as follows:
\begin{IEEEeqnarray}{rcl}
\nonumber
\kappa_{7,1}(\rho')  & \!  \leqslant  \! & \kappa_{1,1}(\rho')+\kappa_{4}(\rho') \\
\nonumber
& \leqslant & \! \frac{1}{2} \! \log \!  \left( \! \overrightarrow{\SNR}_{1} \! +  \! \INR_{12} \! +  \! 1\right) \! +  \! \frac{1}{2}  \! \log  \! \left( \! 2+  \! \frac{\overrightarrow{\SNR}_{1}}{\INR_{21}}  \! \right)\\
\nonumber
& & +\frac{1}{2}\log \left(\overrightarrow{\SNR}_{2}+\INR_{21}+1\right)+1 \\
\nonumber
& \leqslant & \frac{1}{2} \! \log \!  \left( \! \overrightarrow{\SNR}_{1} \! +  \! \INR_{12} \! +  \! 1\right) \! +  \! \frac{1}{2}  \! \log  \! \left( \! 2+  \! \frac{\overrightarrow{\SNR}_{1}}{\INR_{21}}  \! \right)\\
\nonumber
& & +\frac{1}{2}\log \left(\INR_{21}+\INR_{21}+1\right)+1 \\
\nonumber
& \leqslant & \frac{1}{2} \! \log \!  \left( \! \overrightarrow{\SNR}_{1} \! +  \! \INR_{12} \! +  \! 1\right) \! +  \! \frac{1}{2}  \! \log  \! \left( \! 2+  \! \frac{\overrightarrow{\SNR}_{1}}{\INR_{21}}  \! \right)\\
\label{EqprofGapcHIRweightedsumratebb1}
& & +\frac{1}{2}\log \left(\INR_{21}+1\right)+\frac{3}{2}. 
\end{IEEEeqnarray}

From \eqref{Eqdelta3Rproofgap}, \eqref{Eqgap2v1v2} and  \eqref{EqprofGapcHIRweightedsumratebb1}, it follows that in Case~$1.1$:
\begin{IEEEeqnarray} {rcl}
\nonumber
\delta_{3R_1}^* &  \leqslant &  \frac{1}{2} \! \log \!  \left( \! \overrightarrow{\SNR}_{1} \! +  \! \INR_{12} \! +  \! 1\right) \! +  \! \frac{1}{2}  \! \log  \! \left( \! 2+  \! \frac{\overrightarrow{\SNR}_{1}}{\INR_{21}}  \! \right)\\
\nonumber
& & +\frac{1}{2}\log \left(\INR_{21} \! + \! 1\right) \! + \! \frac{3}{2} \! - \!  \frac{1}{2}\log\left(\frac{\overrightarrow{\SNR}_{1}}{\INR_{21}} \! + \! 2\right)\\
\nonumber
& & -\frac{1}{2} \! \log \! \left(\overrightarrow{\SNR}_{1} \! + \! \INR_{12} \! + \! 1\right) \! - \! \frac{1}{2} \! \log \! \left(\INR_{21} \! + \! 1\right)\\
\nonumber
& & +\frac{3}{2} \\
\label{Eqgapweightedsumrate}
&=& 3. 
\end{IEEEeqnarray} 
The same procedure holds in the calculation of $\delta_{3R_2}^*$ and it yields:
\begin{IEEEeqnarray} {rcl}
\label{Eqgapweightedsumrate2}
\delta_{3R_2}^* &  \leqslant & 3. 
\end{IEEEeqnarray} 
Therefore, in Case~$1.1$, from \eqref{Eqdelta}, \eqref{EqGapR2}, \eqref{EqGapR1}, \eqref{EqGapR1R2}, \eqref{Eqgapweightedsumrate} and \eqref{Eqgapweightedsumrate2} it follows that
\begin{IEEEeqnarray} {rcl}
\label{EqdeltaHIRfinal}
\delta& = & \sup\left(\delta_{R_1}^*,\delta_{R_2}^*,\frac{\delta_{2R}^*}{2},\frac{\delta_{3R_1}^*}{3}, \frac{\delta_{3R_2}^*}{3}\right) \leqslant \frac{3}{2}.
\end{IEEEeqnarray}      
\end{enumerate}
This completes the calculation of the gap in Case $1.1$. Applying the same procedure to all the other cases listed above yields that $\delta \leqslant 4.4$ bits. 
\end{appendices}

\balance
\bibliographystyle{IEEEtran}
\bibliography{IT-GT}

\balance

\end{document}